\documentclass[iop, numberedappendix]{emulateapj}

\usepackage{natbib}
\usepackage{multirow}
\usepackage{multirow}
\usepackage{gensymb} 
\usepackage[flushleft]{threeparttable} 
\usepackage{amsmath}
\usepackage{longtable}
\usepackage{threeparttablex} 

\shorttitle{CANDELS EGS photometric catalog}
\shortauthors{Stefanon et al.}

\begin{document}

\title{CANDELS Multiwavelength Catalogs: Source Identification and Photometry in the CANDELS Extended Groth Strip}

\author{Mauro Stefanon\altaffilmark{1,2}, Haojing Yan\altaffilmark{2}, Bahram Mobasher\altaffilmark{3}, Guillermo Barro\altaffilmark{4,5}, Jennifer L. Donley\altaffilmark{6}, Adriano Fontana\altaffilmark{7}, Shoubaneh Hemmati\altaffilmark{8},  Anton M. Koekemoer\altaffilmark{9}, BoMee Lee\altaffilmark{10}, Seong-Kook Lee\altaffilmark{11}, Hooshang Nayyeri\altaffilmark{3,12}, Michael Peth\altaffilmark{13}, Janine Pforr\altaffilmark{14}, Mara Salvato\altaffilmark{15}, Tommy Wiklind\altaffilmark{16}, Stijn Wuyts\altaffilmark{17},
Matthew L. N. Ashby\altaffilmark{18},  Marco Castellano\altaffilmark{7}, Christopher J. Conselice\altaffilmark{19}, Michael C. Cooper\altaffilmark{12},  Asantha R. Cooray\altaffilmark{12},  Timothy Dolch\altaffilmark{20}, Henry Ferguson\altaffilmark{9}, Audrey Galametz\altaffilmark{7,15}, Mauro Giavalisco\altaffilmark{10},  Yicheng Guo\altaffilmark{4}, Steven P. Willner\altaffilmark{18}, 
Mark E. Dickinson\altaffilmark{21}, Sandra M. Faber\altaffilmark{4}, Giovanni G. Fazio\altaffilmark{18},  Jonathan P. Gardner\altaffilmark{22}, Eric Gawiser\altaffilmark{23}, Andrea Grazian\altaffilmark{7}, Norman A. Grogin\altaffilmark{9},  Dale Kocevski\altaffilmark{24}, David C. Koo\altaffilmark{4}, Kyoung-Soo Lee\altaffilmark{25}, Ray A. Lucas\altaffilmark{9}, Elizabeth J. McGrath\altaffilmark{24},  Kirpal Nandra\altaffilmark{15,26}, Jeffrey A. Newman\altaffilmark{27}, Arjen van der Wel\altaffilmark{28}}

\email{Email: stefanon@strw.leidenuniv.nl}

\altaffiltext{1}{Leiden Observatory, Leiden University, NL-2300 RA Leiden, Netherlands}
\altaffiltext{2}{Department of Physics and Astronomy, University of Missouri, Columbia, MO 65211, USA}
\altaffiltext{3}{Department of Physics and Astronomy, University of California, Riverside, CA 92521, USA}
\altaffiltext{4}{UCO/Lick Observatory, Department of Astronomy and Astrophysics, University of California, Santa Cruz, CA 95064, USA}
\altaffiltext{5}{Department of Astronomy, University of California at Berkeley, Berkeley, CA 94720-3411, USA}
\altaffiltext{6}{Los Alamos National Laboratory, Los Alamos, NM 87544 USA}
\altaffiltext{7}{INAF - Osservatorio Astronomico di Roma, via Frascati 33, I-00040 Monte Porzio Catone (RM), Italy}
\altaffiltext{8}{Infrared Processing and Analysis Center, California Institute of Technology, MS 100-22, Pasadena, CA 91125}
\altaffiltext{9}{Space Telescope Science Institute, 3700 San Martin Drive, Baltimore, MD 21218, USA}
\altaffiltext{10}{Department of Astronomy, University of Massachusetts, 710 North Plesant Street, Amherst, MA 01003, USA}
\altaffiltext{11}{Center for the Exploration of the Origin of the Universe (CEOU),  Department of Physics and Astronomy, Seoul National University,   Seoul, Korea}
\altaffiltext{12}{Department of Physics and Astronomy, University of California, Irvine, CA 92697, USA}
\altaffiltext{13}{Department of Physics and Astronomy, The Johns Hopkins University, 366 Bloomberg Center, Baltimore, MD 21218, USA}
\altaffiltext{14}{Aix Marseille Universit\'e, CNRS, LAM (Laboratoire d'Astrophysique de Marseille) UMR 7326, 13388, Marseille, France}
\altaffiltext{15}{Max-Planck-Institut fur Extraterrestrische Physik, Giessenbachstrasse 1, D-85748 Garching bei Munchen, Germany}
\altaffiltext{16}{Department of Physics, Catholic University of America, Washington DC 20064.}
\altaffiltext{17}{Department of Physics, University of Bath, Claverton Down, Bath, BA2 7AY, UK}
\altaffiltext{18}{Harvard Smithsonian Center for Astrophysics, 60 Garden Street, MS-66, Cambridge, MA 02138-1516, USA}
\altaffiltext{19}{School of Physics and Astronomy, University of Nottingham, Nottingham, UK}
\altaffiltext{20}{Deptartment of Physics, Hillsdale College, 33 E. College St., Hillsdale, MI 49242, USA}
\altaffiltext{21}{National Optical Astronomy Observatories, 950 N Cherry Avenue, Tucson, AZ 85719, USA}
\altaffiltext{22}{NASA's Goddard Space Flight Center, Greenbelt , MD 20771, USA}
\altaffiltext{23}{Department of Physics and Astronomy, Rutgers, The State University of New Jersey, 136 Frelinghuysen Road, Piscataway, NJ 08854, USA}
\altaffiltext{24}{Department of Physics and Astronomy, Colby College, Waterville, Maine 04901, USA}
\altaffiltext{25}{Department of Physics, Purdue University, 525 Northwestern Avenue, West Lafayette, USA}
\altaffiltext{26}{Astrophysics Group, Blackett Laboratory, Imperial College London, London SW7 2AZ, UK}
\altaffiltext{27}{Department of Physics and Astronomy and PITT PACC, University of Pittsburgh, Pittsburgh, PA 15260, USA}
\altaffiltext{28}{Max-Planck Institute for Astronomy, D-69117 Heidelberg, Germany}

\begin{abstract}
 
We present a 0.4-8$\mu$m multi-wavelength photometric catalog in the Extended Groth Strip (EGS) field. This catalog is built on the \textit{Hubble Space Telescope (HST)} WFC3 and ACS data from the Cosmic Assembly Near-infrared Deep Extragalactic Legacy Survey (CANDELS), and it incorporates the existing \textit{HST} data from the All-wavelength Extended Groth strip International Survey (AEGIS) and the 3D-HST program. The catalog is based on detections in the F160W band  reaching a depth of F160W=26.62 AB (90\% completeness, point-sources). It includes the photometry for 41457 objects over an area of $\approx 206$ arcmin$^2$ in the following bands: \textit{HST} ACS F606W and F814W; \textit{HST} WFC3 F125W, F140W and F160W; CFHT/Megacam $u^*$, $g'$, $r'$, $i'$ and $z'$; CFHT/WIRCAM $J$, $H$ and $K_\mathrm{S}$; Mayall/NEWFIRM $J1$, $J2$, $J3$, $H1$, $H2$, $K$; \textit{Spitzer} IRAC $3.6\mu$m, $4.5\mu$m, $5.8\mu$m and $8.0\mu$m.  We are also releasing value-added catalogs that provide robust photometric redshifts and stellar mass measurements. The catalogs are publicly available through the CANDELS repository.

\end{abstract}


\section{Introduction}

The Cosmic Assembly Near-infrared Deep Extragalactic Legacy Survey (CANDELS; \citealt{grogin2011,koekemoer2011}) is a 902-orbit Hubble Space Telescope (\textit{HST}) Multi-Cycle Treasury (MCT) program aimed at obtaining deep multi-wavelength photometric data for more than quarter million objects. Observations followed a two-tiered strategy and were distributed over five fields. 

The deeper layer of our survey (CANDELS Deep)  covers $\sim130$ arcmin$^2$ to a depth of $27.6-29.4$ AB (5$\sigma$ limit for point sources) over the North and South fields of the Great Observatories Origins Deep Survey (\citealt{giavalisco2004}). Its photometric depth has already been shown capable of reaching $0.5L^*$ galaxies at $z\gtrsim 8$ (see e.g., \citealt{finkelstein2014,bouwens2015}) and also to cover the low-mass end of the galaxy population at $z\sim2-5$ (e.g., \citealt{grazian2015,duncan2014,mortlock2015,tomczak2014}). The shallower layer of the survey (CANDELS Wide) extends the coverage to a total of $\sim720$ arcmin$^2$ down to a depth of $26.9-28.9$ AB in five fields, namely GOODS-N, GOODS-S,  the Extended Groth Strip (EGS; \citealt{davis2007}), the Cosmic Evolution Survey field (COSMOS; \citealt{scoville2007}), and the Ultra Deep Survey field (UDS; \citealt{lawrence2007, cirasuolo2007}). Furthermore, the combination of photometric depth and covered area allow detection of potential luminous  high-$z$ galaxies, primary targets for follow-up observations with e.g., \textit{HST}, Atacama Large Millimeter/submillimeter Array (ALMA) and \textit{James Webb Space Telescope (JWST)} (see e.g., \citealt{yan2012,oesch2015}).

The multi-wavelength photometric catalogs for the GOODS-S and UDS fields have already been presented by \citet{guo2013}  and \citet{galametz2013}, respectively; photometric redshifts and stellar population parameters for these two fields are described by \citet{dahlen2013} and \citet{santini2015}, respectively. The CANDELS GOODS-S data, combined to the medium bands from Subaru \citep{cardamone2010}, were also used in \citet{hsu2014} to compute photometric redshifts of normal galaxies and AGN. The description of  the multi-wavelength catalogs, photometric redshifts and stellar population parameters for the COSMOS and GOODS-N fields are presented by \citet{nayyeri2015} and \citet{barro2015}, respectively.
This paper presents the CANDELS  multi-wavelength photometric catalog for the EGS field. Companion papers will present the rest-frame luminosities (\citealt{kocevski2016} in prep.) and the probability distribution functions of photometric redshifts (\citealt{kodra2016} in prep.). Since the first observations (\citealt{rhodes2000}, PI Groth 1994), the EGS has been the site of several surveys, notably the All-wavelength Extended Groth Strip International Survey (AEGIS, \citealt{davis2007}) and the DEEP2+3 spectroscopic survey \citep{coil2004,cooper2006,newman2013,cooper2011,cooper2012}. The catalog is complemented by measurements of the photometric redshifts and stellar population parameters.

This paper is organised as follows. Section 2 describes the datasets used to construct our multi-wavelength catalog. Section 3 discusses the World Coordinate System (WCS) for the \textit{HST} mosaics. Section 4 describes the procedures for the flux measurements over the wide wavelength range, while Section 5 presents the validation tests of the multi-wavelength catalog. Section 6 presents the photometric redshift and stellar mass measurements. Our results are summarised in Section 7. Our catalogs are accessible on the primary CANDELS pages at MAST\footnote{https://archive.stsci.edu/prepds/candels}, through the Vizier service\footnote{http://vizier.u-strasbg.fr/viz-bin/VizieR}, and from the CANDELS team project website\footnote{http://candels.ucolick.org}.

All magnitudes are given in the AB system. We adopted a standard cosmology with $H_0=70$km s$^{-1}$ Mpc$^{-1}$, $\Omega_\Lambda=0.7$, $\Omega_m=0.3$.

\begin{figure*}
\includegraphics[width=18cm]{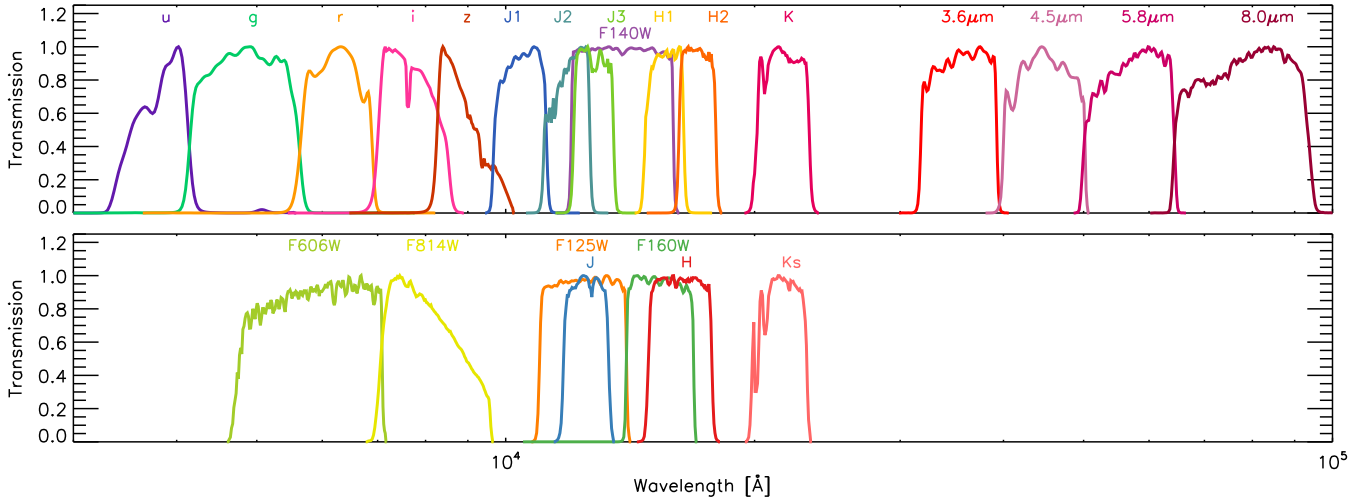}
\caption{Response curves of the 22 bands included in the CANDELS EGS multi-wavelength catalog normalized to a maximum value of unity. The response curves are organised over two vertically stacked panels for ease of presentation. The responses correspond to the filter transmission combined with the detector quantum efficiency as well as atmospheric transmission in the case of ground-based instruments.\label{fig:filters}}
\end{figure*}

\begin{figure*}
\hspace{-1cm}\hspace{2cm}\includegraphics[width=16cm]{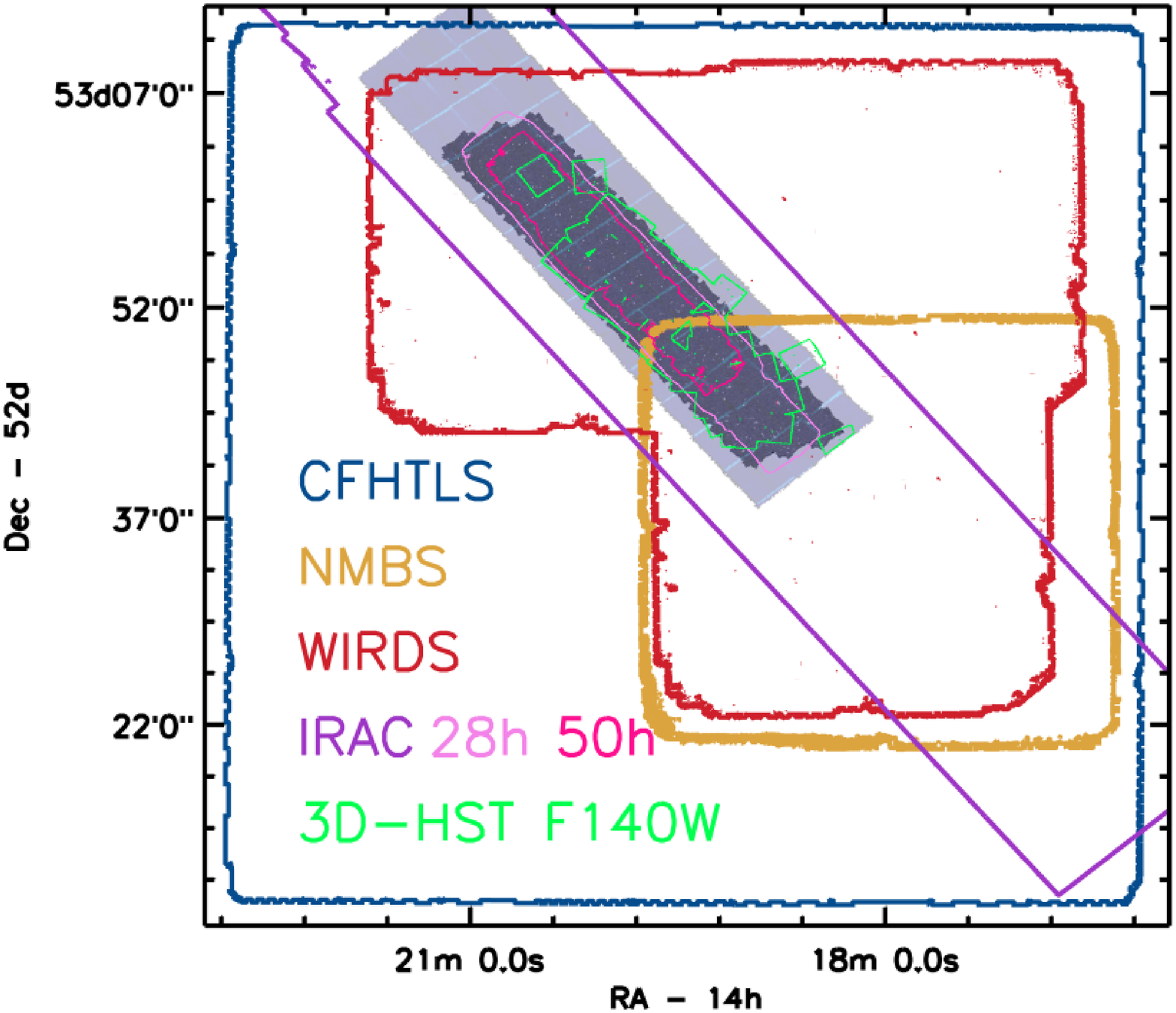}
\caption{Coverage map of the different data sets used in this work. The dark filled area shows the coverage of the \textit{HST} WFC3 F160W band from CANDELS, which has been adopted as the detection band. This area amounts to $\sim200$ arcmin$^2$. The light-blue filled region reproduces the ACS F606W  exposure map from the AEGIS project \citep{davis2007}, which fully covers the F160W map. The other data sets are also outlined: CFHTLS (optical broad-band $u^*$-to-$z'$, blue contour), WIRDS (\citealt{bielby2012}; NIR J,H and Ks bands, red contour), S-CANDELS (\citealt{ashby2015}; dark purple: full coverage of the IRAC 3.6-to-8.0$\mu$m mosaics; pink: 28 hours depth; magenta: 50 hours depth coverages),  NMBS (\citealt{whitaker2011}; yellow contour), and 3D-HST (\citealt{skelton2014}; green contour). \label{fig:coverage}}
\end{figure*}

\section{Data}

The CANDELS EGS field is centered at $\alpha(J2000) = 14h17m00s$ and $\delta(J2000) = +52\degree30' 00''$, corresponding to high Galactic latitudes ($b\sim 60\degree$). The AEGIS project provided deep \textit{HST} ACS imaging data in the F606W and F814W bands to 5$\sigma$ depths of 28.7 and 28.1 mag, respectively. The CANDELS survey adds deep \textit{HST} WFC3 F125W and F160W coverage, and increases the depth of the ACS F606W and F814W mosaics with new data from parallel observations (\citealt{grogin2011,koekemoer2011}). 

The wavelength coverage is complemented by extensive ground- and space-based imaging: UV data from GALEX (PI C. Martin); $u^*,g',r',i,z'$ from CFHT/MegaCam observations;  B, R, and I-band imaging obtained with the CFHT 12K mosaic camera (\citealt{cuillandre2001,coil2004}); near infrared (NIR) broad band $J$, $H$ and $K_\mathrm{S}$ imaging  from the WIRCam Deep Survey (WIRDS;  \citealt{bielby2012}); $J$ and $K$-band data from the Palomar Observatory Wide-Field Infrared Survey \citep{conselice2008}; medium-band NIR filters from the NEWFIRM camera (\citealt{vandokkum2009});  \textit{HST} WFC3 F140W data from the 3D-HST survey (PI van Dokkum; \citealt{skelton2014}); IRAC  maps from \textit{Spitzer} SEDS (\citealt{ashby2013}) and S-CANDELS \citep{ashby2015}; MIPS data from FIDEL (PI: M. Dickinson);  Herschel  PACS 100$\mu$m and 160$\mu$m (\citealt{lutz2011}) and HerMES SPIRE 250$\mu$m, 350$\mu$m, 500$\mu$m (\citealt{oliver2010}) and 1.4 and 4.8 GHz VLA data (\citealt{willner2006,ivison2007}). 

The field also benefits from substantial integration time  ($\sim800$ks) in X-ray by Chandra (\citealt{laird2009,nandra2015})\footnote{See also http://www.mpe.mpg.de/XraySurveys/AEGIS-X/}. The area covered by the CANDELS EGS footprint also has spectroscopic coverage from the DEEP2 survey (\citealt{coil2004,cooper2006,newman2013}). Although spectroscopically incomplete, this survey provides redshift information for $\sim5\times10^4$  objects brighter than $R<24.1$~AB (corresponding to approximately $z<1.4$), $\sim1400$ of which fall within the CANDELS EGS boundaries. The recently concluded DEEP3 survey (\citealt{cooper2011,cooper2012}) increases the number of spectrscopic redshifts in the CANDELS EGS field to a total of  $\sim 2200$. Furthermore, the 3D-HST project provides \textit{HST} G141 slitless grism spectroscopy over the F140W footprint.

The multi-wavelength photometric catalog presented in this work is assembled using data spanning from 0.4 to 8$\mu$m, taken by six different instruments. Specifically, the CANDELS EGS photometric catalog is built using the data from CFHT/MegaCam, NEWFIRM/NMBS, CFHT/WIRCAM, HST/ACS, HST/WFC3 and Spitzer/IRAC. Figure  \ref{fig:filters} presents the response curves of the involved passbands, which are the convolution products of the filter transmissions and the detector throughputs. In case of the ground-based instruments, the atmospheric transmissions are also convolved. Table \ref{tab:photometric_data} summarizes the main properties of each datatset.  Similar to all the rest of the CANDELS fields (\citealt{guo2013,galametz2013,nayyeri2015,barro2015}), object detection was performed on the WFC3/F160W-band image. This mosaic is fully covered by all the adopted data sets, with the exception of the data from the NEWFIRM Medium-Band Survey (NMBS) and the WFC3 F140W mosaic which cover only about 1/3 and 2/3 of the F160W field respectively (see Figure \ref{fig:coverage}).

\begin{table*}
\caption{Summary of photometric data \label{tab:photometric_data}}
  \begin{threeparttable}
\begin{tabular}{cccccccc}
\hline
\hline
Filter & Filter & Filter & PSF &  Depth\tnote{a}  & ZP\tnote{b} & Astrometric & Reference  \\
Name & $\lambda_{\mathrm{eff}}$ &  FWHM & FWHM  & 5$\sigma$ & & System &   \\
           & (\AA) &  (\AA) & (arcsec)  &  (AB) & (AB) &  & \\

\hline
CFHT/MegaCam $u*$       & 383   &  61 & 0.95 & 27.1 & 30.0 & NOMAD+SDSS & \citet{gwyn2012}\\
CFHT/MegaCam $g'$       & 489   &  144 & 0.90 & 27.3 & 30.0 & ... & ... \\
CFHT/MegaCam $r'$       & 625   &  122 & 0.77 & 27.2 & 30.0 & ... & ... \\
CFHT/MegaCam $i'$       & 769   &  138 & 0.71 & 27.0 & 30.0 & ... & ... \\
CFHT/MegaCam $z'$       & 888   &  87 & 0.71 & 26.1 & 30.0 & ... & ... \\

Mayall/NEWFIRM J1              & 1047   &  150 & 1.13 & 24.4 & 23.31 & NOMAD+SDSS\tnote{c} & \citet{whitaker2011}\\
Mayall/NEWFIRM J2              & 1195   &  151 & 1.16 & 24.1 & 23.35 & ... & ... \\
Mayall/NEWFIRM J3              & 1279   &  140 & 1.08 & 24.0 & 23.37 & ... & ... \\
Mayall/NEWFIRM H1              & 1561   &  169 & 1.10 & 23.6 & 23.59 & ... & ... \\
Mayall/NEWFIRM H2              & 1707   &  176 & 1.06 & 23.6 & 23.61 & ... & ... \\
Mayall/NEWFIRM K               & 2170   &  307 & 1.08  & 23.5 & 23.85 & ... & ... \\

CFHT/WIRCAM $J$         & 1254   &  157 & 0.72 & 24.4 & 30.0 & 2MASS & \cite{bielby2012}\\
CFHT/WIRCAM $H$         & 1636   &  287 & 0.68 & 24.5 & 30.0 & ... & ... \\
CFHT/WIRCAM $K_s$       & 2159   &  326 & 0.65 & 24.3 & 30.0 & ... & ... \\

HST/ACS F606W           & 596   &  231 & 0.12 & 28.8 & 26.491 & USNOB1.0\tnote{d} & \citet{koekemoer2011}\\
HST/ACS F814W           & 809   &  189 & 0.12 & 28.2 & 25.943 & ... & ... \\
HST/WFC3 F125W          & 1250   &  301 & 0.19 & 27.6 & 26.250 & ... & ... \\
HST/WFC3 F140W          & 1397   &  395 & 0.19 & 26.8 & 26.465 & ... & \citet{skelton2014},\citet{brammer2012}\\
HST/WFC3 F160W          & 1542   &  288 & 0.20 & 27.6 & 25.960 & ... & \citet{koekemoer2011}\\

Spitzer/IRAC $3.6\mu$m  & 3563   &  744 & 1.80 & 23.9 & 21.581 & 2MASS & \citet{ashby2015}\\
Spitzer/IRAC $4.5\mu$m  & 4511   &  1010 & 1.82 & 24.2 & 21.581 & ... & ...\\
Spitzer/IRAC $5.8\mu$m  & 5759   &  1407 & 1.94 & 22.5 & 21.581 & ... & \citet{barmby2008}\\
Spitzer/IRAC $8.0\mu$m  & 7959   &  2877 & 2.23 & 22.8 & 21.581 & ... & ... \\
\hline
\end{tabular}

\begin{tablenotes}
\item {\bf Notes:}
\item[a] The $5\sigma$-depths correspond to $5\times$ the standard deviation of flux measurements in $\sim5000$ circular apertures, with diameter of $2\times$FWHM of the PSF, randomly placed across each image in the regions that are free of detected objects.
\item[b] Photometric zero-point.
\item[c] The astrometric system was calibrated on the CFHTLS mosaics.
\item[d] This is the original astrometric system adopted for calibration in \citet{lotz2008}. See Sect. \ref{sect:astrometry} for further details.
\end{tablenotes}

\end{threeparttable}
\end{table*}

\subsection{HST}
\label{sect:data_hst}

The \textit{HST} data-set consists of both optical and NIR images, which were taken by
the ACS/WFC and the WFC3/IR instruments, respectively. The WFC3 images were
mostly obtained by the CANDELS program in Cycle 18 and 20. The ACS images consist of those  obtained in the coordinated parallel mode during the CANDELS WFC3 observations and those from the AEGIS program obtained in Cycle 13 (\citealt{davis2007}). A detailed description of \textit{HST} data acquisition and reduction is presented by \citet{grogin2011} and \citet{koekemoer2011}. We briefly summarize the main features below.

  The AEGIS ACS data were taken in a mosaic pattern of contiguous tiles,
covering an effective area of $10\farcm1\times70\farcm5$ in the F606W and the F814W bands. The
major axis of the rectangular area has a position angle of $40\degree$. 
The nominal exposure time was one orbit ($\sim$ 2000 seconds) per filter. 

The CANDELS WFC3 observations were performed within the AEGIS ACS footprint, using
a rectangular grid of $3\times15$ tiles and forming a contiguous field of $6\farcm7\times30\farcm6$
at a position angle of $42\degree$. Each tile was observed in two epochs, 
separated by roughly 52 days and at different orientations (by $22\degree$).
During each epoch a given tile nominally received one
orbit of observing time ($\sim2000$ seconds)\footnote{The integration time in the WFC3 IR bands for nine of the tiles was reduced 
by 410 seconds in each orbit due to the WFC3 UVIS observations for the supernovae search;
see \citet{grogin2011} for details.}.
 Each orbit was shared between the F160W and
the F125W filters such that $\sim2/3$ of the orbit was assigned to the former and
$\sim1/3$ to the latter, respectively. The observation in each filter within
a given orbit was always divided into two sub-exposures. With the existing AEGIS
ACS data in mind, the contemporaneous CANDELS
ACS parallel observations were taken in the F814W band during the first
epoch and split evenly between the F606W and the F814W bands during the second
epoch. As the ACS/WFC has a factor of $1.55\times$ larger field of view than the
WFC3/IR, the CANDELS ACS tiles overlapped heavily due to the abutting WFC3/IR
tiling, and hence effectively resulted in 2-orbit and 1-orbit of exposures per
pixel in F814W and F606W, respectively.

The WFC3 and ACS data were all reduced and combined following the approaches described by  \citet{koekemoer2011}. 
  The image mosaics used in this work have a scale of
60mas/pixel and all have the same world coordinate system. The WFC3 IR mosaics have 
nominal exposure times of  $\sim1300$ and $\sim2700$ seconds in F125W and F160W, 
respectively. The ACS mosaics incorporate the contemporaneous CANDELS data
and the earlier AEGIS data and have nominal exposure times of $\sim$ 6000 and 12,000
seconds in F606W and F814W, respectively, reaching
AB=28.8 and 28.2 mag ($5\sigma$ in apertures of $0\farcs24$ diameter, corresponding to $2\times$ the full-width-half-maximum (FWHM) of the point-spread function (PSF)).

We also incorporated the WFC3 F140W images from the 3D-HST
program (\citealt{vandokkum2011,brammer2012,skelton2014}). 
The 3D-HST survey was an \textit{HST} Treasury program that offered spectroscopic and photometric data in the CANDELS fields over a combined area of $\sim 625$ arcmin$^2$. 3D-HST obtained imaging in the WFC3 F140W filter as well as spatially resolved spectroscopy with the WFC3 G141 grism. In the EGS field, the F140W imaging covered $\sim 2/3$ of the CANDELS F160W footprint (see Figure \ref{fig:coverage}) and reached average 5$\sigma$ depth of 25.8 mag (in a $1''$-diameter aperture). We used the v4.1 F140W mosaic produced by the 3D-HST team\footnote{http://3dhst.research.yale.edu/Data.php},
which has the same WCS as the CANDELS WFC3 mosaics (see also Table 5 in \citealt{koekemoer2011}).

\subsection{Ground-based data}
\label{sect:data_ground}

\subsubsection{CFHTLS}

The catalog incorporates flux measurements from the broad-band $u^*$, $g'$, $r'$, $i'$, and $z'$ images obtained by the Megacam instrument at the 3.6m Canada-France-Hawaii Telescope (CFHT). These data correspond to the D3 field of the Deep component of the CFHT Legacy Survey (CFHTLS); the field is 1 deg$^2$ in size and completely covers the CANDELS EGS field.

For this work, we adopted the mosaics  generated by the MegaPipe pipeline \citep{gwyn2008} as described by \citet{gwyn2012}. The sensitivity limits corresponding to the 50\% completeness are $u^* = 27.5, g' = 27.9, r' = 27.6, i' = 27.3$, and $z' = 26.4$ mag, for the five bands respectively  (\citealt{gwyn2012}). They were obtained by adding artificial point sources to the image after replacing the pixels of all the detected objects with a realisation of noise and recovering sources using SExtractor \citep{bertin1996}.  The image quality is  homogeneous across the bands, with the PSF FWHM average values of $\sim 0\farcs7-0\farcs9$.

The astrometric calibration was based on the Naval Observatory Merged Astrometric Dataset (NOMAD\footnote{http://www.nofs.navy.mil/nomad/}), and was refined using the Sloan Digital Sky Survey (SDSS) DR7 catalog (\citealt{abazajian2009}). The final internal and external astrometric uncertainties are $0\farcs02$ and $0\farcs07$, respectively \citep{gwyn2012}.

\subsubsection{WIRCam Deep Survey}

Deep broad-band J, H and Ks images from the WIRCam Deep Survey (WIRDS; \citealt{bielby2012}) complement the \textit{HST} WFC3 data. The images were obtained with the WIRCam instrument at the CFHT under good seeing conditions (FWHM$\sim0\farcs6$), and they cover 0.4 square degrees centered on the EGS.  The 50\% completeness limit for point sources ranges between 24.6-24.8 mag. The astrometry was calibrated using the Two Micron All Sky Survey  (2MASS) catalog (\citealt{skrutskie2006}) with a final internal accuracy of $\sim 0\farcs1$ (\citealt{bielby2012}).

\subsubsection{NEWFIRM Medium Band Survey}

The overlap of the CANDELS WFC3/F160W-band image with the mosaics of the NEWFIRM Medium Band Survey (NMBS, \citealt{vandokkum2009,whitaker2011}) amounts to $\sim30\%$ in the south-eastern region of the CANDELS footprint (see Figure \ref{fig:coverage}). The CANDELS EGS multi-wavelength catalog incorporates these data. The data were taken with the NEWFIRM camera mounted on the Mayall 4m telescope at Kitt Peak. This NIR imaging survey used one traditional $K_s$ filter and five medium-band filters in place of the usual $J$ and $H$ bands. Specifically, the $J$ band was split 
into three filters $J1$, $J2$, and $J3$, and the $H$ band was split into two filters, $H1$ and  $H2$ (\citealt{vandokkum2009}; see also Figure \ref{fig:filters}). The average seeing FWHM ranged between $1\farcs06$ to $1\farcs16$, and the photometric depth reaches AB=23.5-24.4~mag (5$\sigma$ in $2\times$FWHM apertures), with a 50\% point-source completeness at $K_{s}$=23.6~mag. The NMBS mosaics were aligned to the CFHTLS $i'$-band images with an astrometric precision of $\sim0\farcs1$ over the entire field of view.

\subsection{Spitzer}

We also included the IR data from the \textit{Spitzer} InfraRed Array Camera (IRAC; \citealt{fazio2004}), whose four bands center at 3.6, 4.5, 5.8, and 8.0$\mu$m.

The $3.6\mu$m- and $4.5\mu$m-band mosaics were obtained by combining four different image sets, two from the cryogenic mission and two from the warm mission phases. The first cryogenic data imaged a narrow $2.3\degree \times 0.29\degree$ strip overlapping with the AEGIS \textit{HST} ACS imaging (PID 8, \citealt{barmby2008}). The depth and width of the central $\sim 1\degree$ portion were later increased  (PID 41023, PI: Nandra), providing better overlap with the deep (800 ks) X-ray imaging by Chandra. These 3.6$\mu$m and 4.5$\mu$m data were combined with those from the warm mission phase, namely, those from the \textit{Spitzer} Extended Deep Survey (SEDS, PID 61042; \citealt{ashby2013}) and \textit{Spitzer}-CANDELS (S-CANDELS, PID 80216; \citealt{ashby2015}). The resultant 3.6$\mu$m and 4.5$\mu$m mosaics cover most of the WFC3 F160W area to a depth of at least 50 hours and the rest to at least 28 hours in each band (see Figure \ref{fig:coverage}; we refer to \citealt{ashby2015} for full details). The $5.8\mu$m and $8.0\mu$m mosaics, on the other hand, were made from the cryogenic data of the AEGIS project (\citealt{barmby2008}) and those of PID 41023 (PI Nandra).
All these IRAC data were reprocessed and mosaicked using the same CANDELS \textit{HST} tangent-plane projection and with a pixel scale of $0\farcs6$/pixels, to prepare them appropriately for further photometric analysis (see also \citealt{ashby2015}). The internal astrometry was checked by crossmatching approximately 500 point sources from the 2MASS catalog to the detections in the 3.6$\mu$m and 4.5$\mu$m mosaics. The 1$\sigma$ dispersions in RA and Dec are $\lesssim 0\farcs2$ in both bands.

\subsection{Value-added Data}

The CANDELS EGS field has a large number of spectroscopic redshifts and deep Chandra X-ray data. While they are not a part of the photometry, these data are included in our catalog (see Appendix \ref{appendix:photcat} and \ref{appendix:mstar}).

\begin{figure*}
\includegraphics[width=18cm]{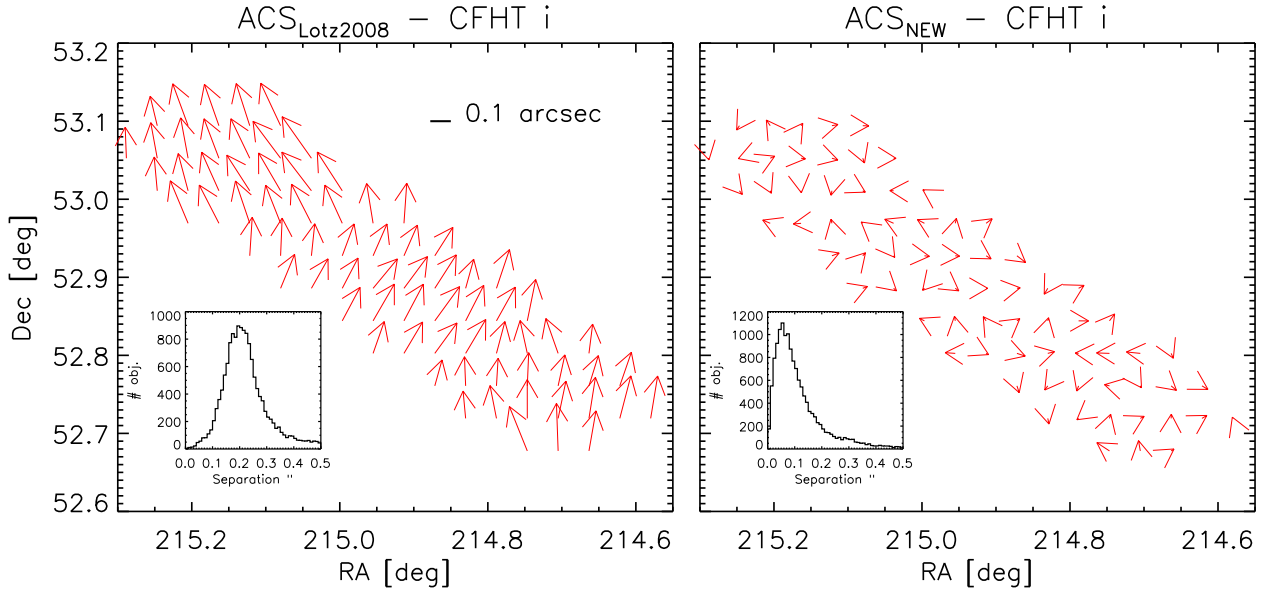}
\caption{{\bf Left panel:} Offsets in RA and Dec positions between the previous CANDELS astrometric system (inherited from the AEGIS system) and the CFHTLS i-band system. Each arrow represents the average of the offsets measured from the matched objects within the rectangle region ($1\farcm5 \times 1\farcm5$  in size) centered on the arrow tail. {\bf Right panel:} Similar to the left panel, but after correcting the CANDELS system to the CFHTLS system. The bands used for the recalibration are the ACS F814W and the CFHTLS iÕ-band. In both panels, the insets show the distribution of the offsets (bin size $0\farcs01$): the median offset of the old astrometric calibration is $0\farcs2$, while that of the offsets after applying the new astrometric solution is $0\farcs04$. \label{fig:astrometry}}
\end{figure*}

\subsubsection{Spectroscopy}

We included the spectroscopic redshifts from the DEEP2+3 Surveys (\citealt{coil2004,willner2006,cooper2011,cooper2012,newman2013})\footnote{http://deep.ps.uci.edu/DR4/home.html and  http://deep.ps.uci.edu/deep3/ for DEEP2 and DEEP3, respectively.}, which were  spectroscopic surveys targeting galaxies brighter than R=24.1 mag. Observations were carried out using the DEIMOS spectrograph on Keck 2 with a resolution of $\lambda/\Delta\lambda\sim 5000 $  at the central wavelength of 7800\AA. In the EGS field, the campaign covered a $\sim 30'\times 120'$ strip centered on the AEGIS ACS mosaic. Within the CANDELS F160W mosaic coverage, there are 2132 unique DEEP2+3 objects that have secure redshifts (quality parameter \texttt{Q}$\geq 3$) within a matching radius of $0\farcs8$ (if more than one object falls within the matching radius, the closest match was adopted).

\subsubsection{X-ray data}

The EGS field was initially observed by Chandra/ACIS with 200ks integration time as part of the AEGIS project (AEGIS-X Wide, \citealt{nandra2005,laird2009}). The observations covered the full AEGIS ACS mosaic with eight pointings, $\sim17'\times17'$ each, for a total area of $\sim 0.7$~deg$^2$. Recently, new Chandra/ACIS data to a nominal depth of 600ks were acquired over a region of approximately 0.29 deg$^2$ centered on three central tiles of the AEGIS-X Wide coverage. These data have been combined with the previous 200ks Chandra/ACIS observations to provide a cumulative depth of 800ks in the three central ACIS fields (AEGIS-X Deep or AEGIS-XD, \citealt{nandra2015}). The astrometry was calibrated using the CFHTLS i-band image as the reference (\citealt{nandra2015}). The AEGIS-XD is currently the third deepest X-ray survey in existence and it covers an area larger than the Chandra Deep Fields (CDFs) by a factor of 3. While being approximately $2-3$ times shallower than the CDFs, it is  sufficient to probe the dust-obscured X-ray galaxy populations at high redshifts (e.g., $L_X \sim 10^{43}$ erg s$^{-1}$ at $z \sim 3$ in the soft X-ray 0.5-2 keV band).

\begin{figure}
\includegraphics[width=8.5cm]{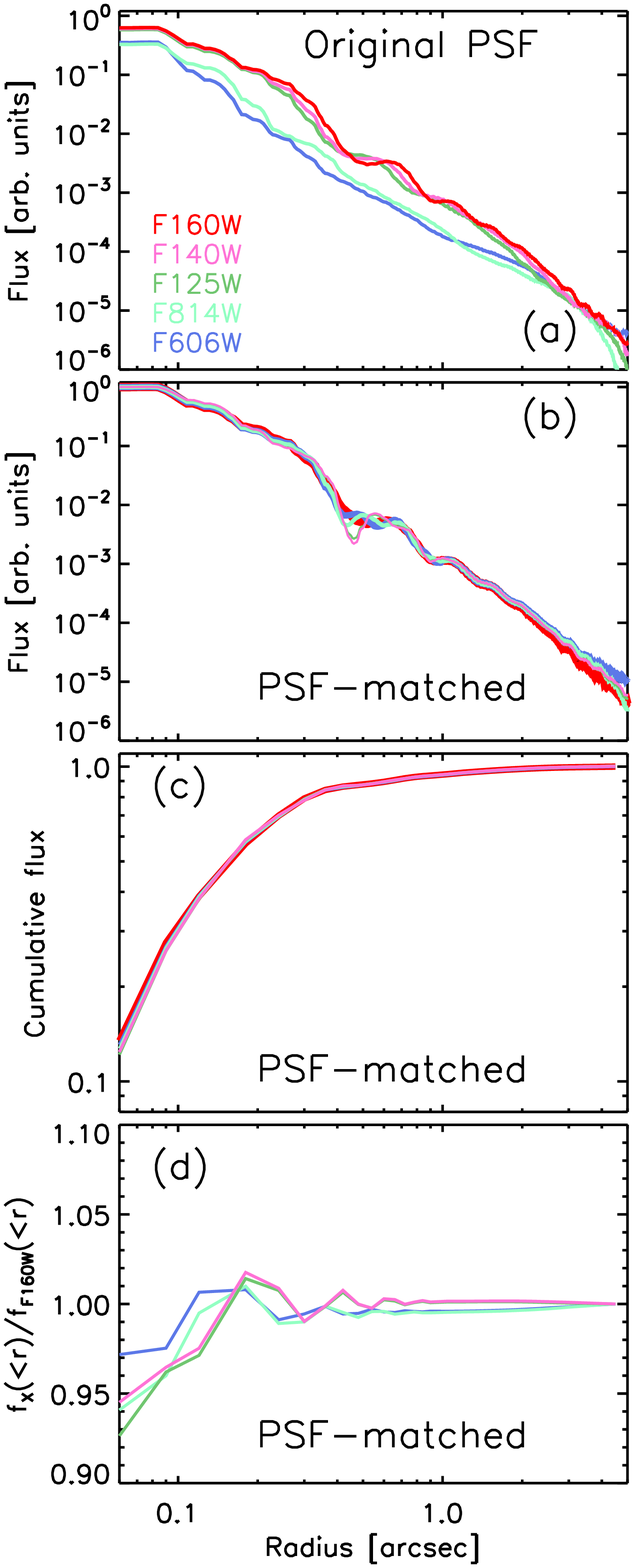}
\caption{Top to bottom panels:  (a) Original PSF profiles of the \textit{HST} ACS F606W, ACS F814W, WFC3 F125W  WFC3 F140W and WFC3 F160W bands.  (b) Similar to the previous plot, but after PSF matching to  F160W. (c) Growth curves in the same four bands after PSF matching. (d) Ratios of the cumulative flux in the four bands after the PSF-matching to that of F160W, shown as a function of radius.  The color-coding scheme is the same in all panels. \label{fig:psfmatch}}
\end{figure}

\section{Astrometric calibration of CANDELS \textit{HST} ACS and WFC3 mosaics}
\label{sect:astrometry}

The catalog is photometrically selected in WFC3 F160W band and hence it 
inherits the astrometry of this image. The earlier astrometry of the CANDELS
EGS WFC3 and ACS mosaics was calibrated using the AEGIS ACS catalog from  \citet{lotz2008}. This calibration was based on the CFH12K mosaic of \citet{coil2004}, 
which was tied to the USNOB1.0 catalog. However, a comparison between the F160W
positions using this earlier astrometric solution and the CFHTLS D3 $i'$-band
positions revealed systematic offsets. Figure \ref{fig:astrometry} presents the measured offsets. Both the amplitude and the direction of the
offset vary across the field, with a median
offset of $\sim0\farcs2$, significant enough to cause potential problems in many
applications. Indeed, previous works already showed that the median dispersion of the positions of a given object imaged on multiple overlapping plates in the USNOB1.0 catalog generated offsets up to $0\farcs2-0\farcs6$ in the position of extended objects compared to the SDSS Early Data Release (\citealt{stoughton2002} - see \citealt{monet2003} ) or to PPMX  \citep{roeser2010}.
  
Because the CFHTLS D3 astrometry implements higher-quality data and a better calibration  (in particular the SDSS calibration), we believe that it is
more trustworthy.  For the sake of internal consistency, we have opted to keep the WFC3 and ACS mosaic images unchanged in this version of the data release and to register all the F160W source positions to the CFHTLS D3 system at the catalog level. 

This astrometric registration was achieved as follows.  At first, we matched the WCS coordinates of the objects in the ACS F814W catalog of \citet{lotz2008} to those in the CFHTLS D3 catalog of Gwyn (2012). A matching radius of $0\farcs5$ was adopted. The full region was then divided into contiguous tiles, each $1.5\times1.5$ arcmin$^2$ in size. This size was chosen as a compromise between obtaining a sufficient number ($\sim 100$)  of objects in each tile to reduce the statistical uncertainties and yet keeping the tiles as small as possible in order not to lose the resolution. In each tile, the average of the displacements between the F814W and the CFHTLS coordinates was computed. These averages were used as the input to the \texttt{IRAF} (\citealt{tody1986,tody1993}) task \texttt{geomap} to generate the surface which converts one astrometric system to the other.  The task was run interactively, adopting a 5th-order Legendre function to represent the surface, as it turned out to provide the best results in terms of residuals.  In this process, we rejected a few objects with residuals larger than 3$\sigma$. Finally we re-registered the F814W coordinates to the new system by applying the best-fit surface to correct the positions.

After applying the correction, the astrometry shows much less discrepancy with
respect to the CFHTLS D3 system, with a median offset of $\sim0\farcs04$, i.e., 
a factor of 5$\times$ smaller than the previous astrometric solution (see the right panel
of Figure \ref{fig:astrometry}). This is in agreement with the average offsets of our catalogs in other CANDELS fields when compared to the external catalogs. Our final catalog adopts
this new astrometric solution, and the World Coordinates System (WCS) positions in the catalog refer to this
improved system. For backward compatibility we also provide the
positions using the earlier AEGIS ACS system (columns \texttt{RA\_LOTZ2008} and \texttt{DEC\_LOTZ2008} in the photometry catalog; see Appendix \ref{appendix:photcat}). In the CANDELS repository we provide for download the versions of the {\it HST} ACS and WFC3 mosaics registered to the \citet{lotz2008} and to the CFHTLS D3 system as well.

\begin{table}
\caption{Main SExtractor parameters in the hot and cold mode\label{tab:se}}
\begin{tabular}{lcc}
\hline
\hline
Parameter Name & Cold Mode & Hot Mode \\
\hline
\texttt{DETECT\_MINAREA} & 5.0 & 10.0 \\
\texttt{DETECT\_THRESH} & 0.75 & 0.7 \\
\texttt{ANALYSIS\_THRESH} & 5.0 & 0.7 \\
\texttt{FILTER\_NAME} & \texttt{tophat\_9.0\_9x9.conv} & \texttt{gauss\_4.0\_7x7.conv} \\
\texttt{DEBLEND\_NTHRESH} & 16 & 64 \\
\texttt{DEBLEND\_MINCONT} & 0.0001 & 0.001 \\
\texttt{BACK\_SIZE} & 256 & 128 \\
\texttt{BACK\_FILTERSIZE} & 9 & 5 \\
\texttt{BACKPHOTO\_THICK} & 100 & 48.0 \\
\hline
\end{tabular}
\end{table}

\section{Photometry}

For the flux measurements in the CANDELS EGS photometric catalog we adopted two different approaches, depending on the angular resolution of the images, as parameterised by the FWHM of the PSF charactersic of each band. For the high-resolution images, i.e., all the \textit{HST} bands, the photometry was performed using \texttt{SExtractor} \citep{bertin1996} in dual-mode with a detection on the F160W band and measurements on the PSF-matched ACS/WFC3 images matched to the F160W resolution, the lowest in our \textit{HST} dataset. For the low-resolution images, i.e., all the ground-based data and the \textit{Spitzer}/IRAC images, the photometry was performed using the \texttt{TFIT} software  \citep{laidler2007}, which uses a morphological template fitting technique. These procedures have been extensively described by \citet{galametz2013} and  \citet{guo2013}.

 \subsection{PSF and convolution kernels}
For the \textit{HST} and IRAC bands, the empirical PSFs were constructed from a set of high S/N and isolated point sources using \texttt{IRAF/DAOPHOT} (\citealt{stetson1987}). Despite the fact that IRAC PSFs vary across the mosaics due to the heterogenous nature of the input data and the intrinsic variations in the instrument itself, we opted for a single IRAC PSF in each band.
For the ground-based optical and NIR images, the PSFs were created by stacking a number ($6-21$) of bright and isolated point sources. The convolution kernels were  obtained using the \texttt{IRAF/lucy} task, for later uses in either PSF-matching or \texttt{TFIT}. Figure \ref{fig:psfmatch} shows the light profile of the empirical \textit{HST} PSFs before and after the PSF-matching process, together with the growth curves of the PSF-matched point sources. The growth curves do not show any significant offset with respect to that of the F160W band. Within the central 2 pixels ($0\farcs12$), the differences amount to $\sim5\%$. By $8.3$ pixels ($0\farcs5$) the differences essentially vanish.

\begin{figure*}
\includegraphics[width=18cm]{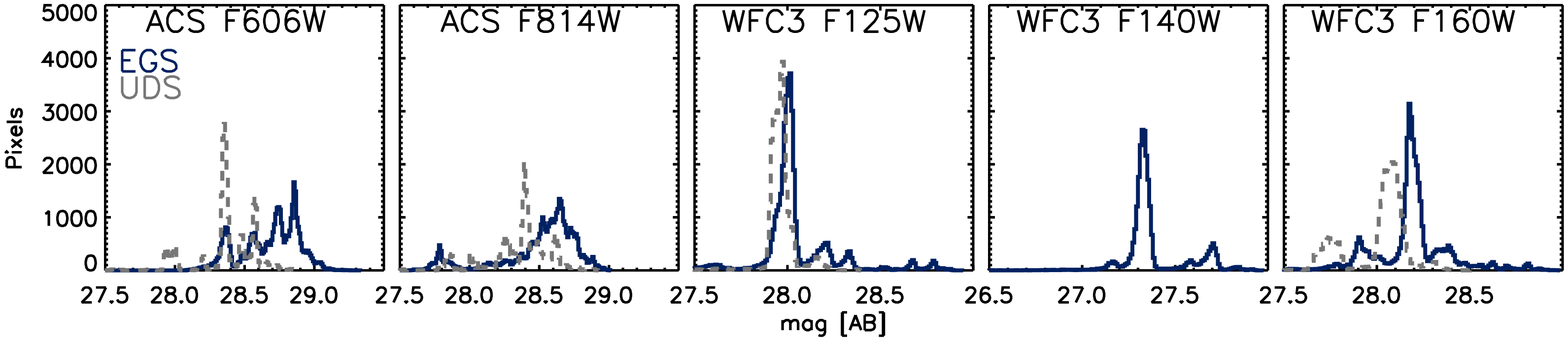}
\caption{Distributions of the pixel-by-pixel 1$\sigma$ magnitude limits in the five \textit{HST} bands in the EGS (blue solid line) and the UDS (grey dashed line) fields. The UDS catalog does not include the F140W band from 3D-HST. The limits were calculated from the RMS maps and are scaled to an area of 1 arcsec$^2$. The bin width is 0.01 mag for all bands.  \label{fig:exposure_map} }
\end{figure*}

\subsection{Photometry of \textit{HST} images}

\begin{figure*}
\includegraphics[width=18cm,bb = 44 307 566 502,clip]{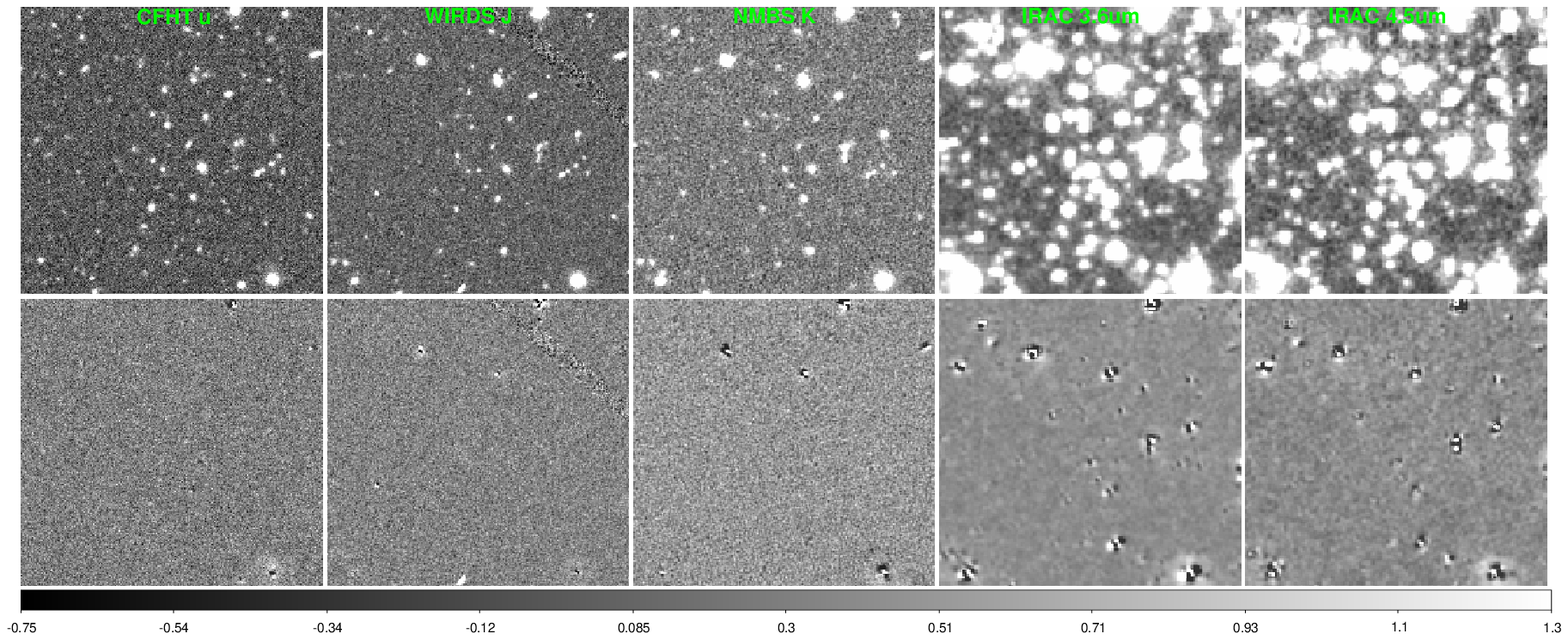}
\caption{Examples of residuals after the second pass of \texttt{TFIT}. Cutouts $\sim 80'' \times 80''$  from the original science images are shown in the top row for CFHT $u^*$, WIRDS $J$, NMBS $K$, IRAC $3.6\mu$m and IRAC $4.5\mu$m bands, respectively, from left to right. The corresponding cutouts of the residual images are shown in the bottom row. \label{fig:residuals}}
\end{figure*}

Both source detection and photometry were performed using a modified version of \texttt{SExtractor} v2.8.6, which was already used in the construction of the CANDELS UDS (\citealt{galametz2013}) and GOODS-S (\citealt{guo2013}) catalogs. This modified version provides a better measurement of the local background by imposing a minimum inner radius of $1''$ for the sky annulus (see \citealt{grazian2006,galametz2013}) and refines the cleaning process by  rejecting non-detections that are often merged to real sources \citep{galametz2013}.

\begin{figure*}
\includegraphics[width=18cm]{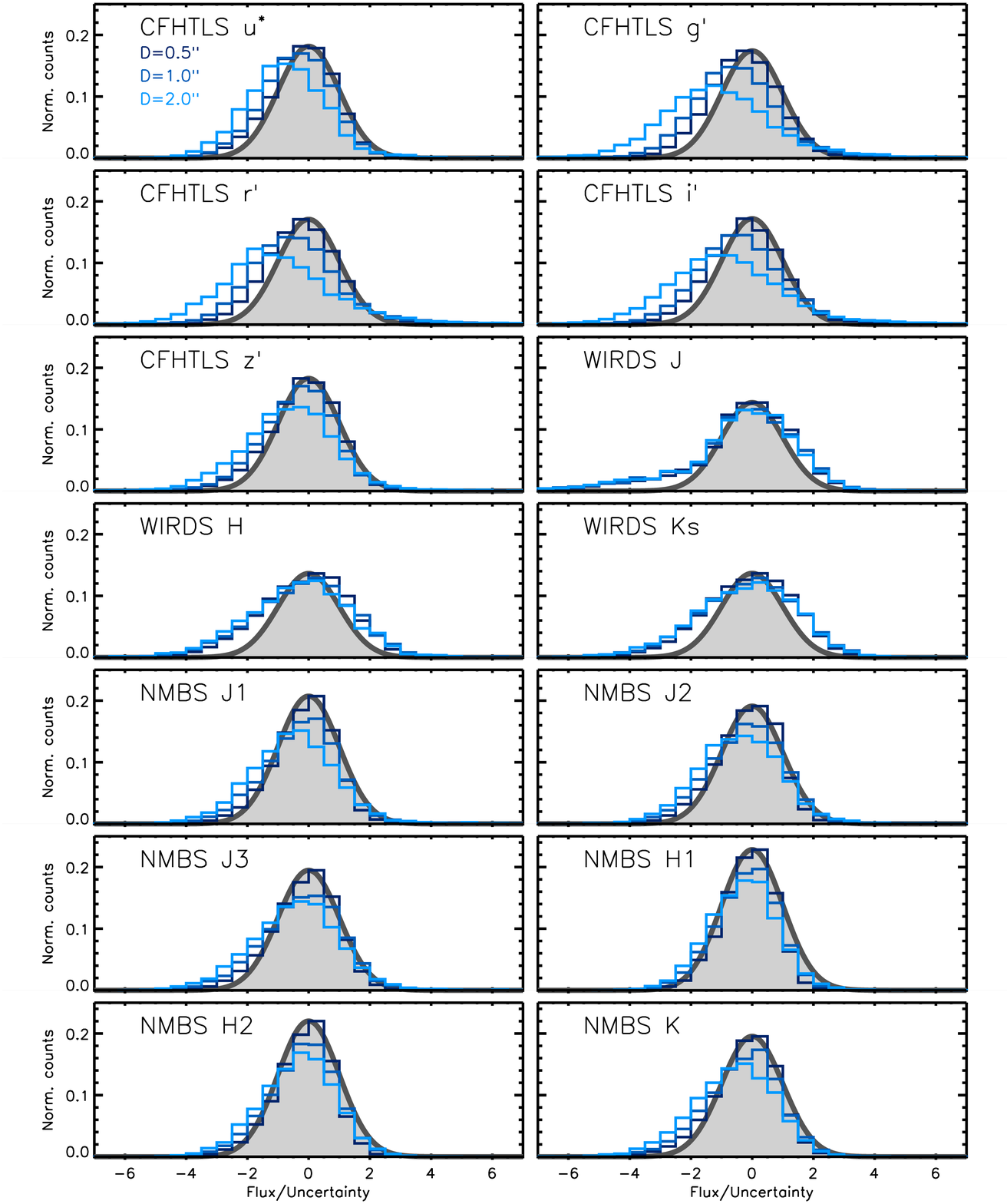}

\caption{Histograms showing the  background level in units of the local noise level, for the background-subtracted ground-based bands. The bin width is 0.5.The noise value was computed from the rms maps. The three different blue shades refer to empty apertures randomly positioned on the residual map avoiding any already detected source and adopting three different apertures: $0\farcs5$, $1\farcs$ and $2\farcs$, as indicated by the legend. These histograms have been renormalised to the total number of used apertures ($\sim10^4$). For reference, we also plot as a grey curve a standard normal distribution, whose peak value has been normalised to that  of the $D=0\farcs5$ distribution. Most of the distributions are centered on S/N $\sim 0$, and they are all consistent with zero at $3-\sigma$ level, validating the background correction procedure.  We relied on \texttt{SExtractor} background measurement capabilities for the \textit{HST} bands.\label{fig:back_all}}
\end{figure*}

A single configuration of \texttt{SExtractor} usually does not provide the most optimal detection of all the objects in an image. Indeed it is difficult to reach the balance that can achieve high completeness for faint objects without a large number of spurious detections caused by oversplitting bright and/or extended objects. Therefore, we adopted the same two-step approach used for the CANDELS GOODS-S and CANDELS UDS fields (\citealt{guo2013,galametz2013}). First, \texttt{SExtractor} was run in the so-called \emph{cold} mode: the relaxed values of the detection parameters allow for a reliable extraction  and deblending of the brighter sources. In the second run, called the \emph{hot} mode, the detection parameters were tuned to optimize the detection of faint and blended objects. Table \ref{tab:se} presents the main Sextractor parameter values adopted for the cold and hot mode  (see Appendix A in \citealt{galametz2013} for the full list of \texttt{SExtractor} parameters). These two initial catalogs were then merged to keep the objects that are detected in at least one catalog. However, those objects in the \emph{hot} catalog falling within the \citet{kron1980} ellipse of the \emph{cold}-mode sources were rejected, as these are likely the result of an excessive source shredding  (see also e.g. \citealt{caldwell2008}, \citealt{gray2009}, \citealt{barden2012}).

The photometry in both the cold and the hot modes was performed in the dual-image mode of \texttt{SExtractor}, using the F160W band for the detection. As shown in Table \ref{tab:photometric_data}, the FWHM of \textit{HST} PSFs differ from band to band, and the F160W-band PSF has the widest FWHM. To ensure the most accurate color measurements,  we used the \texttt{IRAF/psfmatch} task to homogenize the PSFs of the mosaics in the F606W, F814W, F125W, and F140W bands to the PSF of the F160W band. Figure \ref{fig:psfmatch} shows the profiles of the \textit{HST} PSFs before and after the PSF-matching procedure. The after-matching ones deviate from the F160W PSF at only a few percent level, which validates our matching process.

Different flux measurements were derived. Specifically, fluxes were measured using  \citet{kron1980} elliptical apertures (\texttt{SExtractor} \texttt{FLUX\_AUTO}), isophotes (\texttt{FLUX\_ISO}) and a set of 11 circular apertures (\texttt{FLUX\_APER}). These individual values are reported in the supplemental catalogs that accompany our main catalog.

The flux measurement provided by SExtractor inside the Kron aperture is within $6\%$ of the total flux \citep{bertin1996}, and it is thus often regarded as the measurement of total flux (but see e.g., \citealt{labbe2003} and \citealt{graham2005} for discussions on the deviation of SExtractor \texttt{FLUX\_AUTO} from the total flux). However, this often is not ideal for a faint source in terms of S/N because the large aperture needed to capture the total flux necessarily includes noise from many background pixels, potentially swamping the signal from the targets.  On the other hand, the isophotal flux that maximizes the S/N for faint sources could underestimate the total flux because of the smaller aperture in use.  If the flux measurements of a given object in different bands are done through the same aperture (i.e., through the dual-image mode as we did here), the isophotal fluxes will give the best measurements of  colors. This kind of measurement is crucial for most of the possible applications of our catalog such as SED fitting. 

In our main catalog, the \texttt{FLUX\_AUTO} values were adopted as the total flux measurements in the F160W band. In any other \textit{HST} bands, the quoted total fluxes were derived through the flux ratio with respect to the F160W band in terms of \texttt{FLUX\_ISO}, i.e.,

\begin{equation}
f_{\mathrm{tot},b} = f_{\mathrm{iso},b} \times \frac{f_{\mathrm{auto},\mathrm{F160W}}}{f_{\mathrm{iso},\mathrm{F160W}}}
\end{equation}

\noindent where $f_{\mathrm{tot},b}$  and $f_{\mathrm{iso},b}$ are the total and isophotal flux for band $b$, respectively, while $f_{\mathrm{auto},\mathrm{F160W}}$ is the Kron-aperture flux in the F160W band.  The area adopted for the measurement of the isophotal flux $f_{\mathrm{iso},b}$ is defined from the F160W mosaic and it is thus the same for all bands. $f_{\mathrm{tot},b}$ is therefore a Kron-like flux measurement recovered from the higher S/N isophotal flux estimate. The validity of Equation 1 relies on the fact that the morphologies of the galaxies in our catalog do not vary with wavelength. Indeed, the majority of sources in our catalog have small sizes; furthermore, the smoothing introduced by the PSF-matching to the F160W-band resolution further acts in the direction of homogenizing the morphology of each source across the different bands. Similar approaches have been widely adopted in the literature (e.g. \citealt{whitaker2011,guo2013,galametz2013,muzzin2013,skelton2014}). In particular \citet{vandesande2013} showed that stellar masses from \citet{whitaker2011} are consistent with dynamical mass measurements recovered from absorption lines, increasing the confidence on physical parameter estimates from total flux measurements based on Eq. 1.

Figure \ref{fig:exposure_map} presents the distribution of the $1\sigma$ depth in each \textit{HST} band. The limiting magnitudes in the WFC3 bands roughly follow a tri-modal distribution, which is due to the different degrees of tile overlapping within the mosaics: the heavily overlapped regions among adjacent tiles have higher sensitivities than what is typical, while the boundary regions of less coverage have lower sensitivities.  The distribution for ACS limiting magnitudes is broader and more complex than that of the WFC3 ones. One possible explanation for this is that  the ACS mosaics are the result of two different datasets being combined together (AEGIS and CANDELS), each one likely with its own tri-modal distribution. Another possible explanation could be the different native pixel scale of the ACS camera compared to WFC3. The smaller pixel scale of ACS compared to WFC3 would result in lower S/N for each pixel (for the same cumulative S/N), introducing the uncertainty. Overall, the depths corresponding to the peak of the distributions in 1$\sigma$ magnitudes reported in Figure~\ref{fig:exposure_map} for each band are generally within $\approx$0.1~mag from the rescaled 5$\sigma$ depths presented in Table 1 (obtained from randomly placed circular apertures), supporting our rms maps generation. Finally, the plots show that the average depth of the EGS \textit{HST} mosaics is slightly deeper than that of  UDS.

\subsection{Photometry of low resolution bands}

\label{sect:phot_lowres}

\subsubsection{TFIT flux measurement}

The flux measurements for the low-resolution bands in the CANDELS multi-wavelength catalogs are  based on \texttt{TFIT} \citep{laidler2007}. A detailed description of this software is provided by  \citet{fernandez-soto1999}, \citet{papovich2001} and \citet{laidler2007}, while  \citet{lee2012} presented a set of simulations aimed at validating this template-fitting technique and quantifying its uncertainties. Briefly, the brightness profile of the source in the high-resolution image, identified from the segmentation maps, is convolved with the kernel required to match the low-resolution image PSF. The result is a template of the object in the low resolution image. Its total flux can then be obtained via best fit. In this way, it is possible to use all the information from the high-resolution image to deblend the objects in the low-resolution bands. This procedure assumes that the morphology of each object does not depend on the wavelength. A wavelength-dependent morphology could result in the outer regions  of some objects being excluded from their segmentation map. Furthermore, the outskirts of fainter and/or smaller objects could fall below the detection threshold and hence be missing from the segmentation map generated during the detection stage. In order to limit the potential loss of flux due to these effects, the area associated with each object in the segmentation map was expanded following the empirical relation of \citet{galametz2013}. The fluxes measured by \texttt{TFIT} have been proven to be very close to total fluxes \citep{lee2012}, and hence no further aperture correction was applied to them.

Template fitting techniques require good alignment between the high-resolution and the low-resolution images. However, the astrometric calibration of the low-resolution images could be different from that of the high-resolution one in the method and/or the reference catalog, which could result in slight offsets in alignment. Furthermore, geometric distortions, if not perfectly corrected, could also produce local misalignments among images.  All this could result in catastrophic failures in template fitting. To overcome this problem, \texttt{TFIT} also measures small position offsets between the high and low resolution images.  \texttt{TFIT}  is then run for a second time, using this information as part of the input to adjust the alignment locally by allowing a slight freedom of the centroid during the fitting process. Figure \ref{fig:residuals} presents an example of the second-pass residual images of a small section in the EGS field. The clean residual images indicate that the fitting procedure was successful.

\begin{figure*}
\includegraphics[width=18cm]{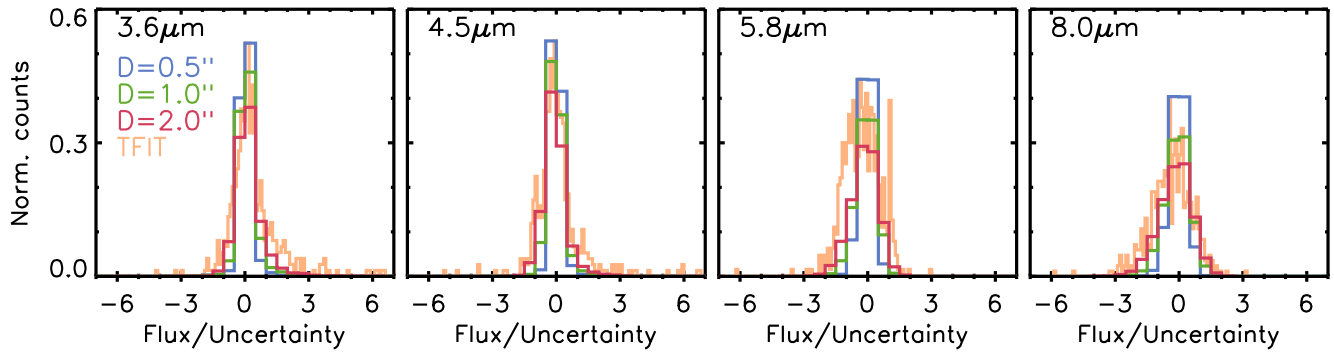}
\caption{Histograms showing the  background level in units of the local noise level, for the background-subtracted ground-based bands. The orange histograms present the data from the TFIT recovery of the simulated sources, while the blue, green and red ones refer to the apertures as indicated by the legend (see Sect. \ref{sect:phot_lowres} for more details). These histograms have been renormalised to the total number of used apertures. The distributions are centered on zero, which further validate our background subtraction procedure. \label{fig:back_irac}}
\end{figure*}

\subsubsection{Background assessment}

A key factor in reliable flux measurement is the determination of the background. \texttt{TFIT} does not attempt any measurement of the background, but instead assumes a zero-background everywhere in the mosaic. Therefore, the background must be subtracted in advance. The full procedure of background estimate and subtraction for the CANDELS multi-wavelength catalog construction was described by \citet{galametz2013}. 

We assessed the goodness of our background subtraction by measuring the fluxes and the associated uncertainties in about $10^4$ empty apertures (i.e., placed at locations free of known sources) across the residual images created by TFIT after subtracting off all the sources detected in F160W.  The statistical distribution of the ratio between the flux measurements and their associated uncertainties should then be described by a standard normal distribution (i.e. zero mean and unit standard deviation). We performed the flux measurements adopting three different aperture values ($0\farcs5$, $1\farcs$0 and $2\farcs$0). Figure \ref{fig:back_all} shows the distribution of the S/N measurements for the ground-based data.  The distributions very closely match the standard normal distribution, as expected. 

Since the background subtraction in the IRAC bands is even more challenging, for these bands we complemented the above test with a Monte Carlo simulation as follows. A set of 500 positions were randomly chosen in regions free of any objects detected in the F160W band. We simulated the presence of the objects by adding these 500  positions to the input catalog for TFIT.  As TFIT needs to refer to the segmentation map for the shape information, we also simulated exponential and \citet{devaucouleurs1948} profiles through the \texttt{IRAF mkobject} task and added their footprints to the segmentation map. The science mosaics were not altered. The simulated objects have circularised effective radii uniformly distributed between $0\farcs6$ and $2\farcs4$ (after taking into account the PSF), which fully encompass the distribution of apparent sizes of the objects detected in the WFC3 F160W mosaic.  \texttt{TFIT} was then re-run with this new catalog, adopting the same configuration used for the actual photometry, with the noise values computed from the rms maps.  The forced flux measurements at the positions of the simulated sources, which are not actually present on the science mosaics, should statistically be zero.   

The results from this test, in terms of S/N distribution, are shown in Figure \ref{fig:back_irac} together with the S/N from the empty aperture measurements. Most of the distributions are centered on S/N $\sim 0$, and they are all consistent with zero at $3-\sigma$ level, validating the background correction procedure. We relied on \texttt{SExtractor} background measurement capabilities for the \textit{HST} bands.

\subsection{Multi-wavelength photometric catalog creation}
\label{sect:cat_creation}

Flux measurements from the PSF-matched bands and from the \texttt{TFIT} process were finally merged together to produce the final catalog. This step was trivial because each object kept its unique ID throughout all the processes, and therefore the matching was done based on the ID rather than on the coordinates. 

The final catalog includes photometry in 22 bands for 41457 objects detected to a depth of 27.6 mag (5$\sigma$) in F160W. For each object, the catalog contains the flux measurement and the associated uncertainty. The catalog includes two distinct measurements for the IRAC 5.8$\mu$m and $8.0\mu$m bands. One was obtained with an older kernel and was used for the computation of photometric redshifts and stellar masses. The second measurement was obtained with an improved kernel (see Appendix \ref{appendix:photcat}). The plots presented in this paper refer to the improved kernel version of the IRAC 5.8$\mu$m and $8.0\mu$m photometry. The catalog also includes all the objects with reliable (quality flag $Q\ge 3$) spectroscopic redshifts from DEEP2+3 \citep{coil2004,newman2013,cooper2011,cooper2012}.

\begin{figure*}
\begin{tabular}{cc}
\includegraphics[width=8cm]{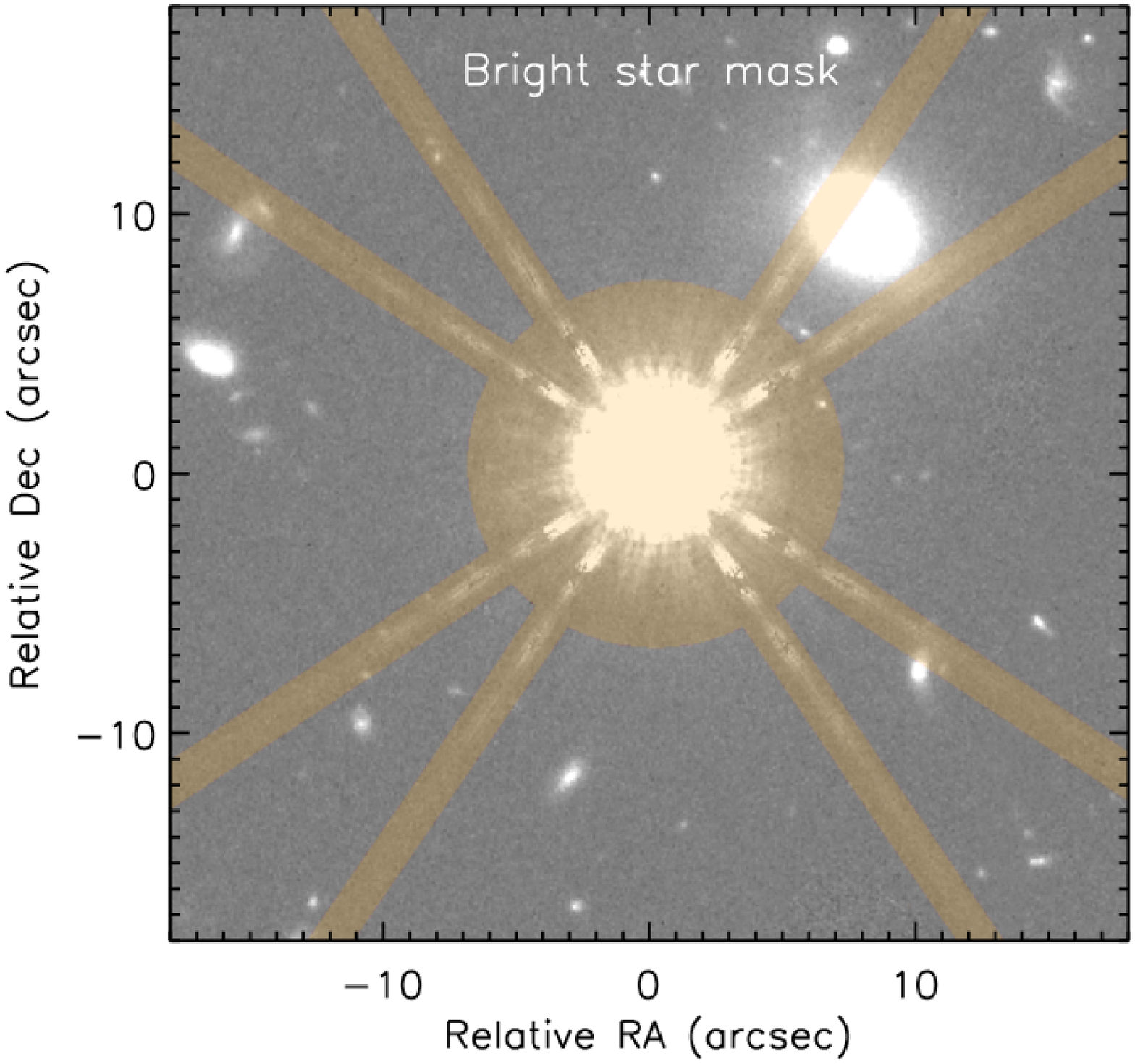} & \includegraphics[width=8cm]{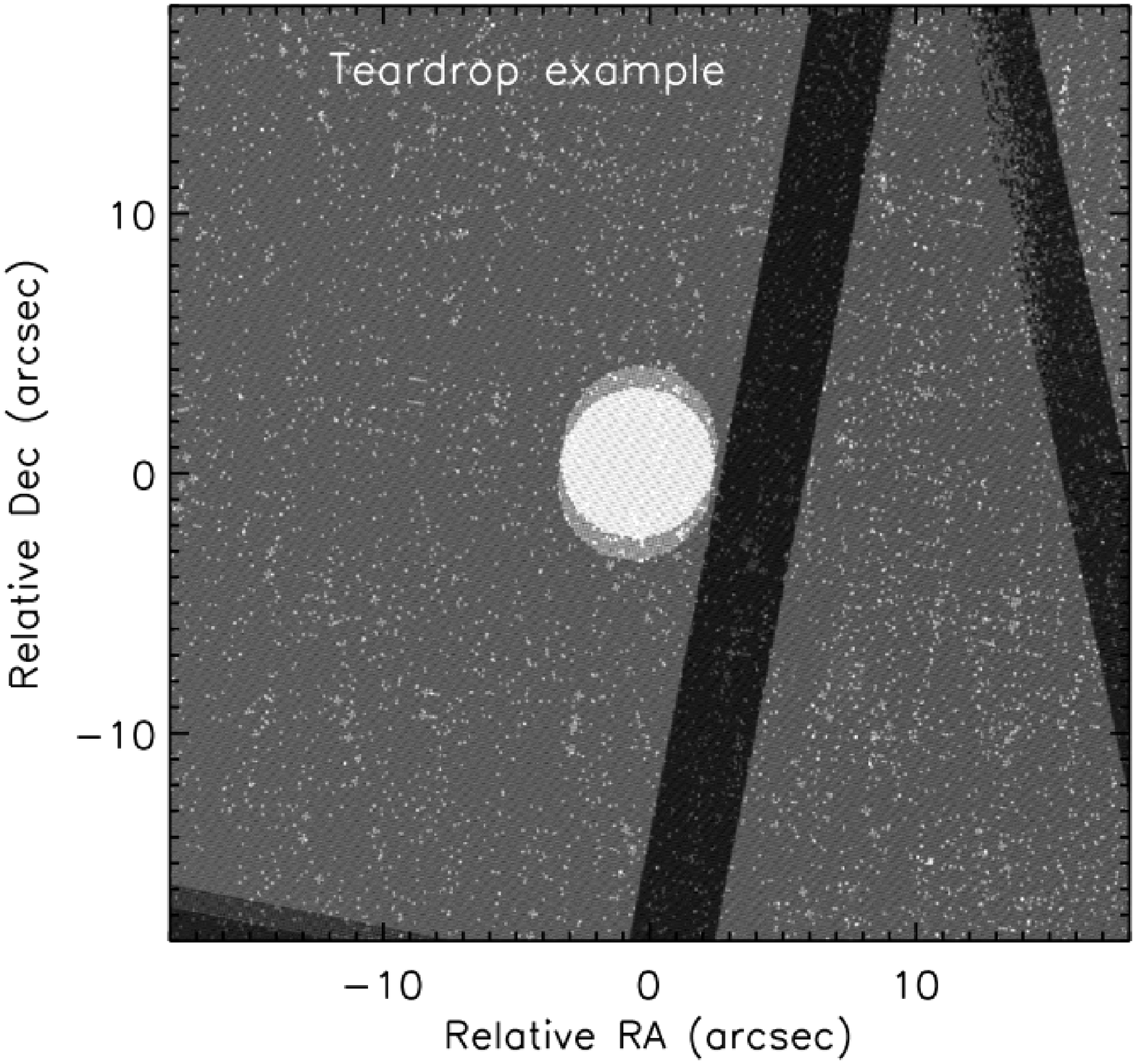}
\end{tabular}
\caption{\emph{Left panel:} Example of the mask adopted for brights stars and their spikes. The grey-scale image shows the cutout of the WFC3/F160W mosaic centered on a bright star arbitrarily picked from the photometric catalog.  The yellow area marks the region set in the flag map to enclose the bright star and its spikes, as these could either contaminate the photometry and/or generate spurious detections. \emph{Right panel:} Example of teardrop. The figure shows the cutout of the rms map associated to the WFC3/F160W mosaic, centered on a region of higher rms signal (a teardrop, the white spot in the center). Regions like the one shown have also been detected and masked. The darker regions correspond to lower rms values from the overlap of contiguous exposures.  \label{fig:flag_mask}}
\end{figure*}

A flag mask, built from the F160W science mosaic, weight and rms maps, is included with the catalog in this data release to indicate if the object is very close to a bright star, falls within a defect region such as the \emph{teardrops} (circular regions of dead pixels in the WFC3 array, $\sim50$ pixels in diameter), or is in a location of higher noise than usual.  Figure \ref{fig:flag_mask} shows an example of mask for a bright star and an example of a teardrop. Details on how the flag mask was generated are given in Appendix B of \citet{galametz2013}.

The EGS field is at Galactic latitude $b\sim60\degree$; therefore the Galactic foreground extinction is expected to be minimal. Furthermore, the region of sky covered by the EGS footprint is small compared to the characteristic scale over which the Galactic foreground extinction varies, hence a single value for the Galactic foreground extinction in each band for all the objects in the catalog should suffice. Considering also that the exact value of the extinction in each band depends on the adopted dust model (e.g., for EGS $A_V=0.025$ mag for a \citet{cardelli1989} model or $A_V=0.022$ mag for a \citet{fitzpatrick1999} model) we opted not to apply any Galactic foreground extinction correction.

The main multi-wavelength photometric catalog is accompanied by three additional catalogs containing estimates of the weight, based on the median exposure time inside the segmentation map, an estimate of the limiting magnitude, obtained from the median of the values in the rms maps from those pixels inside the segmentation map and a number of  morphological and photometric quantities recovered from \texttt{SExtractor} on a per-object basis. We refer the reader to the README file for full details on the content of each of these catalogs. Together with the multi-wavelength photometric catalog, with this work we also release the catalogs of photometric redshifts, stellar masses and physical parameters for all the detected objects, derived following the procedures of \citet{dahlen2013} and \citet{mobasher2015a}. Appendixes A through D detail the content of each catalog.  Two companion papers will present the rest-frame luminosities (\citealt{kocevski2016} in prep.) and the probability distribution functions of photometric redshifts (\citealt{kodra2016} in prep.).

For the CANDELS EGS multi-wavelength catalog, we also matched the positions of objects detected in the F160W mosaic to the objects detected in the Chandra 800ks maps of \citet{nandra2015}. 
The most reliable counterparts to the X-ray sources were identified using a maximum-likelihood technique using the redder and deeper bands available (see \citealt{nandra2015} for details). The matching procedure resulted in 246 objects of the F160W-based-multi-wavelength catalog being likely X-ray emitters.

\begin{figure*}
\begin{tabular}{cc}
\hspace{-1cm}\includegraphics[width=8.5cm]{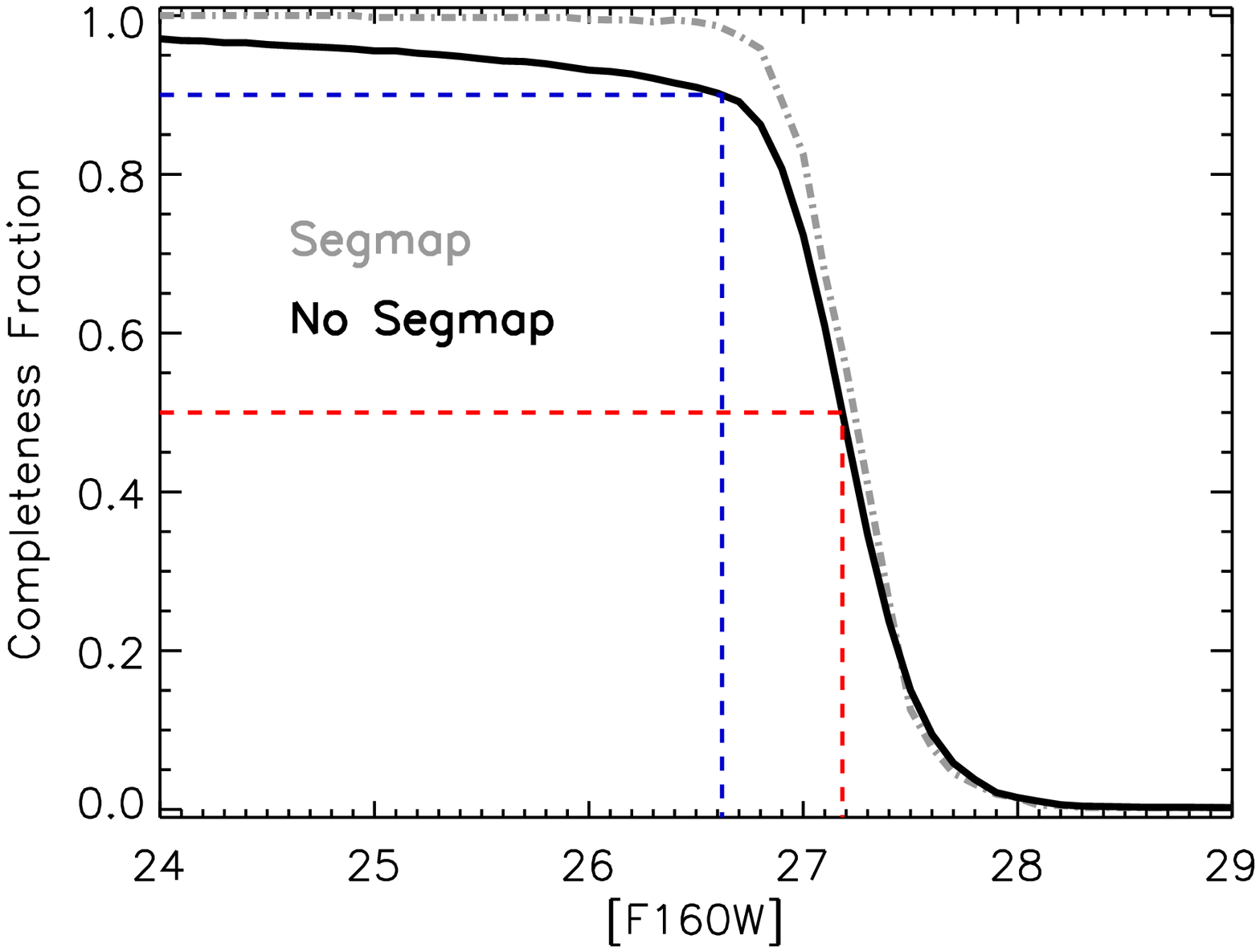}&\hspace{-1cm}\vspace{0.3cm}\includegraphics[width=11cm]{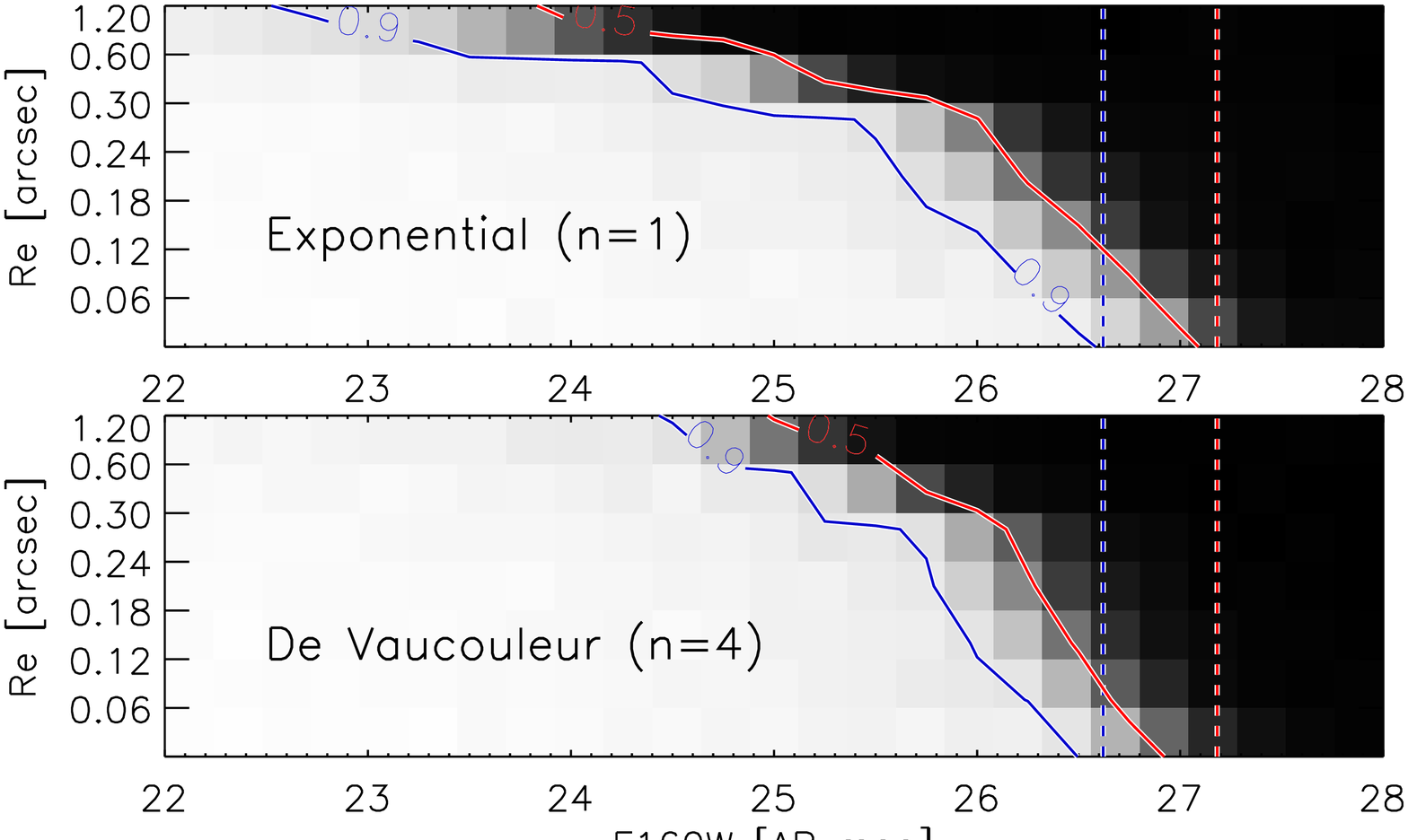}
\end{tabular}
\caption{Detection completeness. The left panel shows the detection completeness on the CANDELS F160W image for a point-source, obtained from a Monte Carlo simulation.  The dot-dashed grey curve is calculated masking all the already detected sources, while the solid black curve is calculated keeping all the sources. The blue and red dashed lines indicate the  90\% and 50\% completeness levels and corresponding magnitudes. The two panels on the right show the completeness for extended objects, for the two cases of an exponential (top panel) and a De Vaucouleur profile (bottom panel) with circularised effective radii in the range $0\farcs06-1\farcs2$, obtained from  a Monte Carlo simulation. The solid red and blue curves mark the 50\% and 90\% completeness limits, while the vertical red and blue dashed lines represent the 50\% and 90\% completeness limits from the point source completeness simulation.\label{fig:completeness}}
\end{figure*}

\begin{figure*}
\includegraphics[width=18cm]{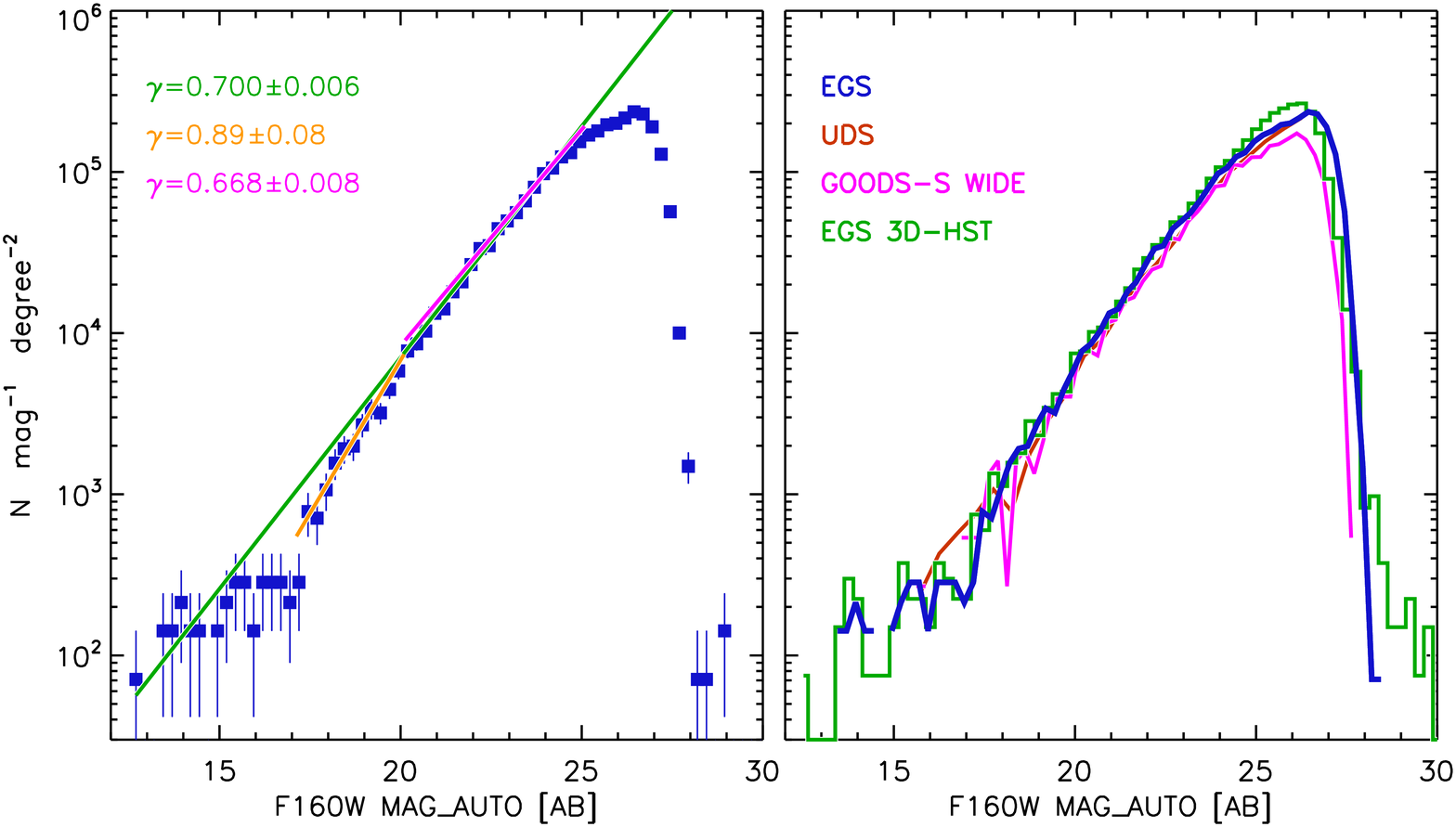}
\caption{\emph{Left panel:} Number counts based on the detection in WFC3 F160W band for CANDELS EGS (filled blue squares with poisson errorbars). The green, magenta and yellow lines represent the best-fitting power-law over the ranges 17~mag $\lesssim$ F160W $\lesssim$ 25~mag, 17~mag $\lesssim$ F160W $\lesssim$ 20~mag, and 20~mag $\lesssim$ F160W $\lesssim$ 25~mag, respectively. The measured power-law slopes are also indicated by the legend. \emph{Right panel:} The CANDELS EGS number counts are compared to other measurements from the literature: CANDELS UDS (solid brown line), CANDELS GOODS-S (solid magenta line) and 3D-HST EGS (green histogram).  The four measurements  are barely distinguishable from each other for 19~mag $\lesssim$ F160W $\lesssim$ 24~mag. The decrease in number counts starts at $\approx24.7$~mag, while at  F160W$\sim26.5$~mag we observe an abrupt decay in number counts,  which is consistent with measurements from the other studies, and is likely a consequence of the detection image completeness for extended objects. \label{fig:numcounts}}
\end{figure*}

\section{Assessment of catalog properties}

In the following subsections we present the tests that we performed to check the consistency of the photometry. These tests consist of both comparisons to external catalogs and of the assessments of the internal self-consistency.

\subsection{Detection completeness}
\label{sect:completeness}

The assessment of the completeness of sources in a catalog is a complex task, which ultimately depends on the class of objects considered and on the specific physical property under analysis. In this subsection we estimate the completeness in detection of sources, while in Section 6.4 we show an example of the completeness assessment in stellar mass.

We  evaluated the detection completeness for both point and extended sources. For point sources, 100 PSFs randomly distributed across the whole field were added to the detection image without any restriction in their positions, and then we detected them using the same \texttt{SExtractor} configuration (two passes) adopted in the actual photometry. The process was repeated 20 times in each magnitude bin (width 0.25 mag) to increase the statistical significance. Although the above procedure provides robust completeness measurements for point sources, to allow for a more direct comparison with completeness measurements from other surveys, the procedure was repeated excluding from the possible random positions those regions of the image which were already occupied by other sources, as identified by the segmentation map. This second method provides a strict upper limit to the completeness measurement. The two curves are presented in the left panel of Figure \ref{fig:completeness}.  The curve corresponding to the detection completeness recovered ignoring any restriction on object position (i.e. the \emph{no-segmap} case) shows lower completeness values than those obtained excluding already detected sources  (\emph{segmap} case) for most of the magnitude range. This difference highlights the impact of source confusion. The 90\% and 50\% completeness limits for point sources are  F160W=26.62~mag and 27.18~mag respectively, when the random positions are not checked against the segmentation map. Excluding from the simulation the position of all the detected objects, the limits  are F160W=26.91~mag and 27.23~mag, respectively.

The completeness for extended sources is sensitive to two main factors: 1) Compared to a point source of the same total flux, an extended source suffers from higher background noise because the extraction area must be larger, which results in a lower S/N. This means that the S/N of extended objects falls below the detection threshold at a brighter total flux level as compared to that of point sources, resulting in a lower completeness for extended sources. 2) Extended objects are more prone to the source blending problem, and hence the difficulty to properly de-blend sources causes a further decrease of completeness.

In order to better model the effect of extended sources in the completeness estimate, the completeness for extended sources was computed adopting two different brightness profiles: an exponential disk (i.e. Sersic index $n=1$), typical of disky galaxies, and a \citet{devaucouleurs1948} profile (Sersic index $n=4$), which characterises elliptical galaxies. For each profile, a grid in apparent magnitude and circularised effective radius was constructed. Successively, using the \texttt{IRAF  mkobject} task, 100 galaxies with the morphological properties drawn from the grid were added to the science image at random places across the image with no constraints on their position\footnote{The uniform distribution of random positions still neglects the clustering of galaxies and thus provides an upper limit to the completeness. However, given the low number of added sources compared to the total number of objects in the catalog, the completeness recovered in this way should still reflect a reliable estimate of the completeness for extended sources.}, and photometry was performed with SExctractor. The process was repeated ten times to increase the statistical significance. The results are shown in the panels on the right side of Figure \ref{fig:completeness}. The 90\% completeness limits for the exponential and the de Vaucoulers profiles for $R_e=0\farcs3$ (roughly corresponding to the median value of $R_e$ of the objects in the catalog with F160W brighter than 26.5~mag; the $R_e$ of objects with magnitudes fainter than this value become close to those of point-sources) is 25.40~mag and 25.62~mag respectively.

As expected, the completeness limits for point sources reach fainter magnitudes than the corresponding limits for extended objects. Furthermore, objects with disky morphologies ($n\sim1$) and larger effective radii ($R_e \gtrsim 0\farcs3$) tend to be missed  by the detection algorithm at brighter apparent magnitudes than objects with a more pronounced bulge (i.e., $n\sim4$).  For $R_e\lesssim0\farcs3$ the detection completeness curves on the $R_e-$F160W plane, however, roughly coincide. This is not unexpected, since  less extended and/or more compact sources  are less sensitive to the differences in the observed (i.e., PSF-convolved) light profile.

\subsection{Number counts}

The distribution of detected objects as a function of their apparent flux densities, the source number counts, is one of the most basic tests for the assessment of a sensitivity-limited catalog. The WFC3 F160W band number counts of our EGS catalog are presented in Figure \ref{fig:numcounts}.

For magnitudes brighter than $\sim25$~mag (which roughly corresponds to the completeness limit for our catalog when extended objects are taken into account), the number counts can be fitted by a power-law with slope $\gamma=0.700\pm0.006$. However, a better description of the data can be obtained by considering two power-laws: one in the range 17~mag $\lesssim$ F160W $\lesssim20$~mag and a second for 20~mag $\lesssim$ F160W $\lesssim25$~mag. In this case, the measured  slopes are $\gamma=0.89\pm 0.08$ and $\gamma=0.668\pm 0.008$, for the brighter and fainter regimes, respectively.  This double power-law behaviour is qualitatively consistent with double power-laws from previous analysis of number counts in NIR bands (see e.g., \citealt{gardner1993,ashby2015}).

In the right panel of Figure~\ref{fig:numcounts} we compare our measurements to three recent F160W number count measurements from the literature, namely, the 3D-HST team's measurement of the EGS data  (\citealt{skelton2014}), the CANDELS UDS field (\citealt{galametz2013}) and the CANDELS GOODS-S (\citealt{guo2013}).  All the measurements are consistent at least up to F160W $\sim24$ mag. The number counts from the UDS field are consistent with the EGS ones up to F160W $\sim26.5$~mag.  In the range $24$~mag $<$ F160W $<26.5$~mag the GOODS-S number counts are below the UDS and EGS measurements by a factor up to ~1.2 at F160W = 26~mag, which could be due to the slight shallower depth (0.2~mag) of GOODS-S compared to EGS. The number counts from 3D-HST agree completely with the measurements from the CANDELS EGS catalog up to F160W $\sim24$~mag, while at 24~mag $<$ F160W $<26.5$~mag they show an excess with respect to those from the CANDELS UDS and the CANDELS EGS up to a factor $\sim1.2$ at F160W $\sim26$~mag. One possible explanation for this excess could be that the detection for the 3D-HST catalog was performed on the noise-equalized combination of WFC3 F125W, F140W and F160W bands, which could help in the detection of fainter sources. At magnitudes fainter than F160W $\sim26.5$~mag we observe an abrupt decay in the number counts in all three catalogs, likely a consequence of the completeness for extended sources in the F160W band (adopted for the source detection).

\subsection{Color-color plots}

\begin{figure*}
\begin{tabular}{ccc}
\includegraphics[width=5.5cm]{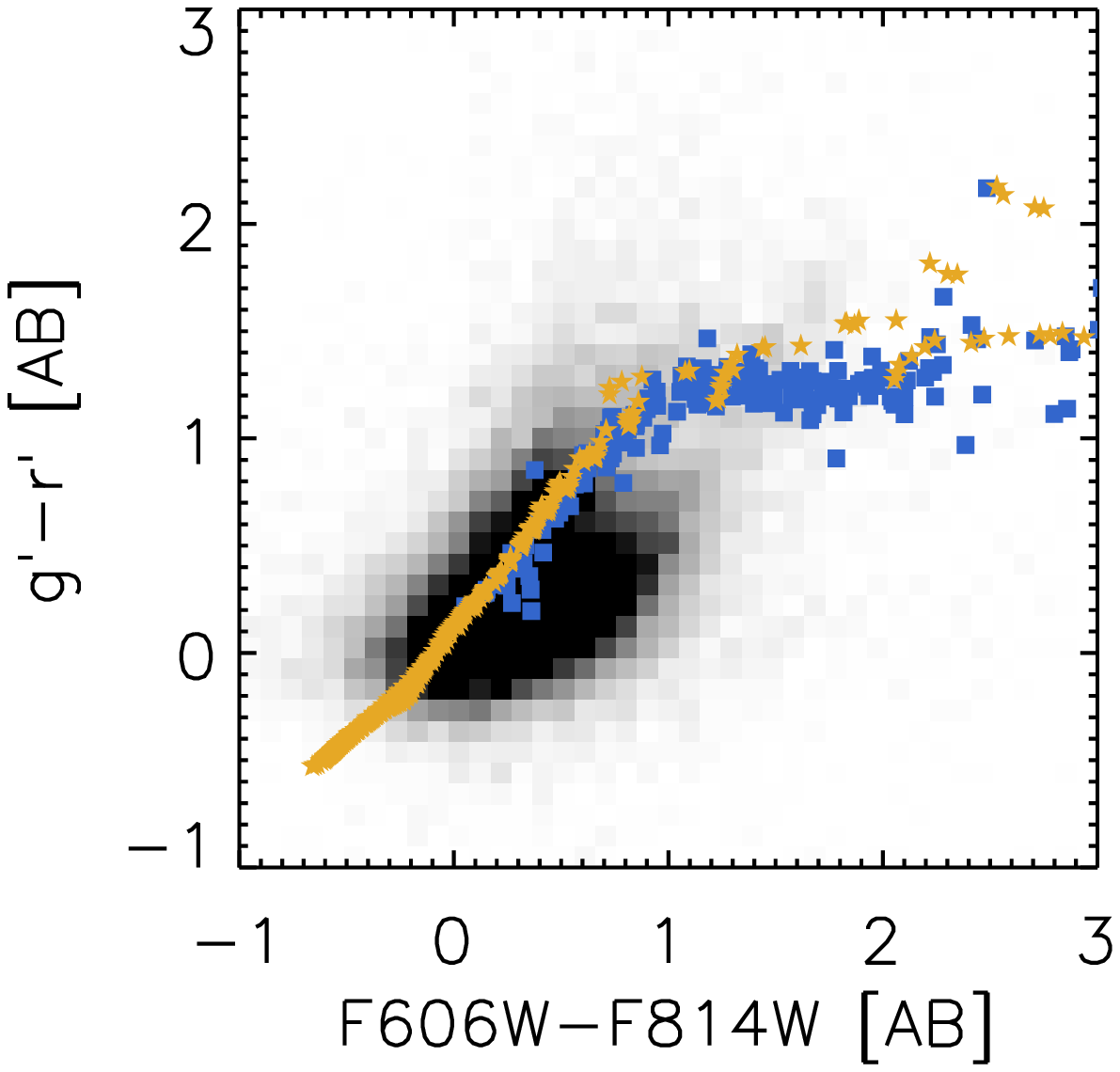} & \includegraphics[width=5.5cm]{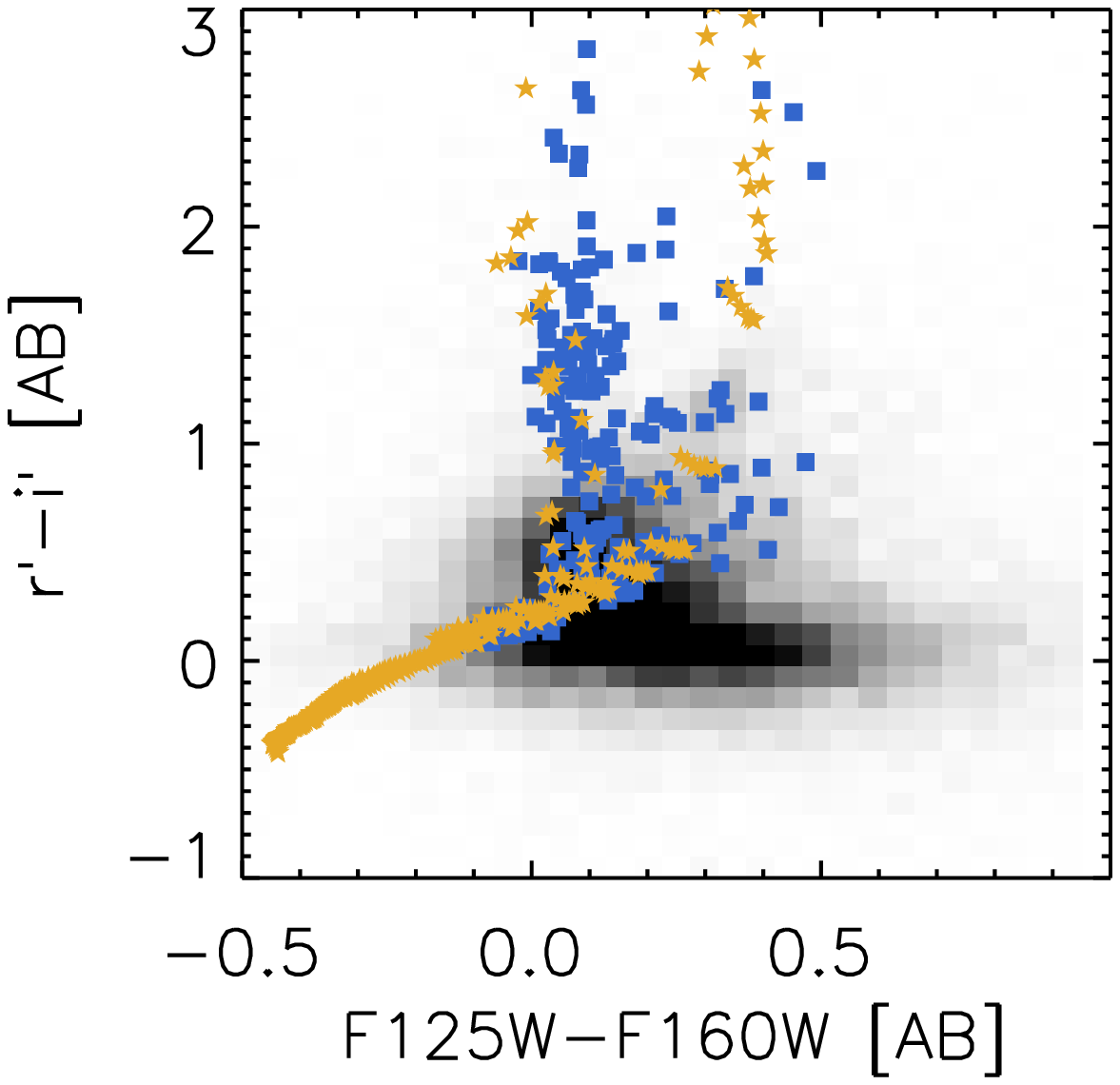} & \includegraphics[width=5.5cm]{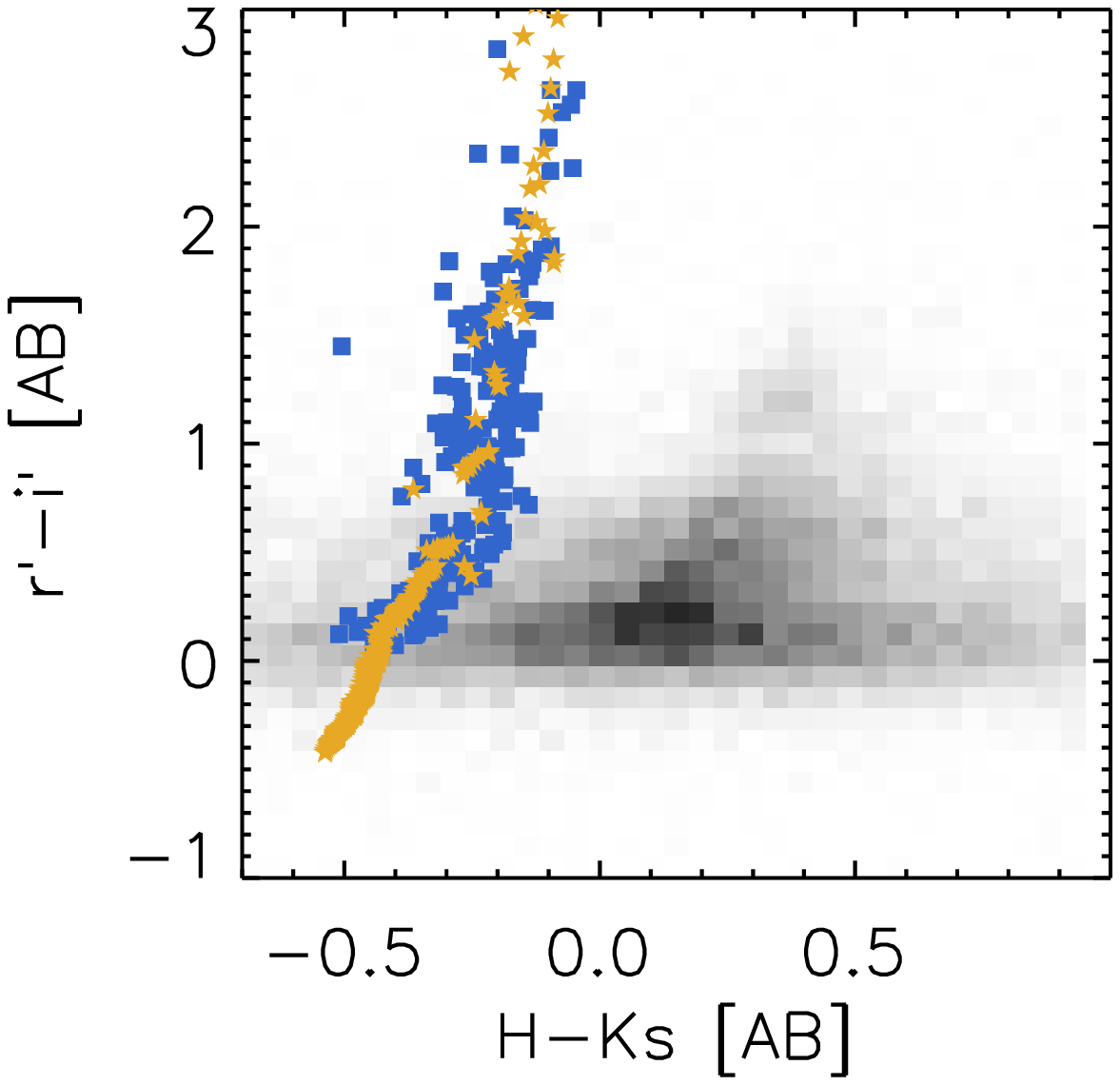} \\
\includegraphics[width=5.5cm]{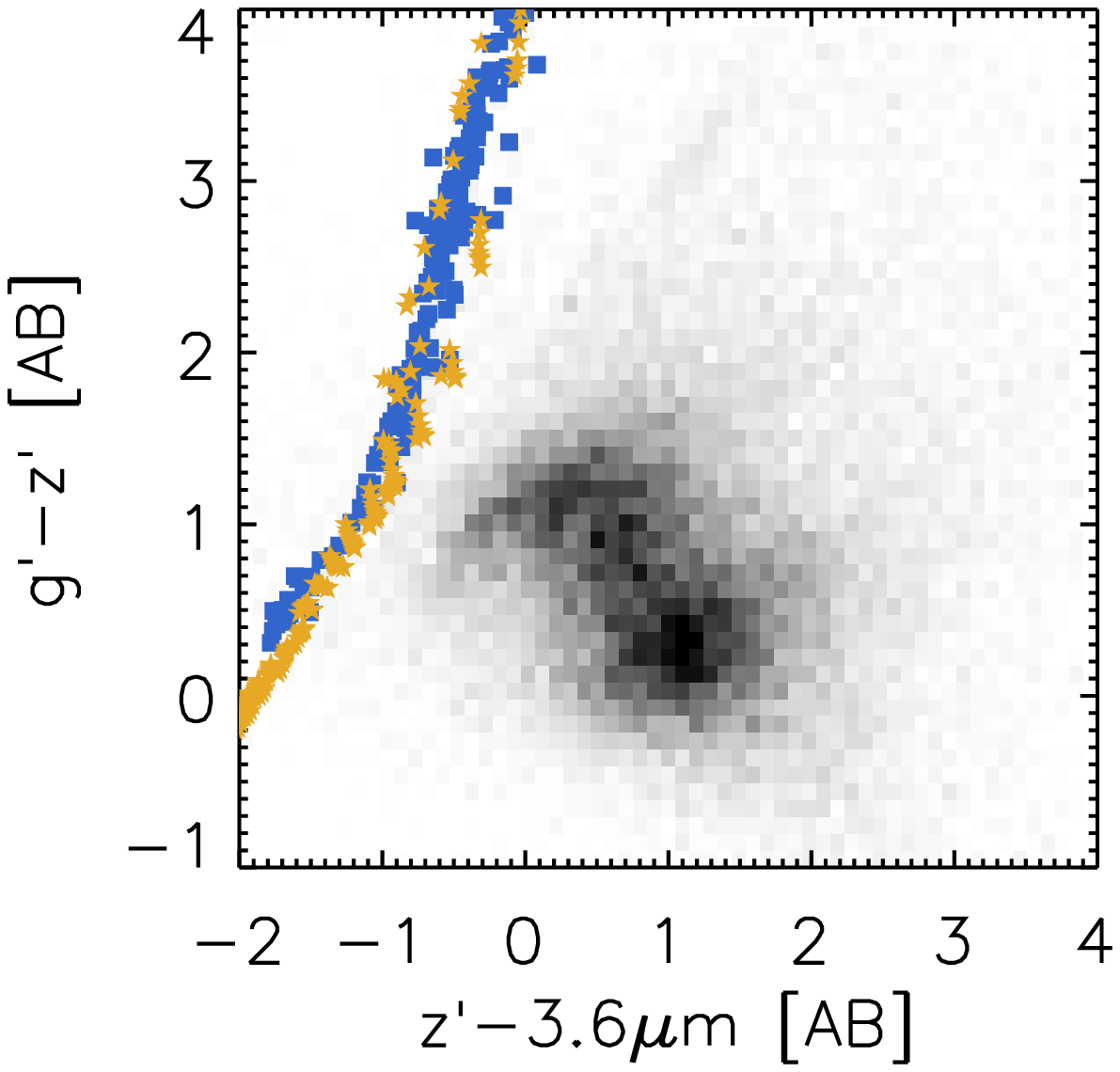} & \includegraphics[width=5.5cm]{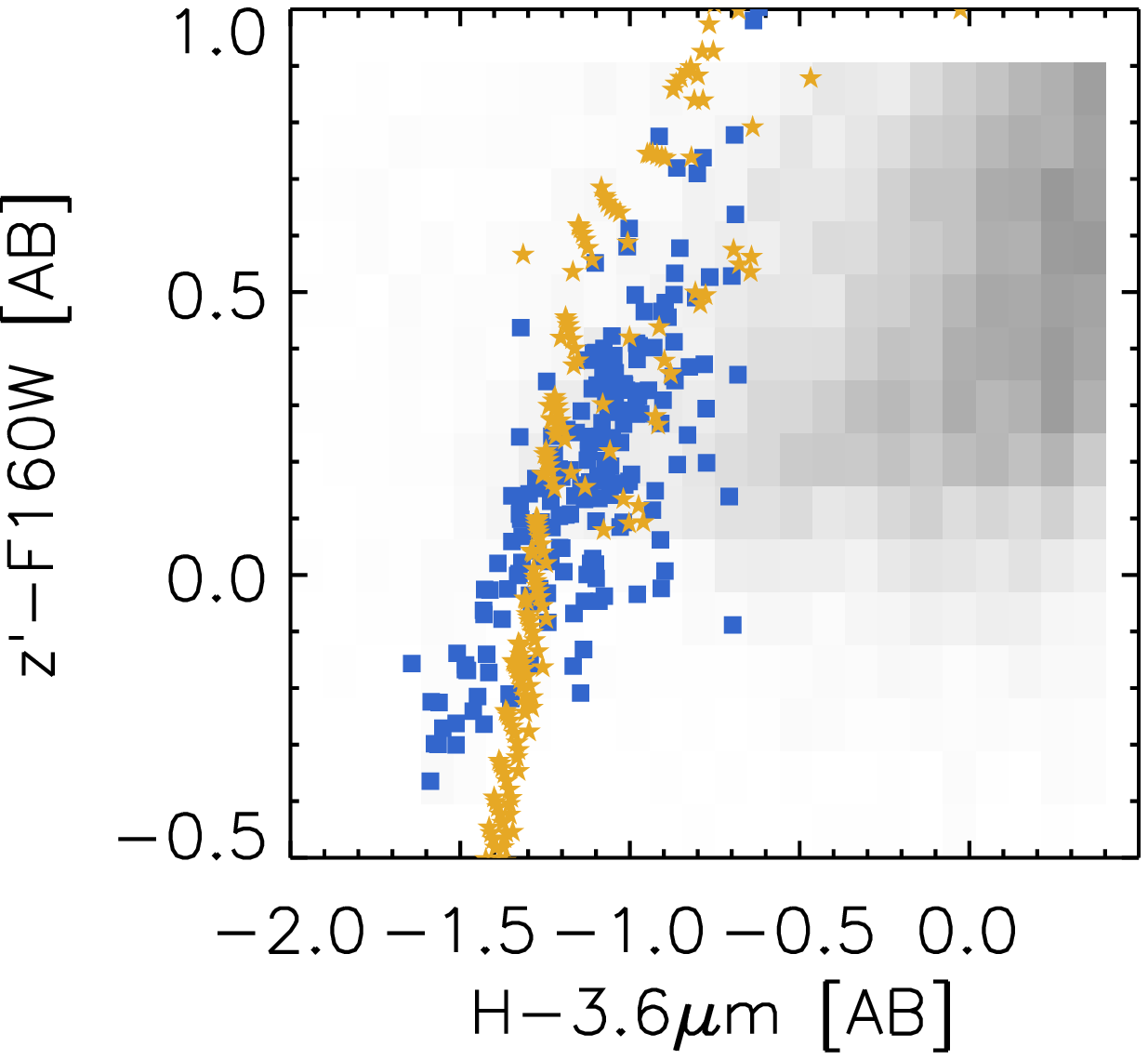} & \includegraphics[width=5.5cm]{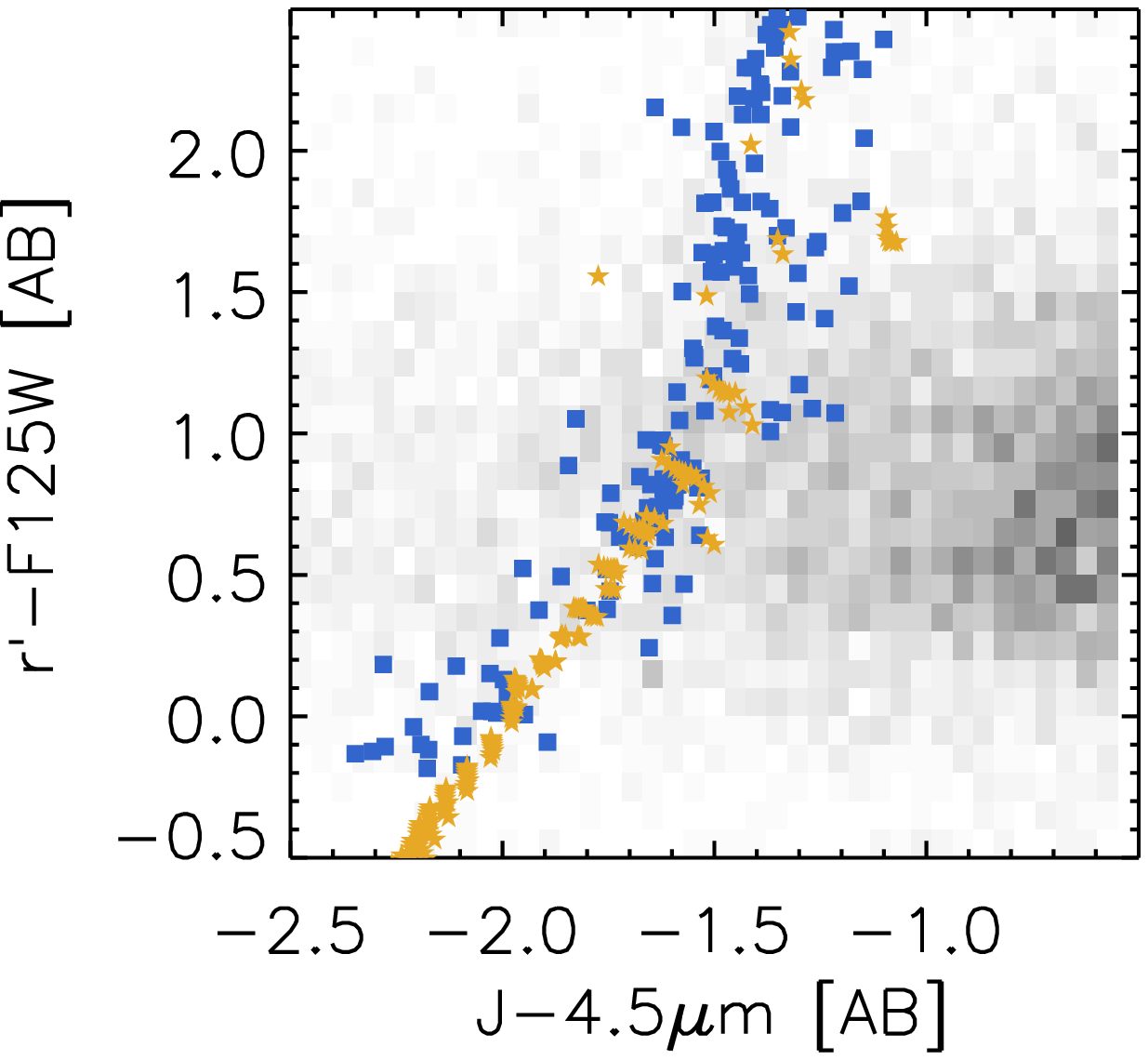}
\end{tabular}
\caption{Color-color diagrams. Objects brighter than F160W$=26.5$~mag in the EGS catalog are presented as a grey-scale density plot, with darker color marking denser regions. Objects brighter than  F160W=22~mag and considered to be stars are marked by filled blue squares (see main text for details). The yellow stars mark the colors of the model stars based on the BaSeL library, which have [M/H]=-0.5 and are typical of the Galactic halo stars. \label{fig:colcol}}
\end{figure*}

Figure \ref{fig:colcol} presents six color-color diagrams. We selected stars to be those objects with \texttt{SExtractor} \texttt{CLASS\_STAR} $>0.9$ and satisfying the following color-color criteria:
\begin{align}
 [z'-3.6\mu\mathrm{m}] < 0.73\times[g'-z']-1.8 & \mathrm{~for }& [g'-z'] \le 1.5  \\
[z'-3.6\mu\mathrm{m}] < 0.40\times[g'-z']-1.3  & \mathrm{~for }& [g'-z'] > 1.5
\end{align}

\begin{figure}
\hspace{-1cm}\includegraphics[width=9.5cm]{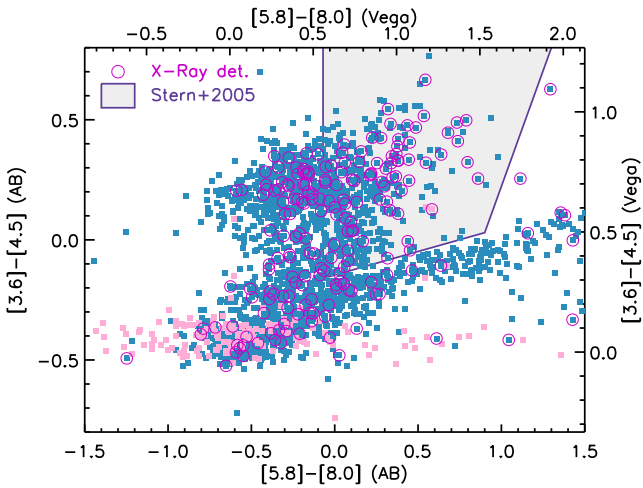}
\caption{IRAC color-color plot. The blue points represent sources in the catalog with S/N$>5$ in all four IRAC bands, while the pink points mark objects with \texttt{CLASS\_STAR}$>0.95$ and F160W$<22$~mag. The shaded region marks the AGN selection box defined by \citet{stern2005};  magenta open circles identify those objects with detection in the X-rays.  \label{fig:colcol_irac}}
\end{figure}

  Most of the point sources occupy a well constrained region across the plots, which should correspond to the stellar locus. We compared the stellar locus to the synthetic colors of stars from stellar synthesis models. Following what has been done for the CANDELS GOODS-S multi-wavelength catalog (\citealt{guo2013}), and considering that the stars in the EGS field should mainly be the halo population (due to its high latitude of $b\sim60\degree$) and thus should be metal-poor, we adopted a set of stellar models of low metallicities ([M/H]=-0.5) from the BaSeL stellar synthesis models (\citealt{lejeune1997,lejeune1998,westera2002}). All the plots show a good agreement between the observed colors of point sources with the colors from the synthetic library. Figure \ref{fig:colcol_irac} shows a color-color diagram built with IRAC fluxes with S/N$>5$ in all four IRAC bands. The point sources (most of them are likely stars) have approximately zero color in the Vega system, consistent with models of stellar atmospheres. The selection box, from \citet{stern2005}, identifies AGNs whose SEDs can be largely represented by a power-law (\citealt{donley2012}); the X-ray sources constitute the majority of the objects inside the  selection box, supporting our color measurements. The X-ray sources outside the selection box are likely AGNs whose host galaxy outshines the active nucleus.

\subsection{Comparison with publicly available catalogs}
\label{sect:public_photometry}

\begin{figure*}
\begin{tabular}{lr}
\hspace{-1cm}\includegraphics[width=9cm]{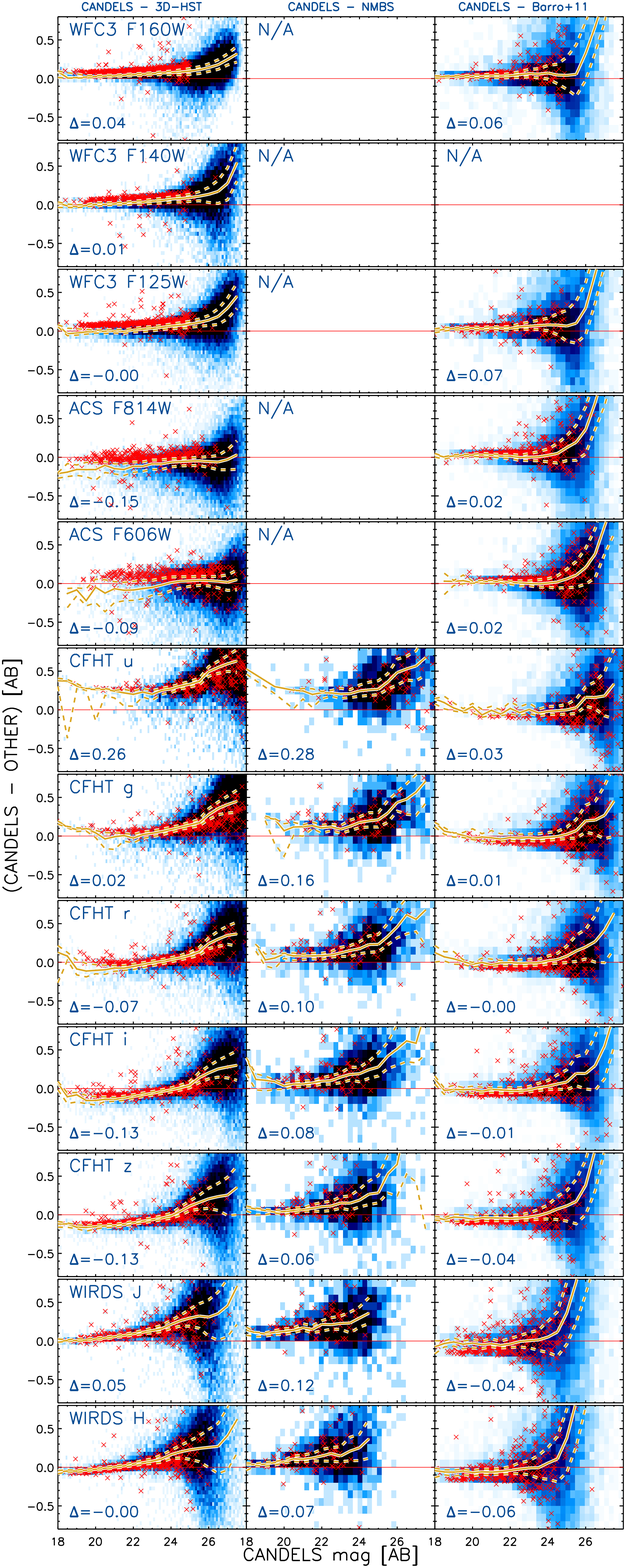} & \includegraphics[width=9cm]{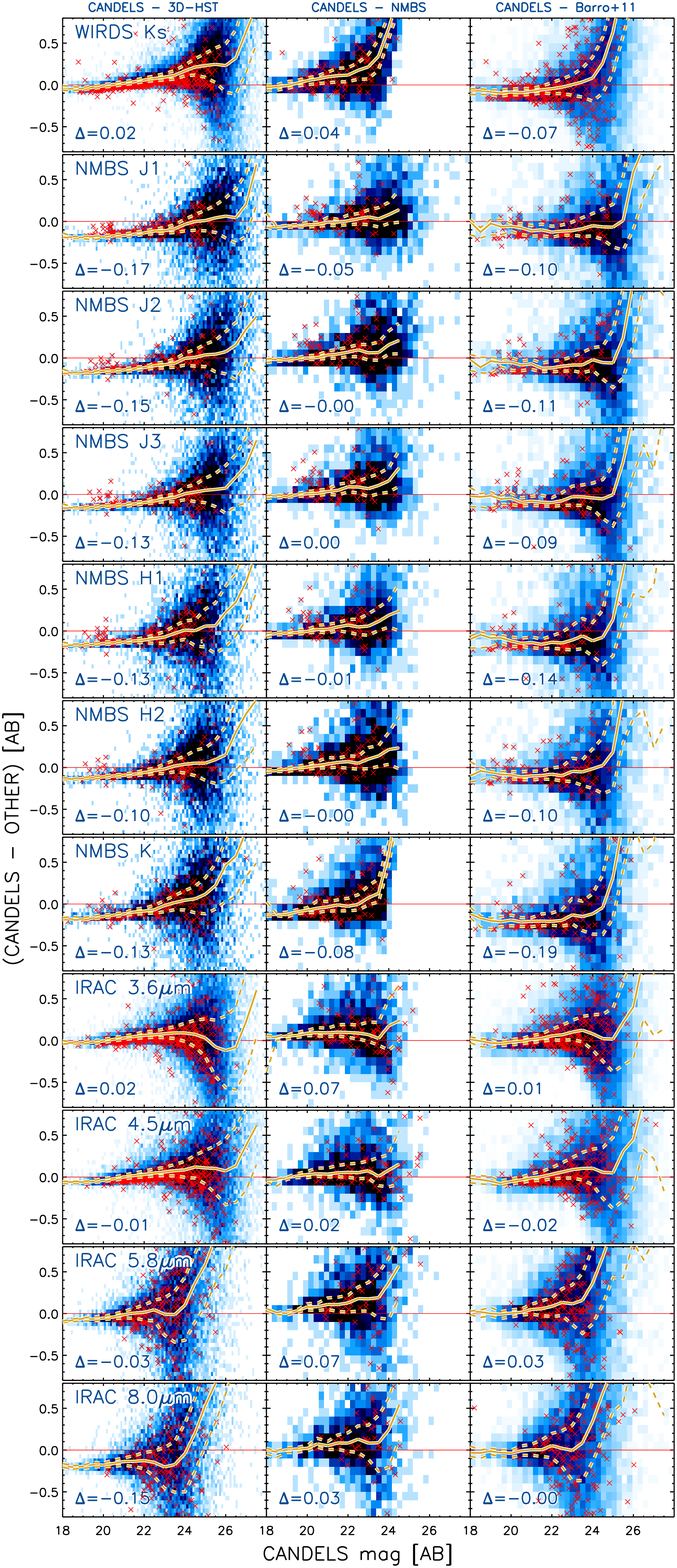}
\end{tabular}
\caption{Comparison of CANDELS EGS 22-band photometry (as labeled in the left-side panels) to three publicly available multi-wavelength catalogs for EGS: 3D-HST, NMBS and Barro+2011 (left to right columns, respectively). The blue density map represents each full matching dataset, while the red crosses mark objects with \texttt{CLASS\_STAR}$>0.95$ and F160W$<25$~mag. The solid yellow curve marks the running median, while the dashed curves encompass the 68\% of points. The number reported in the lower-left corner of each plot represents the median of the offset for the bright-end of the distribution (arbitrarily chosen to be $m<22$ mag, $m<21$ mag  and $m<22$ mag for 3D-HST, NMBS and Barro+2011, respectively). \label{fig:comp_phot}}
\end{figure*}

\begin{figure*}
\begin{tabular}{lr}
\hspace{-1cm}\includegraphics[width=9cm]{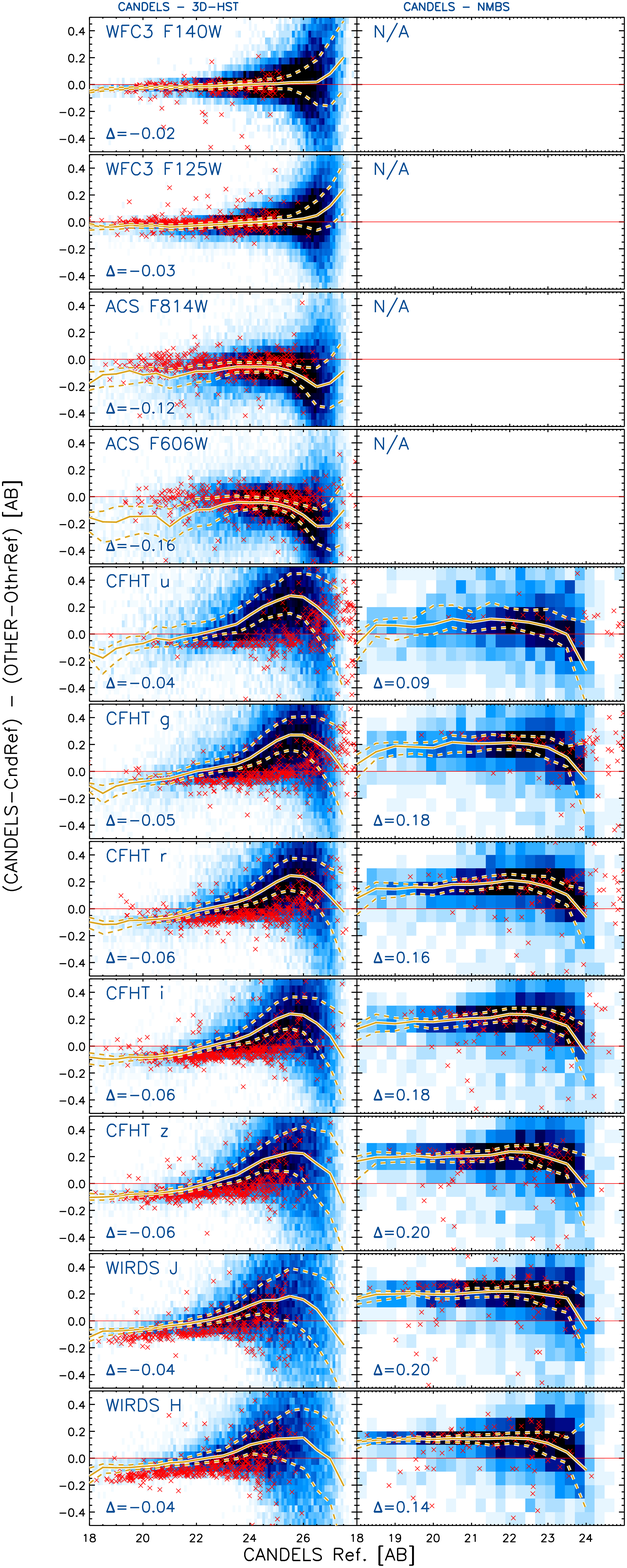} & \includegraphics[width=9cm]{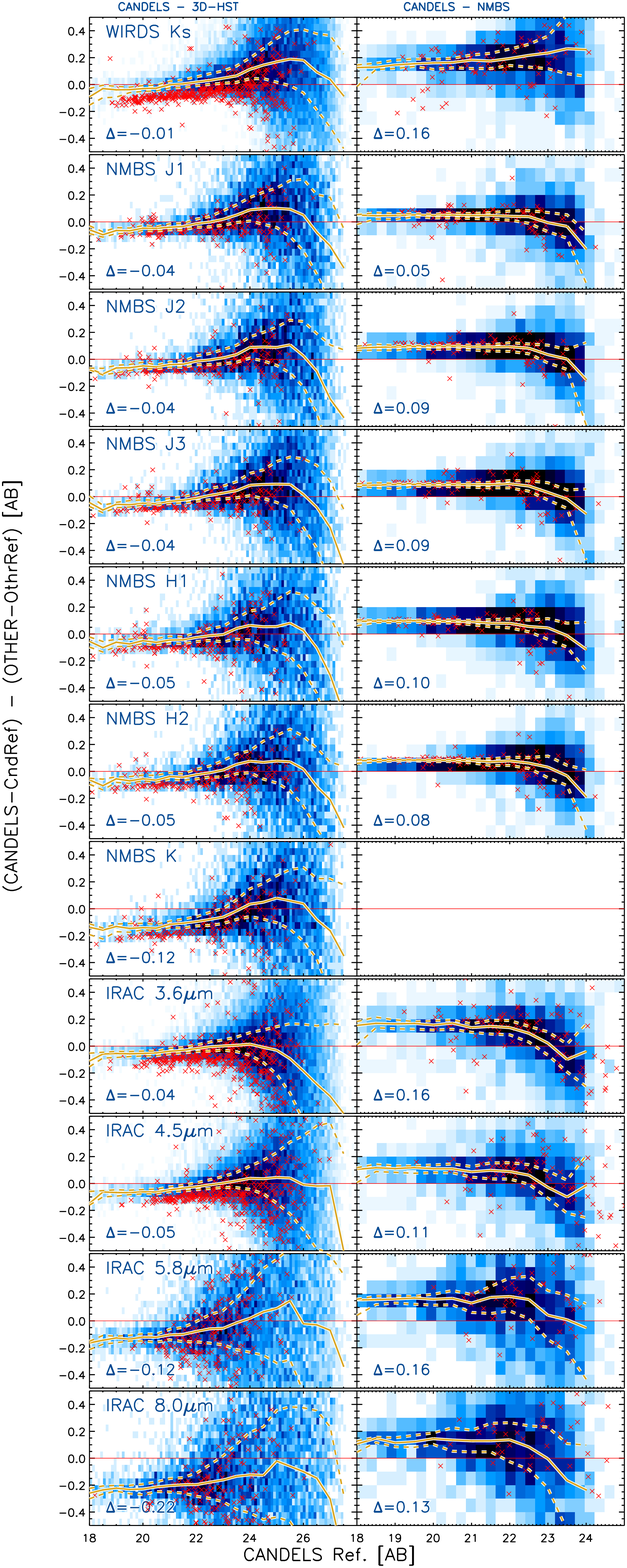}
\end{tabular}
\caption{Comparison of CANDELS EGS colors to the publicly available multi-wavelength catalogs from 3D-HST and NMBS. Zero-point offsets and Galactic extinction corrections have been removed from the 3D-HST and NMBS photometry. The plots present the quantity  $(m_\mathrm{x}-m_\mathrm{ref})_\mathrm{CANDELS}-(m_\mathrm{x}-m_\mathrm{ref})_\mathrm{Other}$ as a function of $m_\mathrm{x,CANDELS}$, where $m_\mathrm{x}$ is the magnitude in band x, $m_\mathrm{ref}$ is the magnitude in the  band which has been used by each team to recover total fluxes and the suffix $\mathrm{Other}$ refers to either 3D-HST or NMBS. Specifically, for 3D-HST $m_\mathrm{ref}$ corresponds to the magnitude in the WFC3/F160W band, while for NMBS it corresponds to the NEWFIRM K band. Other plotting conventions as in Figure \ref{fig:comp_phot}. \label{fig:comp_phot_color}}
\end{figure*}

A number of multi-wavelength photometric catalogs have been produced in the EGS field, (\citealt{ilbert2006}, \citealt{bundy2006}, 
\citealt{whitaker2011}, \citealt{barro2011} and \citealt{skelton2014}). This Section compares the CANDELS multi-wavelength photometry  to the three publicly available catalogs that have a broad wavelength coverage and include the IRAC bands. Specifically, we consider the catalogs from the 3D-HST survey (\citealt{skelton2014}), the catalog from the NMBS (\citealt{whitaker2011}), and the catalog presented by \citet{barro2011}. The comparison is done on a per-filter basis. Figure \ref{fig:comp_phot} shows the comparison of the CANDELS EGS photometry to that directly available from the public catalogs. However, as we explain in Sect. \ref{sect:comp_phot_3DHST}, the total fluxes for the 3D-HST and NMBS catalogs were the result of a number of \emph{corrections} (e.g, zero-point offsets, Galactic extinction, curve-of-growth). For this reason, Figure \ref{fig:comp_phot_color}  presents a comparison after removing those corrections, as this should provide flux measurements as they were originally recovered from the mosaics.

\subsubsection{3D-HST}
\label{sect:comp_phot_3DHST}

The 3D-HST multi-wavelength photometric catalog is described by \citet{skelton2014}. For the \textit{HST} data, the measurements were done using \texttt{SExtractor} on the mosaics PSF-matched to the F160W band. The measurements for the ground based data and the \textit{Spitzer} data were done by using \texttt{MOPHONGO} \citep{labbe2005,labbe2006,labbe2013}, a procedure similar to TFIT but with the difference that the photometry is done with a circular aperture after the target source is cleaned of its neighbours.

The measured fluxes of \citet{skelton2014} were subject to a number of \emph{corrections} with the aim of providing accurate measurements of the total flux. Based on Labb\'e et al. (2003; see e.g., their Figure 5) the \texttt{SExtractor} \texttt{FLUX\_AUTO} flux for faint sources can be systematically lower than the intrinsic total flux.  \citet{skelton2014} derived the offset between the two (i.e., the aperture correction) from the growth curve constructed from bright point sources, and calculated the fraction of the light enclosed by the circular aperture that has the same area as the Kron ellipse used in determining \texttt{FLUX\_AUTO}. The measurement of the total flux in the F160W band was then obtained by applying the corresponding aperture correction to the \texttt{FLUX\_AUTO}. As the size of the Kron ellipse depends on the flux and is smaller for a fainter source, the applied correction depended on the source brightness as well, and it was larger for a fainter source. Total fluxes for the other bands  were recovered as

$$f_{\mathrm{tot},b} = f_{\mathrm{ap},b} \times \frac{f_{\mathrm{tot},\mathrm{F160W}}}{f_{\mathrm{ap},\mathrm{F160W}}}$$

\noindent where $f_{\mathrm{tot},b}$  and $f_{\mathrm{ap},b}$ are the total and aperture flux for band $b$, respectively, while $f_{\mathrm{tot},\mathrm{F160W}}$ is the total flux in the F160W band. UV-to-K-band fluxes were corrected for the Galactic extinction following the  extinction law of Cardelli et al. (1989; see Tab. 4 in \citealt{skelton2014}). Finally, in  \citet{skelton2014} the photometric zeropoints were iteratively adjusted by computing the differences between the measured fluxes and the expected ones from the best-fit  galaxy SEDs; the applied zeropoint offsets ranged from -0.22 mag in the $u^*$ band to +0.17 mag in the NMBS/J1 band, although for 16 out of the 22 bands the zeropoint corrections are within 0.05 mag.

Figure \ref{fig:comp_phot} shows the flux comparison between the CANDELS/EGS and 3D-HST catalogs. In general, the agreement is good to excellent. The offsets vary from $\sim0.02$ mag (e.g., \textit{HST}/WFC3, WIRCAM/WIRDS) to  $\sim0.2$ mag (e.g., CFHT $u^*$, NMBS J1 and IRAC 5.8$\mu$m). However, in most cases the difference between the two catalogs is not a simple offset but has a dependence on the flux, i.e., the difference increases at fainter magnitudes.

The differences and their flux-dependent behavior likely stem from the various systematic corrections that the 3D-HST catalog has applied, namely, the total flux recovery on an object-by-object basis, the photometric zeropoint adjustments and the Galactic foreground extinction corrections. 

A comparison of colors is more straightforward in this context, as this largely (although still not completely) circumvents the differences in the total flux recovery. This comparison is shown in Figure 15, where we use the colors relative to F160W, i.e., we consider $(m_\mathrm{x}-m_\mathrm{F160W})_\mathrm{CANDELS}-(m_\mathrm{x}-m_\mathrm{F160W})_\mathrm{3D-HST}$, where $m_\mathrm{x}$ is the magnitude in band x. The colors based on the 3D-HST catalog were computed after the removal of the zeropoint adjustments and the Galactic foreground extinction corrections. Indeed, our limited knowledge on the galaxy SEDs, especially at cosmological distances, prevents us from a determination of zeropoint adjustments to levels better than $\sim20\%$ (see e.g., the comparison of galaxy SEDs with observed photometry of \citealt{brown2014}, but also e.g. \citealt{brammer2008} for an attempt to deal with this problem using a  template error function). Furthermore, as we already pointed out in Sect. \ref{sect:cat_creation}, the amount of extinction correction depends on the specific model adopted. For these reasons we believe that a more direct comparison between different catalogs would be more meaningful when made without either of these corrections.

The agreements in all bands are now much improved as compared to Figure \ref{fig:comp_phot}, which supports our interpretation mentioned above. Specifically, the $\sim0.2$~mag offsets observed for some of the CFHT and NMBS bands have been largely reduced. 

Considering that the flux comparisons done using the colors relative to the F160W band have the main effect of removing any dependence on aperture correction, the reduced offset and the flattening of the color difference as a function of magnitude suggest that the offsets and trend observed in Figure \ref{fig:comp_phot} for these bands are likely the result of the corrections applied to the 3D-HST photometry to convert aperture into total fluxes. The agreement at the faint end is not as good as in the bright regime, which can be attributed to the smaller S/N for fainter sources and larger background flux. Nevertheless, all this suggests that the flux measurements in both catalogs have been performed in a self-consistent manner.

\subsubsection{NMBS}

\label{sect:comp_phot_NMBS}

The NMBS survey \citep{whitaker2011} was a medium-band NIR survey over the AEGIS and COSMOS fields. Source detection was performed on the K-band.
Flux measurements for the optical and NIR images were performed by using \texttt{SExtractor} in dual-image mode on the mosaics PSF-matched to the broadest PSF (i.e., the PSF of the H1 band). The K-band \citet{kron1980} fluxes were converted to total fluxes using a prescription similar to that adopted for the 3D-HST F160W total fluxes (see \citealt{whitaker2011} for details).  The aperture fluxes for the \textit{Spitzer} IRAC bands were measured using a procedure very similar to that used for the 3D-HST photometry and converted to total flux using the ratio between the total and aperture flux in the K-band. The NIR medium-band filters pin-point the Balmer/4000\AA~break at $1.5\lesssim z\lesssim3$, allowing for accurate photometric redshifts measurements of objects in this range of redshifts. Because NMBS was carried out before the CANDELS project started, the multi-wavelength photometric catalog does not include \textit{HST}/WFC3 data. 

Figure \ref{fig:comp_phot} shows that the fluxes from both catalogs agree reasonably with each other, with the absolute differences generally being within  $\sim0.1$ mag in the bright regime. Similar to the case for 3D-HST,  the differences show a flux-dependent trend, which is likely due to the systematic corrections applied in the NMBS catalog. Comparison of the difference in colors (Figure \ref{fig:comp_phot_color}) shows much improved agreement in the sense that the flux-dependent behavior is largely removed. The amplitudes of the systematic offsets are similar to those found by the 3D-HST team  (see e.g. Figure 34 of \citealt{skelton2014}).

\subsubsection{Barro+2011}

The multi-wavelength photometric catalog of \citet{barro2011} was assembled by crossmatching the \texttt{SExtractor} detections in IRAC $3.6\mu$m and $4.5\mu$m to the detections performed independently in each band, using a search radius of $2\farcs0$ Subaru R-band imaging was used to reduce the multiple matches arising from the larger IRAC PSF. Photometry was carried out in each band using the \citet{kron1980} elliptical aperture obtained from the Subaru R-band mosaic. When multiple counterparts to IRAC sources were found,  IRAC fluxes were measured in $0\farcs9$ apertures after deblending using a template fitting algorithm similar to that implemented in \texttt{TFIT}. Total magnitudes in the IRAC bands for such sources were finally calculated by applying the aperture corrections derived  from the PSF growth curves.

The comparison between the CANDELS EGS photometry and that of \citet{barro2011}  (Figure \ref{fig:comp_phot}) shows a very good agreement for most of the bands, with offsets of $\lesssim 0.1$~mag for the brighter sources. We do not show any color comparisons in Figure \ref{fig:comp_phot_color} corresponding to what we do in Section \ref{sect:comp_phot_3DHST} and \ref{sect:comp_phot_NMBS};  indeed such a comparison would not have the advantages as in the previous two cases because no further corrections were applied to the \texttt{SExtractor flux\_auto} measurements for all bands.

\subsubsection{Summary}

In this Section we compared the flux measurements in all bands from our catalog to the corresponding ones of the matching objects of three public catalogs: \citet{skelton2014}, \citet{whitaker2011} and \citet{barro2011}. Since each team assembled their catalog using different tools and/or configurations and implemented different ways of measuring \emph{total} fluxes, we considered two different approaches: 1) we compared the \emph{total} fluxes as directly provided by each team, and 2) we compared fluxes as closely as possible to those initially recovered from the mosaics, removing any further correction that was successively applied (e.g., zero point adjustments, galactic extinction, aperture corrections). As such, comparisons in this second case should provide a more reliable check on whether systematics in flux measurements exist between two different catalog. We presented the comparisons in Figure \ref{fig:comp_phot} and Figure \ref{fig:comp_phot_color}, respectively. 

Figure \ref{fig:comp_phot} shows that overall the flux measurements in our catalog systematically differ from those in the other three catalogs by up to $\sim10\%$, with only few cases of systematic differences reaching $\sim20\%$. However, trends with magnitude are also present. These trends are almost totally absent when comparing to \citet{barro2011}, while they are more visible when comparing to \citet{skelton2014} and \citet{whitaker2011}. This flux-dependent behavior likely arise from the systematic corrections that were applied, namely, the total flux recovery on an object-by-object basis, the photometric zeropoint adjustments and the Galactic extinction corrections. 

In order to provide a first test to the above hypothesis, in Figure \ref{fig:comp_phot_color} we presented a comparison after removing zero point corrections, galactic extinction and considering a comparison in colors relative to the F160W flux. Indeed, the trends with magnitude decreased sensibly or even disappeared. The average offsets are within 0.1~mag for the \citet{skelton2014} case, which increases the confidence on our flux measurements, although for \citet{whitaker2011} we register an increase in offset values. As the main effect of considering differences between colors relative to the detection band is to cancel any systematics from aperture correction, the reduced trends strongly suggest that the disagreement observed for some of the bands in Figure \ref{fig:comp_phot} are the result of the aperture corrections applied by the other teams.

The origins of systematic errors of the order of up to $10\%$ are very difficult to track as they could be a consequence of different mixtures of fine-tuning parameters for mosaic creation (including zero point determinations), background subtraction, analysis thresholds and other corrections to `total'  fluxes. Although in this section we presented the comparison to public catalogs, the assessment of the origin of such systematics goes beyond the scope of this paper.

\section{Photometric redshifts and stellar masses}

Using the multi-wavelength photometry presented in the previous Sections, we derived photometric redshifts and stellar masses for all the sources in the catalog, measured following the methods of \citet{dahlen2013} and \citet{mobasher2015a}, respectively. We remind the reader that our catalog also contains a match to 246 X-Ray sources, which are likely AGN. Appropriate measurement of their photometric redshifts and stellar masses requires the inclusion of SED templates which take into account the AGN contribution, resulting in otherwise unreliable values. For this reason, for these sources we include the photometric redshift values from   \citet{nandra2015} computed  adopting specific priors and SED templates (see also Appendix \ref{appendix:mstar})\footnote{These photometric redshifts and the associated P(z) are also available from http://www.mpe.mpg.de/XraySurveys/AEGIS-X}. Also, stellar masses for such objects should be computed taking into account the presence of the AGN affecting optical and NIR data, which could otherwise boost the stellar mass measurement. At this stage this part of the computation is not ready and for these sources masses should be considered upper limits.

\subsection{Photometric redshifts}
\label{sect:photoz}
The multi-band photometric data were independently analysed by ten different groups within the CANDELS collaboration. Each group adopted a different code and/or set of SED templates\footnote{Overall, our adopted SED templates (\citealt{dahlen2013, mobasher2015a}) did not include heated dust emission, which will only impact the IRAC 5.8 and 8.0$\mu$m bands when the objects are at very low redshifts ($z\lesssim 0.3$). Only a very small number of objects in our catalog could be impacted, however. Only 187 objects ($\sim 0.4$\% of the full catalog) are expected to be at $z\lesssim 0.3$ and might be impacted by dust emission (having S/N$>3$ in 5.8 $\mu$m).}. The NMBS J1, J2, J3, H1, and H2 data were excluded because of their relatively shallow depth and the limited overlap with the F160W footprint. The WIRCAM J and H data were also excluded because these two bands are similar to the WFC3 F125W and F160W, respectively, but the data are much shallower than the latter two. As a training set, 840 spectroscopic redshifts from the DEEP3 program were also provided to each team. As a common practice, the photometric zeropoints were fine-tuned by each group during the process in order to minimise the overall residuals between the measured flux densities and those expected from the best-fit templates. Such fine adjustments varied slightly among groups due to the differences in their methods and their adopted template libraries.  For this reason, the multi-wavelength photometric catalog does \emph{not} include such offsets. However, the average photometric zeropoint offsets adopted by each group are reported in Table \ref{tab:zpoint_corr}.

\begin{figure}
\includegraphics[width=9cm]{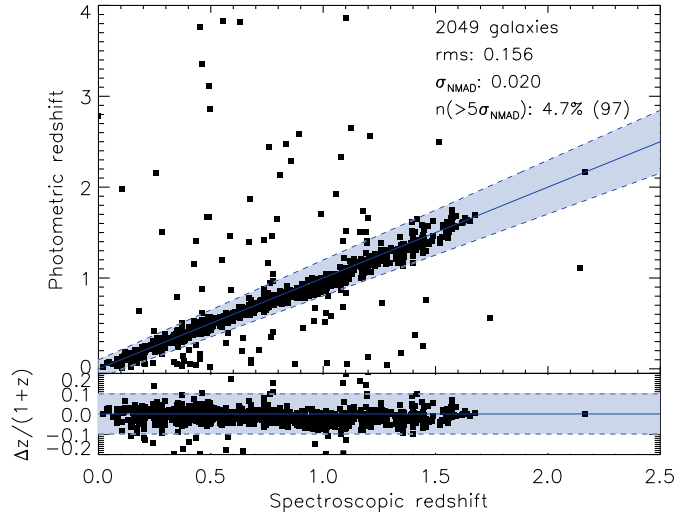}
\caption{Comparison between the spectroscopic redshifts from DEEP2+3 and the CANDELS photometric redshifts. Sources with  detection in the X-ray and \texttt{CLASS\_STAR}$>0.85$ were excluded from the sample. The top panel shows the direct comparison between the spectroscopic redshifts and photometric redshifts, while the bottom panel presents $(z_{\mathrm{phot}}-z_{\mathrm{spec}})/(1+z_{\mathrm{spec}})$. In both panels, the solid blue line indicates the 1:1 correspondence while the filled light blue region encompasses the region within $5\times \sigma_{\mathrm{NMAD}}$. \label{fig:zphot_zspec}}
\end{figure}

\begin{figure*}
\includegraphics[width=18cm]{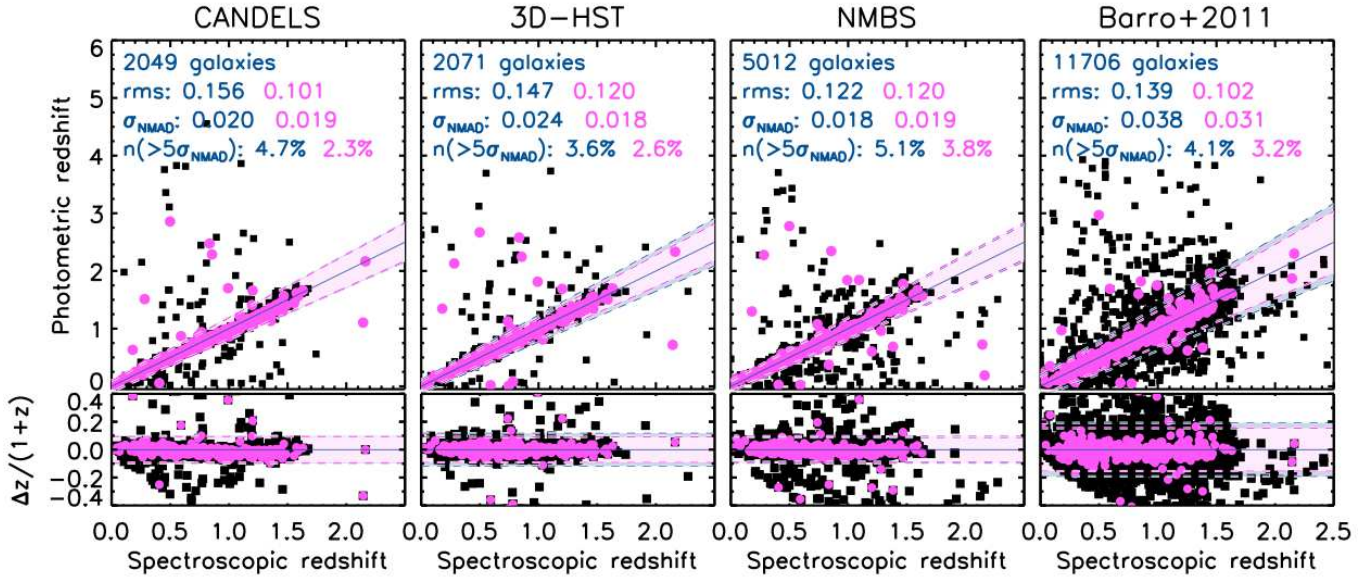}
\caption{Comparison between the spectroscopic redshifts from DEEP2+3 and the photometric redshifts from this work, the 3D-HST survey, NMBS and Barro et al. (2011), left to right, respectively. The magenta points mark the  sources in common among the three catalogs with DEEP2+3 spectroscopic redshift, not detected in the X-ray and with \texttt{CLASS\_STAR}$<0.85$ ($ 558$ objects). The labels present the dispersion and fraction of outliers for the full sample and for the subsample in common among the three surveys (blue and magenta labels, respectively). The 5$\sigma_\mathrm{NMAD}$ limits for the full and for the common samples are indicated by the light blue and pink regions respectively.  \label{fig:zphot_zspec_external}}
\end{figure*}

\begin{table}
\caption{List of average zeropoint offsets applied to the photometric catalog. \label{tab:zpoint_corr}}
\begin{threeparttable}
\begin{tabular}{cc}
\hline
\hline
Band & ZP$_\mathrm{factor}$\tnote{a} \\
\hline
CFHT $u^*$    &  1.05249 \\
CFHT $g'$    &  0.988473 \\ 
CFHT $r'$    &  0.998439 \\
CFHT $i'$    &  0.991876 \\
CFHT $z'$    &  0.993045 \\
ACS F606W     &  0.936776 \\
ACS F814W     &  0.972712 \\
WFC3 F125W     &  1.02849 \\
WFC3 F140W     &  1.02231 \\
WFC3 F160W     &  1.03405 \\
WIRCam $K_\mathrm{S}$  &  0.964757 \\
NEWFIRM $K$ &  0.883957 \\
IRAC 3.6$\mu$m     &  1.00648 \\
IRAC 4.5$\mu$m     &  0.993963 \\
IRAC 5.8$\mu$m     &  1.0 \\
IRAC 8.0$\mu$m     &  1.0 \\
\hline
\end{tabular}
\begin{tablenotes}
\item {\bf Notes:}
\item[a]  The zero-point offsets are such that Flux$_\mathrm{corrected} = $ZP$_\mathrm{factor} \times$ Flux$_\mathrm{original}$. The zeropoint offsets were applied only for photometric redshift estimates. No zeropoint correction is present in the multi-wavelength photometric catalog. 
\end{tablenotes}
\end{threeparttable}
\end{table}

A number of codes for the measurements of photometric redshifts are available today (see e.g., \citealt{dahlen2013} for a list). However, discrepancies in the redshift measurements still exist among themselves and with respect to spectroscopic redshifts. Using spectroscopic redshifts as reference, \citet{dahlen2013} showed that the median of their 13 sets of photometric redshifts  provide the best measurements. This is most likely because systematic effects among these 13 groups have canceled out when taking the median.  For this reason, and because most of the ten configurations adopted in this work for the measurement of photometric redshifts coincide with those of \citet{dahlen2013}, we adopted the median of the ten photometric redshift estimates as the final measurements in our catalog. Figure \ref{fig:zphot_zspec} presents the comparison of these photometric redshifts to the  DEEP2+3 spectroscopic redshifts when available. Our photometric redshifts are tightly distributed around the spectroscopic redshifts, with a low dispersion of $\sigma=0.020$ and only $\sim $5\% catastrophic outliers, defined as $>5\sigma$ difference between photometric and spectroscopic redshifts. The distribution of catastrophic outliers shows a peak at $z_\mathrm{phot}\sim0.1$.  We selected the outliers with $z_\mathrm{phot}<0.2$. This subsample included 20 galaxies. Visual inspection of their SEDs revealed that 17 sources are characterised by strong emission lines, while the photometry of the remaining 3 objects shows inconsistent measurements in several bands.

\begin{figure*}
\begin{tabular}{ccc}
\includegraphics[width=5.5cm]{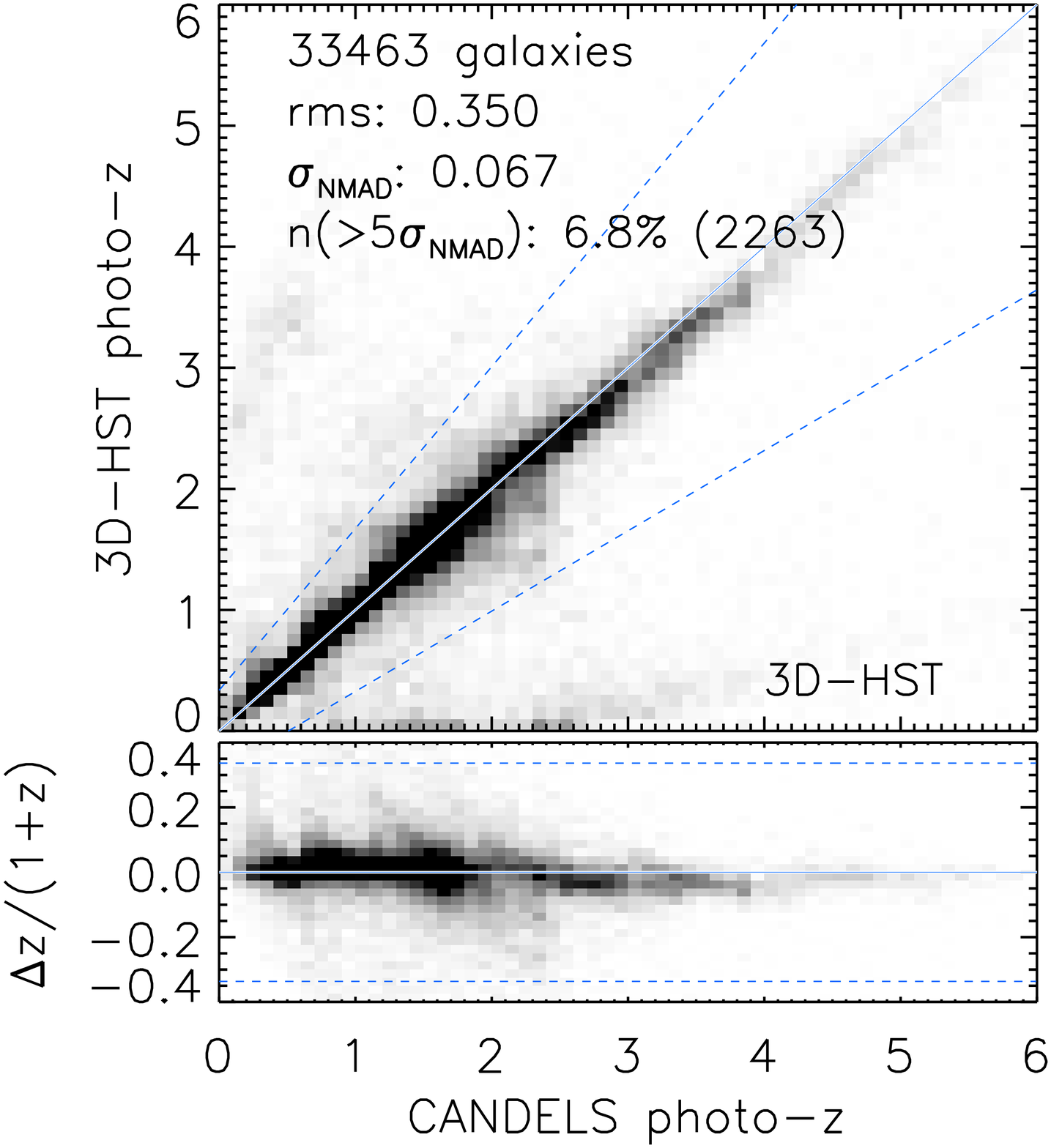} & \includegraphics[width=5.5cm]{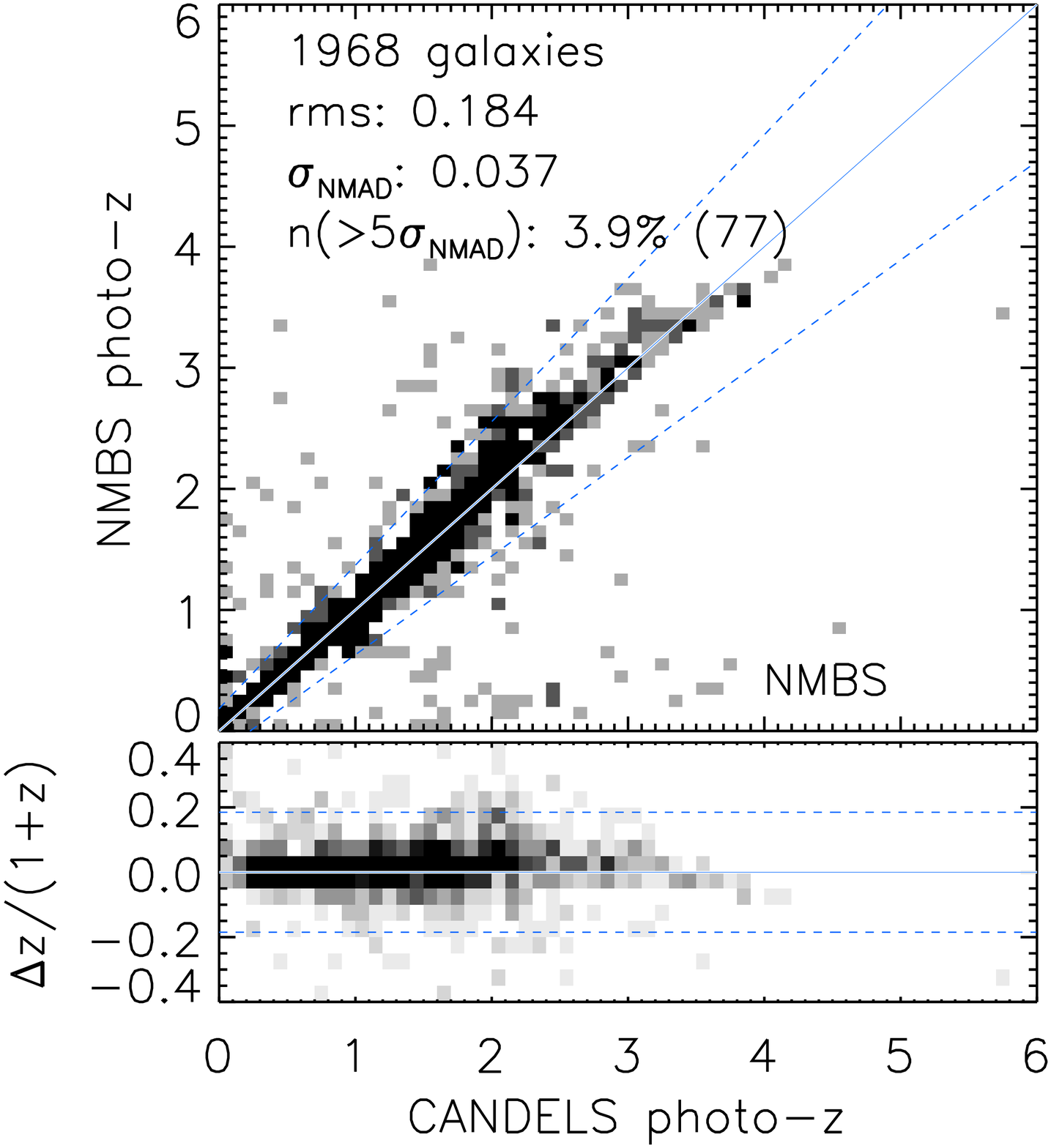} & \includegraphics[width=5.5cm]{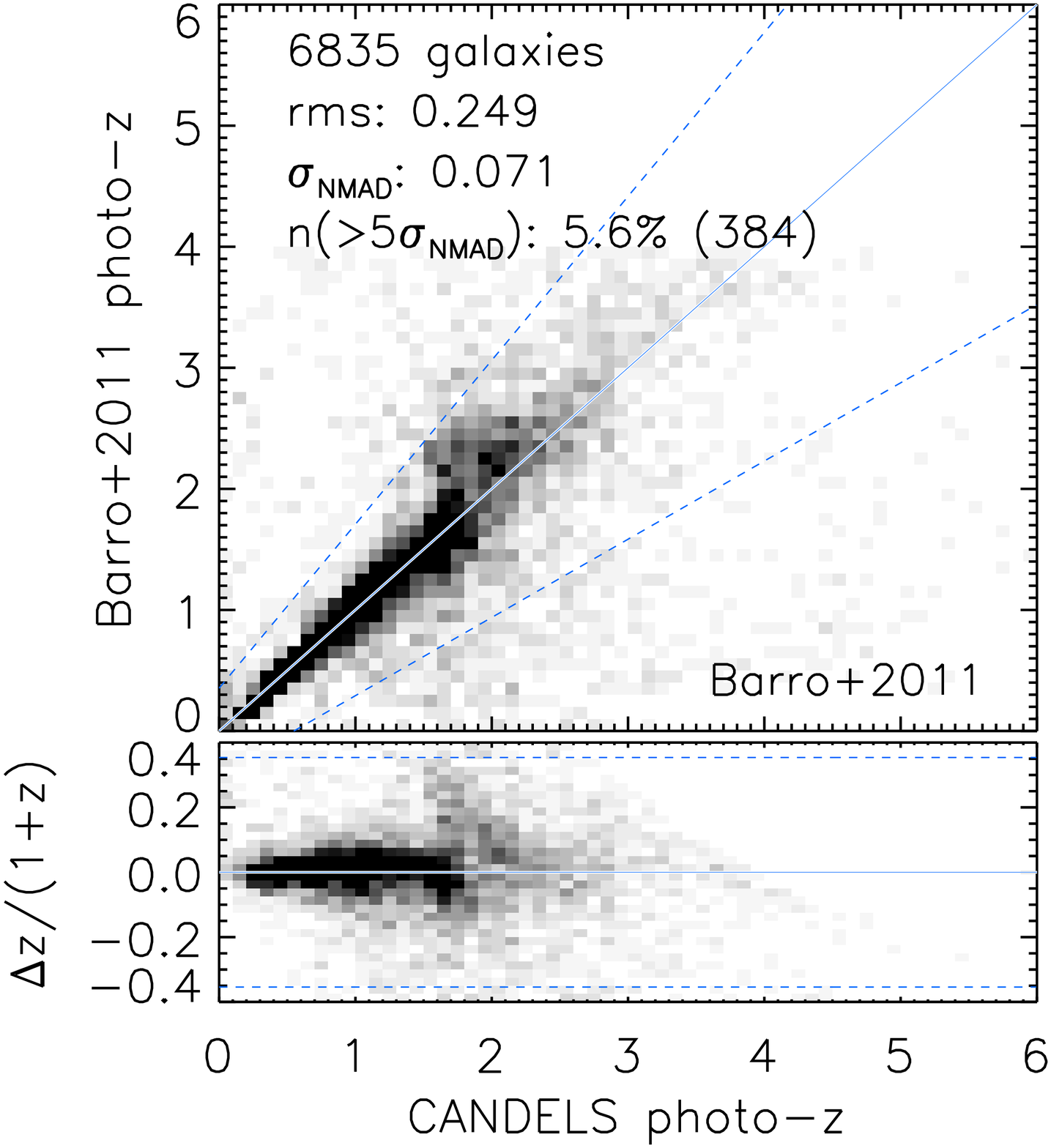}
\end{tabular}
\caption{\emph{Top panels:} Comparison between the photometric redshifts measured using the CANDELS multi-wavelength photometric catalog and the photometric redshifts of the matching sources from three public catalogs: 3D-HST (\citealt{skelton2014}), NMBS (\citealt{whitaker2011}) and \citet{barro2011b}, left to right panel, respectively. The comparison is shown in the form of a density plot, with darker regions corresponding to an underlying higher density of population. The solid blue line marks the 1:1 relation, while the dashed lines delimit the $5\times\sigma_{\mathrm{NMAD}}$ region. Indicated by the labels are the $\sigma_{\mathrm{NMAD}}$ and the fraction of objects lying outside the $5\times\sigma_{\mathrm{NMAD}}$ region ($n(>5\sigma_{\mathrm{NMAD}})$); the number in parenthesis indicates the total number of outliers. \emph{Bottom panels}: Same as above, but showing $(z_{\mathrm{phot,X}}-z_{\mathrm{phot,CND}})/(1+z_{\mathrm{phot,CND}})$ as a function CANDELS photometric redshifts, where  $z_{\mathrm{phot,X}}$ is the photometric redshift from the catalog X. Most of the galaxies show consistent photometric redshift measurements across the three catalogs, with negligible offset. The dispersion is characterised by a $\sigma_{\mathrm{NMAD}}\lesssim0.07$ and a fraction of catastrophic outliers below 7\%. \label{fig:zphot_3d}}
\end{figure*}

\begin{table*}
\caption{Configurations adopted by each team for the measurement of stellar masses. \label{tab:mstar_par}}
  \begin{threeparttable}
\begin{tabular}{ccccccc}
\hline
\hline
 Label\tnote{a} &                Code               &               SSP\tnote{b}         &          SFH\tnote{c}            &           $Z/Z_\odot$         &          IMF         & Neb. lines \\
\hline
M2 &  FAST \citep{kriek2009} & BC03  & $\tau$ & 1 & \citet{chabrier2003} & no \\  
M6 & own (PI: Fontana) & BC03  & $\tau$ & 1 & \citet{chabrier2003} & no \\  
M10 & HyperZ \citep{bolzonella2000} & MA05 & $\tau$,  const., trunc. & 0.2-2.5 & \citet{chabrier2003} & no \\ 
M11 & Le Phare \citep{ilbert2006} & BC03 & $\tau$\ & 0.4, 1 & \citet{chabrier2003}& yes\\ 
M12 & WikZ \citep{wiklind2008} & BC03 & del-$\tau$ & 0.2-2.5 & \citet{chabrier2003} & no \\
M13 & FAST \citep{kriek2009} & BC03 & $\tau$ & 1 & \citet{chabrier2003} & no \\ 
M14 & SpeedyMC \citep{acquaviva2012} & BC03 &  $\tau$, del-$\tau$, const., lin. incr., incr.-$\tau$ & 1 & \citet{chabrier2003}& yes \\ 
M15 & own \citep{lee2010} & BC03 & del-$\tau$ & 0.2-2.5 & \citet{chabrier2003} & no \\  
\hline
\end{tabular}
\begin{tablenotes}
\item {\bf Notes:}
\item[a] Labels as defined in \citet{mobasher2015a}
\item[b] Simple stellar population models are: BC03$\equiv$\citet{bruzual2003}; MA05$\equiv$\citet{maraston2005}
\item[c] Star-formation histories are: $\tau\equiv$ Exponentially declining; const. $\equiv$ Constant; trunc. $\equiv$ Exponentially decreasing with truncation; del-$\tau\equiv$ Delayed exponential (SFH$\propto t \times \exp(-t/\tau)$); lin. incr. $\equiv$ Linearly rising; incr.-$\tau\equiv$Exponentially rising.
\end{tablenotes}
\end{threeparttable}
\end{table*}

\begin{table*}
\caption{Configuration adopted by three external programs for the computation of stellar masses$^{\mathrm{a}}$.\label{tab:mstar_other}}
\begin{threeparttable}
\begin{tabular}{ccccc}
\hline
\hline
Catalog & Code & SSP & SFH& IMF   \\
\hline
3D-HST & FAST \citep{kriek2009} & BC03 & del-$\tau$  & \citet{chabrier2003} \\
NMBS & FAST \citep{kriek2009} & BC03 & $\tau$   & \citet{chabrier2003}\\
\citet{barro2011b} & Rainbow \citep{perezgonzalez2008} & P\'EGASE & $\tau$  & \citet{salpeter1955}\\
\hline
\end{tabular}
\begin{tablenotes}
\item{\bf Notes:}
\item[a]  Same abbreviations as for Table \ref{tab:mstar_par}
\end{tablenotes}
\end{threeparttable}
\end{table*}

\subsection{Comparison to other photometric redshift catalogs}

A number of different groups have derived photometric redshifts in the EGS field (\citealt{ilbert2006}, \citealt{bundy2006}, 
\citealt{whitaker2011}, \citealt{barro2011b} and \citealt{skelton2014}). Here we compare our CANDELS photometric redshifts to those from three other public catalogs, namely,  the 3D-HST \citep{skelton2014}, the NMBS \citep{whitaker2011} and \citet{barro2011b} measurements. As mentioned in Section \ref{sect:public_photometry}, these three programs have covered a wide wavelength range and also include IRAC data, which are key for more accurate redshift measurements.

The photometric redshifts for the 3D-HST catalog were determined with the \texttt{EAzY} software \citep{brammer2008} using linear combinations of a set of seven templates from the P\'EGASE models \citep{fioc1997} with the addition of a young and dusty template and of a red and old template from \citet{whitaker2011}. For the NMBS catalog, photometric redshifts were measured using \texttt{EAzY} and adopting its default template set, generated from the P\'EGASE population synthesis code (see \citealt{brammer2008} for details) with the addition of a young and dusty template and of an old, red galaxy (\citealt{whitaker2011}). The photometric redshifts of \citet{barro2011b} were derived using the \texttt{rainbow} code (\citealt{perezgonzalez2008}) and the P\'EGASE model templates. Figure \ref{fig:zphot_zspec_external} shows the accuracies of these three sets of photometric redshifts by comparing to the spectroscopic redshifts from the DEEP2+3 catalogs. The different number of objects with spectroscopic redshifts in each catalog is the result of the different detection band, depths and overlap of the detection band with the DEEP2+3 footprint. In this comparison, the NMBS results show slightly lower dispersion around the spectroscopic redshifts with respect to the other two sets, which could be due to its five medium-band NIR filters that can better  pinpoint the Balmer/4000\AA~break at $z\lesssim 2.5$. The higher dispersion of \citet{barro2011b} photometric redshift measurements compared to the other catalogs is likely the result of the absence of deep NIR data bracketing the Balmer/4000\AA~break\footnote{Although in Figure \ref{fig:comp_phot} we compare to the \textit{HST} WFC3 bands data from \citet{barro2011}, these were not included by \citet{barro2011b}  in the computation of the photometric redshifts and stellar masses.}, and it highlights the importance of the inclusion of deep NIR data in the measurement of photometric redshifts.

Figure \ref{fig:zphot_3d} compares these three sets of photometric redshifts to CANDELS.  The agreement shows $< 7$\% of catastrophic outliers. At $z\gtrsim 2$, the 3D-HST photometric redshifts appear to be systematically lower than ours by $\Delta z/(1+z)\sim 0.04$. The agreement with the NMBS ones is significantly better than the other two sets  (a factor $\gtrsim1.5\times$). This is partly because NMBS is the shallowest among all, and thus the objects going into this comparison are predominantly the brighter ones in our catalog, which have higher accuracies. The systematically higher photometric redshifts of \citet{barro2011b} for $z \gtrsim 2$ are likely the result of the lower S/N NIR data available in the catalog of \citet{barro2011}. Indeed, the Balmer break enters the $J$ band at $z\sim1.7$. Shallower data in the $J$ band can favour photometric redshift solutions where the Balmer break has actually already left the $J$ band and has entered the $H$ band, even in those cases where the Balmer break still lie in the $J$ band. The net effect is thus to bias the redshift measurement towards higher values. The observed offset values in photometric redshift measurement are consistent with this hypothesis. At $z\gtrsim2.5$ the effect is less marked as the SED immediately blue-ward of the Balmer break is probed by both the $J$ and $H$ bands, increasing the effective S/N of the break. These plots indicate that the main source of discrepancy in photometric/spectroscopic redshifts between different methods is the 
sources that are not in common. For these other things might be wrong as well, such as spectroscopic redshifts, matching and identification.

\begin{figure*}
\begin{tabular}{cc}
\includegraphics[width=8.5cm]{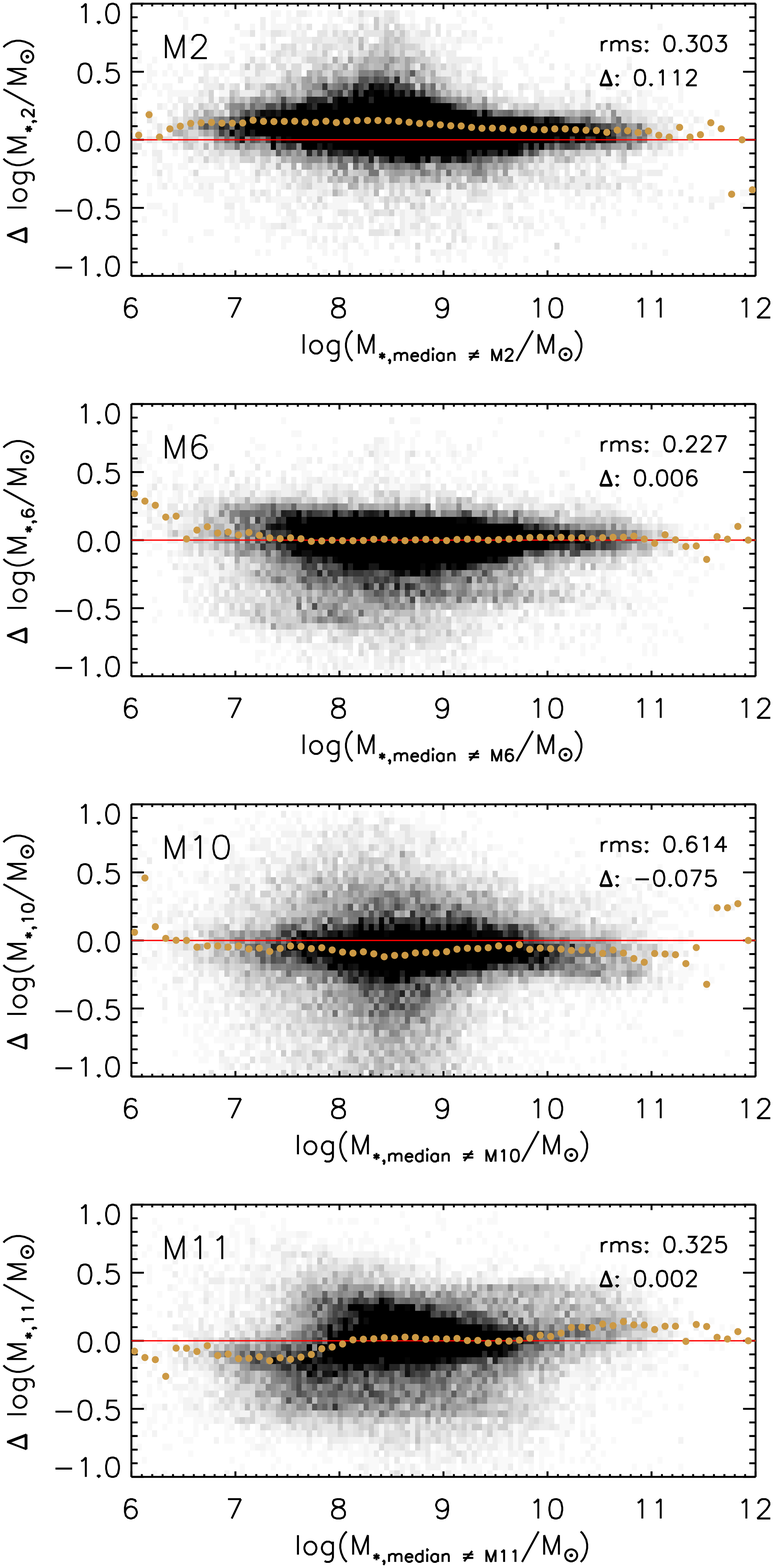} & \includegraphics[width=8.5cm]{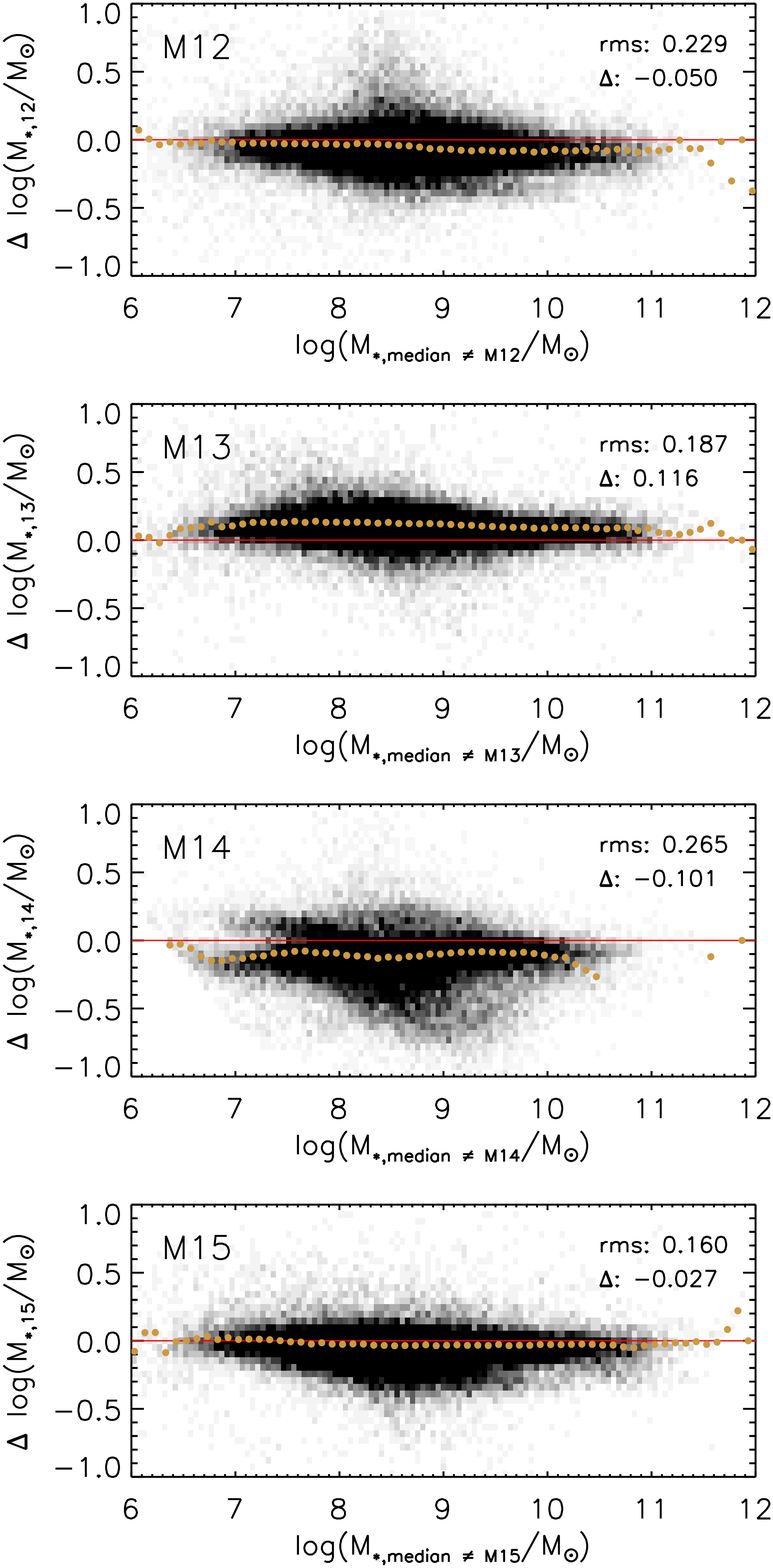}
\end{tabular}
\caption{Comparison of the stellar mass measurements to the median of the measurements excluding that specifically considered in each plot. The vertical axis presents $\Delta \log(M_{*,n}/M_\odot) \equiv \log(M_{*,n}/M_\odot)-\log(M_{*,\mathrm{median} \neq n}/M_\odot)$, where $\log(M_{*,n})$ is the stellar mass from group $n$, while $\log(M_{*,\mathrm{median} \neq n}/M_\odot)$ is the median of the stellar mass measurements over the different groups excluding group $n$. Mass measurements are identified in Table \ref{tab:mstar_par}. Yellow points are the running median. Indicated are the rms (in dex) and the median offset $\Delta$ (in dex) for the sample over the full range in stellar masses. \label{fig:mstar_median}}
\end{figure*}

\subsection{Stellar masses}
Stellar masses have been calculated by eight groups. For each object, the redshift was fixed to either the photometric redshift adopted by our catalog or the DEEP3 spectroscopic redshift if available.  Each group adopted their choices of the fitting code, the template set, the metallicity, the SFH, the IMF and  the extinction law. The ranges and the grid step size of the free parameters also varied from one group to another. Nebular emission lines can bias the stellar mass estimates (e.g, \citealt{schaerer2009}). For this reason, three  groups computed the stellar masses separately with and without taking the nebular lines into account. Table \ref{tab:mstar_par}  summarises the set of configurations adopted by each group. We refer the reader to \citet{mobasher2015a} for full details. Figure \ref{fig:mstar_median} shows the comparisons of the stellar masses obtained by each individual group to the median of the results from other groups, excluding measurement from that one group. In this case the two axes would be independent and the resulting comparison is free from bias. The scatter is  about 20\%-25\% around the 1:1 relation in most cases with a median offset of $\sim-0.024$~dex.

The final stellar masses adopted in our catalog were computed as the median of the results from the six groups who adopted an exponentially declining SFH (with $\tau$ as free parameter) and the \citet{chabrier2003} IMF.

\begin{figure}
\includegraphics[width=8.6cm]{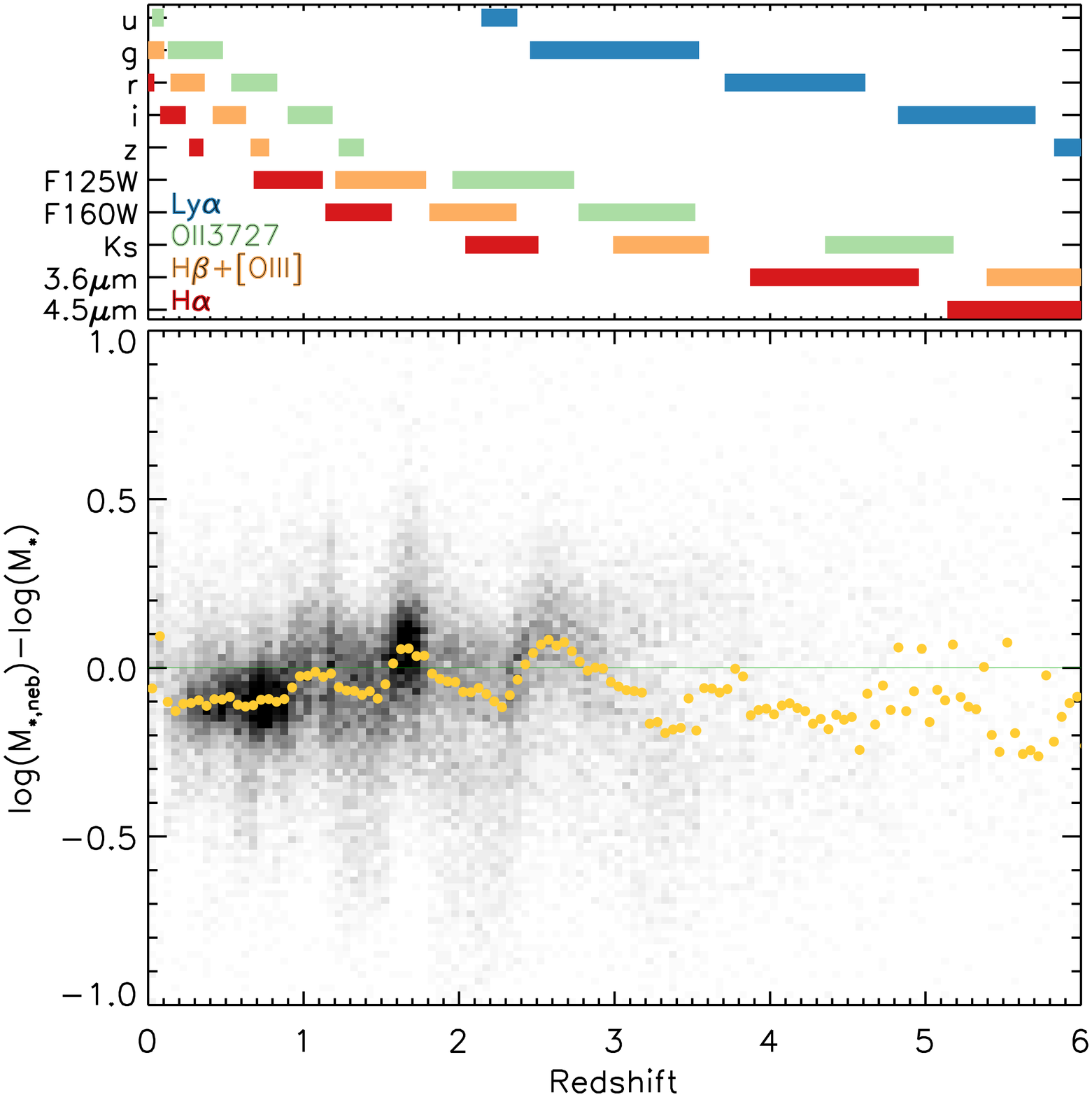}
\caption{\emph{Top panel:} Cartoon of the presence of four of the stronger emission lines (Ly$\alpha$, OII$\lambda3727$, H$\beta$+[OIII] and H$\alpha$, see legend) in a subset of bands  from the CANDELS multi-wavelength photometric catalog (labeled on the vertical axis), as a function of redshift. \emph{Bottom panel:} Comparison between the stellar mass measurements with and without applying correction for nebular emission contamination.  The yellow points mark the median of the difference of the logarithm of stellar mass across redshift, while the horizontal green line marks the 1:1 relation. \label{fig:mstar_neb_noneb}}
\end{figure}

Two sets of stellar mass values are quoted in the catalog, one with the nebular line contributions taken into account and the other one without. Figure \ref{fig:mstar_neb_noneb} compares the two measurements. Overall, the values taking into account the nebular line contaminations are $\sim 0.1$ dex smaller than the ones without, with a dispersion of $\sim 0.25$ dex. However, the exact trend depends on the redshift. In particular, at some specific redshift intervals,  stellar mass values with the nebular line corrections are systematically larger than those obtained without this correction. One possible reason is the over-correction of the nebular line contamination in dusty and/or old stellar population SEDs (see, e.g., Figure 9 of \citealt{stefanon2015}). For example, the over-correction of the [OII]$\lambda$3727 at $z\sim 2.7$ could mimic a deeper Balmer/4000\AA~break, which will lead to best-fit templates being shifted to older and less luminous populations, requiring higher stellar masses.

\begin{figure}
\includegraphics[width=9cm]{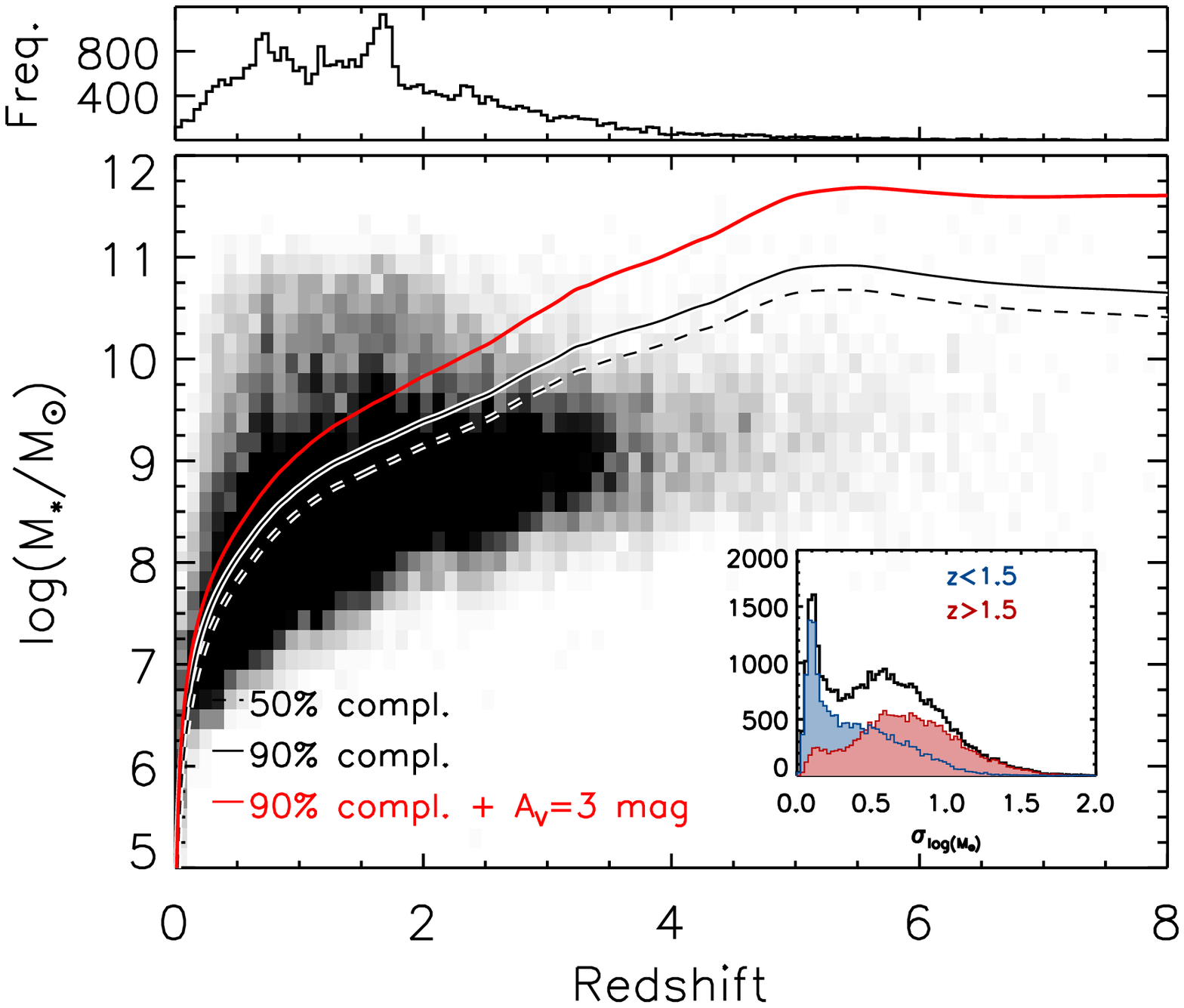}
\caption{\emph{Top panel}: Photometric redshift distribution of the extended sources (\texttt{SExtractor} \texttt{CLASS\_STAR} $<0.7$) in the catalog in redshift bin of 0.05. \emph{Bottom panel}: Distribution of the stellar mass as a function of redshifts for the full sample of extended sources (\texttt{SExtractor} \texttt{CLASS\_STAR} $<0.7$) in the CANDELS stellar mass catalog. Overplotted are also the 50\% and 90\% completeness for a (point-source) passively evolving simple stellar population (SSP) from \citet{bruzual2003} together with the 90\% completeness of an SSP subject to a dust extinction of $A_V=3$~mag, as indicated by the legend. The inset presents the distribution of the spread in $\log{M_*}$ measurements from the different adopted methods for the full sample (black line)  and for objects selected to be at $z<1.5$ and $z>1.5$ (blue and red filled histograms, respectively, bin width 0.025 dex). The dispersions of the full sample follow a bi-modal distribution. The peak with lower stellar mass measurement dispersion is mainly composed of objects at $z<1.5$.  \label{fig:zmstar}}
\end{figure}

The distribution of stellar mass with redshift is presented in Figure \ref{fig:zmstar}. For a passively evolving simple stellar population (SSP) model  from \citet{bruzual2003} with $A_\mathrm{V}=3$~mag, the 90\% completeness in point source detection corresponds to stellar masses $\log(M_*/M_\odot)\sim9, 10, 11 $ at $z\sim1, 2, 4$, respectively; for $A_\mathrm{V}=0$ the stellar mass limits become $\log(M_*/M_\odot) \sim 8.7, 9.2, 10.5$ at $z\sim1, 2, 4$. Most galaxies lie below the completeness limit from the passively evolving SSP model. Indeed,  the majority of galaxies are star-forming, implying mass-to-light ratios lower than for the quiescent galaxies.  In the inset of Figure \ref{fig:zmstar} we show the distribution of the spread in measurement in stellar mass resulting from the different methods. The distribution is characterised by a bi-modality, around $\sigma_{\log(M_*)}\sim 0.3$ dex, with the peak corresponding to the lower dispersion regime mostly populated by objects at $z<1.5$.

\subsection{Comparison to other stellar mass catalogs}

\begin{figure*}
\includegraphics[width=18cm]{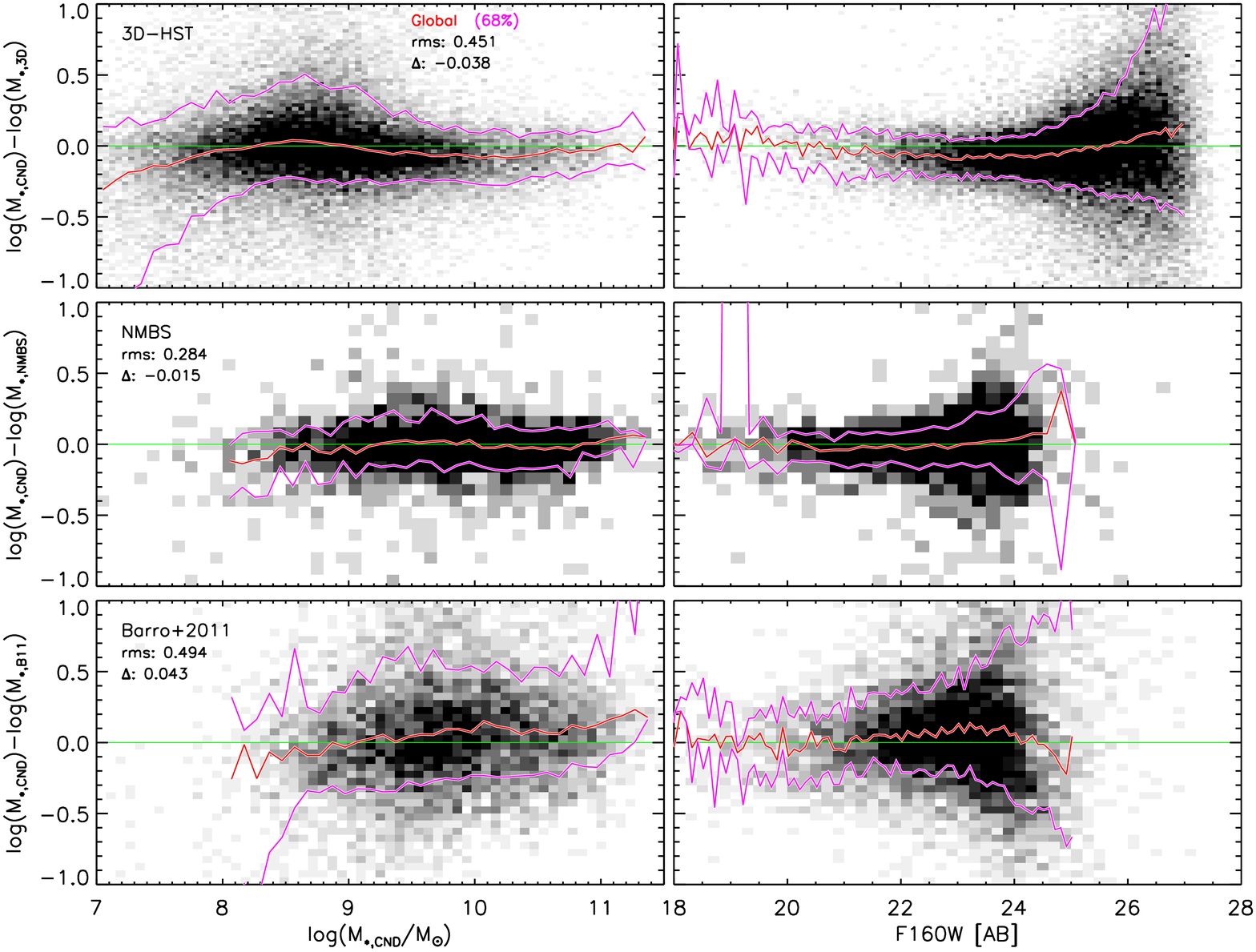}
\caption{\emph{Left-side column:} Comparison of the stellar mass measurements from the CANDELS photometric catalog to the stellar mass measurements of the matching objects in three public catalogs: 3D-HST (\citealt{skelton2014}), NMBS (\citealt{whitaker2011}) and \citet{barro2011b}, top to bottom, respectively. The identity is marked by the horizontal green line. The solid red line marks the median of the difference between the logarithm of the stellar mass, while the two magenta curves encompass 68\% of the points. Indicated are also the rms (in dex) and the median offset $\Delta$ (in dex). \emph{Right-side column:} Difference in stellar mass as a function of the CANDELS F160W magnitude. Same plotting conventions as above. \label{fig:mstar_comp}}
\end{figure*}

\begin{figure*}
\includegraphics[width=18cm]{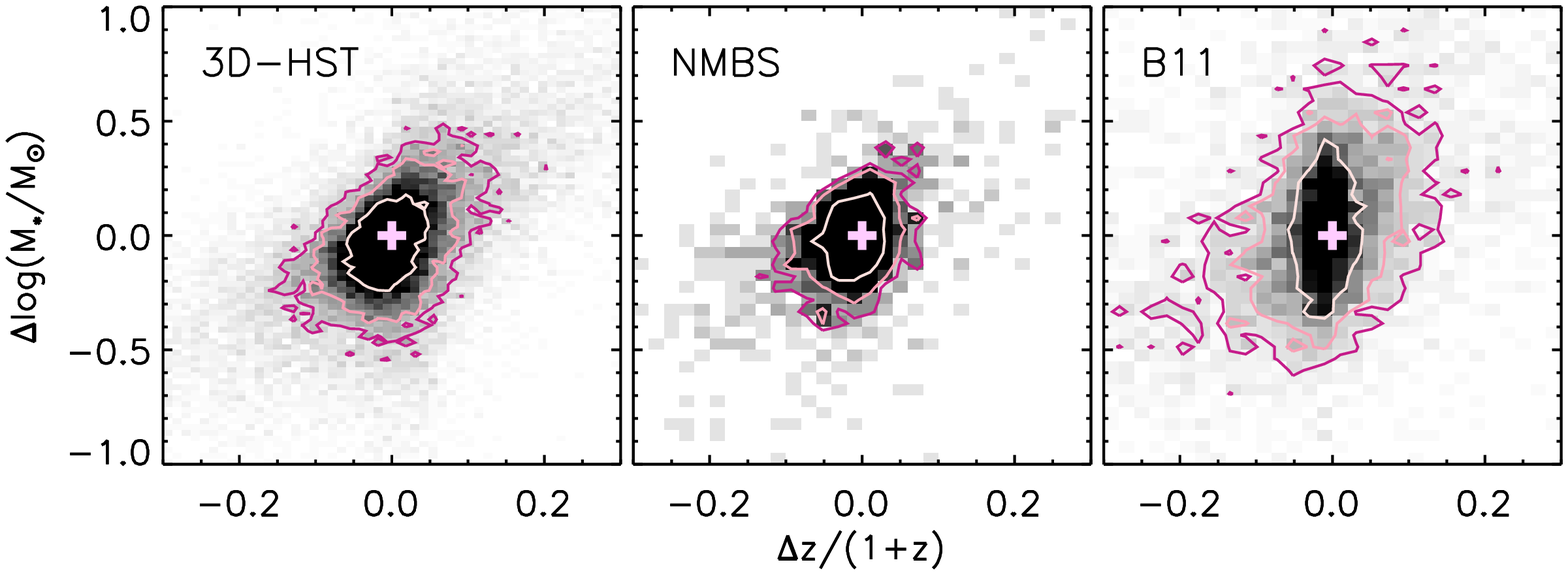}
\caption{The panels show $\Delta\log(M_*/M_\odot)\equiv \log(M_*/M_\odot)_\mathrm{CANDELS}-\log(M_*/M_\odot)_\mathrm{Other}$ versus $\Delta z/(1+z)\equiv (z_\mathrm{CANDELS}-z_\mathrm{Other})/(1+z_\mathrm{CANDELS})$, where Other refers to the  survey indicated by the label on the top of each panel, and $z$ is the photometric redshift. Overplotted are the contours corresponding to 0.05, 0.1 and 0.3 times the peak value of the distribution. For reference,  a cross marks the [0,0] position.   \label{fig:dz_dm}}
\end{figure*}

The configurations adopted by each program for the computation of stellar masses are summarised in Table \ref{tab:mstar_other}. In addition, the three programs adopted solar metallicity, \citet{calzetti2000} dust extinction law and did not apply any correction for nebular emission contamination.
Figure \ref{fig:mstar_comp} shows the comparisons of their stellar mass measurements to CANDELS, which suggest reasonable agreements among these different derivations.  The offsets measured over the stellar mass range $8<\log(M_*/M_\odot)<11$ vary between $\sim-0.40$~dex to $\sim0.40$~dex, with rms values $\lesssim 0.5$ dex.  In comparison to the 3D-HST results, the agreement at $\log(M*/M_\odot)<9$ is excellent. At $\log(M*/M_\odot)>9$, the 3D-HST stellar masses are systematically higher by $\gtrsim 0.1$~dex.  Comparing to the NMBS values, similar to the case of the photometric redshift comparison (Section \ref{sect:photoz}), the agreement is excellent (with an rms of $\sim0.3$  and $\lesssim0.1$ dex offset). The comparison to the stellar masses from \citet{barro2011b} is made after applying an offset of -0.25 dex to their values to convert from the Salpeter IMF to the Chabrier IMF. The agreement is reasonable; at stellar masses $\log(M*/M_\odot)\lesssim9$ there is indication of an increasing trend with stellar masses, although this can be the result of the lower S/N data adopted by \citet{barro2011b}.

Since each collaboration adopted different measurements of photometric redshifts for the same object, the systematic differences in stellar mass registered in Figure \ref{fig:mstar_comp} could be, at least in part, the result of the different input redshifts. Figure \ref{fig:dz_dm} therefore compares the offsets in photometric redshift to the offsets in stellar mass between the CANDELS measurements and the corresponding measurements from 3D-HST, NMBS and \citet{barro2011b}. The plots show that for 3D-HST and NMBS there is a correlation between the offset in photometric redshifts and the offset in stellar mass. This could indeed explain the lower values in stellar mass in our catalogs compared to those in 3D-HST and NMBS, as for a given observed SED, a lower redshift (as measured by CANDELS) must correspond to a lower luminosity and hence stellar mass.

\section{Conclusions}
The core of the multi-wavelength photometric catalog produced by the CANDELS team for the AEGIS/EGS field was built on the CANDELS \textit{HST} data in the WFC3 F125W and F160W and the ACS F606W and F814W bands.  Altogether, these data provide photometry in 22 bands that cover a wavelength range of 0.4-8.0$\mu$m. Source detection was done in the WFC3 F160W band with an improved version of \texttt{SExtractor}, which optimises the exclusion of contaminants. 

We have discovered $0\farcs2$ position-dependent offsets between the earlier CANDELS system (based on AEGIS ACS) and that of the CFHTLS. The offsets are reduced to $0\farcs04$ after re-calibrating the astrometry to the CFHTLS system at catalog-level, although we have opted to maintain the WCS of the CANDELS \textit{HST} mosaics in this version. We release \textit{HST} mosaic versions with both tangential planes of the AEGIS and CFHTLS astrometric systems.

The comparison of our photometry to that from the 3D-HST, NMBS and \citet{barro2011} public catalogs shows good to excellent agreement after the various systematic corrections have been taken into account (median offsets are -0.04 mag and 0.14 mag comparing to 3D-HST and NMBS, respectively). Nevertheless, there are still flux-dependent offsets between our measurements of the fluxes and those of the 3D-HST after the zeropoint adjustments and the Galactic extinction corrections have been removed from the latter. Such flux-dependent offsets are also present when compared to the NMBS results, although to a lesser degree. We argue that these differences are due to the flux-dependent aperture corrections in the other catalogs. 

We also present the catalog of photometric redshifts and stellar mass measurements. Photometric redshifts were independently measured by ten groups within the CANDELS team, each one adopting their choices of the fitting code, the algorithm and the template library. The individual measurements were then combined together using a bayesian approach on the P(z) described by \citet{dahlen2013}. The comparison to the spectroscopic redshifts from the DEEP2+3 survey show a dispersion $\sigma_\mathrm{NMAD}\sim 0.013$ and fraction of catastrophic outliers $n_{>5\sigma_\mathrm{NMAD}}\sim 7\%$.

Stellar masses were also computed independently by eight teams. The final measurement of the stellar mass is then taken to be the median of the measurements, done on a per-object basis.  We estimated a conservative stellar mass limit with a \emph{maximally red} SED template (i.e., a passively evolving population with $A_\mathrm{V}=3$~mag). The depth of the catalog corresponds to a stellar mass completeness of 90\% for $\log(M_*/M_\odot)\sim9 $ at $z\sim1$, $\log(M_*/M_\odot)\sim10 $ at $z\sim2$ and $\log(M_*/M_\odot)\sim11 $ at $z\sim4$. 

Comparison of the photometric redshifts and stellar masses to the measurements provided by publicly available catalogs show that the agreement is good and does not have any strong bias.

The covered area and the photometric depth reached in the CANDELS EGS field will be of invaluable aid in the measurements of the luminosity and stellar mass functions at $1\lesssim z\lesssim 5 $. Furthermore, the full coverage and deep X-ray data make the EGS field unique for the study of the evolution of the AGN.  Our catalogs are accessible on the primary CANDELS pages at MAST\footnote{https://archive.stsci.edu/prepds/candels}, through the Vizier service\footnote{http://vizier.u-strasbg.fr/viz-bin/VizieR}, and from the CANDELS team project website\footnote{http://candels.ucolick.org}.\\

Facilities: \facility{{\it Hubble Space Telescope} (ACS, WFC3), {\it Spitzer Space Telescope} (IRAC)}

\acknowledgments
We thank the referee for his helpful comments which improved the paper readability.
This work is based on observations taken by the CANDELS Multi-Cycle Treasury Program with the NASA/ESA HST, which is operated by the Association of Universities for Research in Astronomy, Inc., under NASA contract NAS5-26555.
This study makes use of data from AEGIS, a multiwavelength sky survey conducted with the Chandra, GALEX, Hubble, Keck, CFHT, MMT, Subaru, Palomar, Spitzer, VLA, and other telescopes and supported in part by the NSF, NASA, and the STFC.
This work is based in part on data obtained with the Spitzer Space Telescope, which is operated by the Jet Propulsion Laboratory, California Institute of Technology under a contract with NASA. 
This work is based on observations taken by the 3D-HST Treasury Program (GO 12177 and 12328) with the NASA/ESA HST, which is operated by the Association of Universities for Research in Astronomy, Inc., under NASA contract NAS5-26555.
Funding for the DEEP2 Galaxy Redshift Survey has been provided by NSF grants AST-95-09298, AST-0071048, AST-0507428, and AST-0507483 as well as NASA LTSA grant NNG04GC89G.
Funding for the DEEP3 Galaxy Redshift Survey has been provided by NSF grants AST-0808133, AST-0807630, and AST-0806732.
Based on observations obtained with MegaPrime/MegaCam, a joint project of CFHT and CEA/IRFU, at the Canada-France-Hawaii Telescope (CFHT) which is operated by the National Research Council (NRC) of Canada, the Institut National des Science de l'Univers of the Centre National de la Recherche Scientifique (CNRS) of France, and the University of Hawaii. This work is based in part on data products produced at Terapix available at the Canadian Astronomy Data Centre as part of the Canada-France-Hawaii Telescope Legacy Survey, a collaborative project of NRC and CNRS.
Based on observations obtained with WIRCam, a joint project of CFHT,Taiwan, Korea, Canada, France, at the Canada-France-Hawaii Telescope (CFHT) which is operated by the National Research Council (NRC) of Canada, the Institute National des Sciences de l'Univers of the Centre National de la Recherche Scientifique of France, and the University of Hawaii. This work is based in part on data products produced at TERAPIX, the WIRDS (WIRcam Deep Survey) consortium, and the Canadian Astronomy Data Centre. This research was supported by a grant from the Agence Nationale de la Recherche ANR-07-BLAN-0228.
This study makes use of data from the NEWFIRM Medium-Band Survey, a multi-wavelength survey conducted with the NEWFIRM instrument at the KPNO, supported in part by the NSF and NASA.
This work is based in part on observations made with the Spitzer Space Telescope, which is operated by the Jet Propulsion Laboratory, California Institute of Technology under a contract with NASA. IRAF is distributed by the National Optical Astronomy Observatories, 
which are operated by the Association of Universities for Research 
in Astronomy, Inc., under cooperative agreement with the National 
 Science Foundation.

\bibliographystyle{apj}

\appendix
\section{Content of the photometric catalog}
\label{appendix:photcat}
List of columns in the multi-wavelength photometric catalog.  Our catalogs are accessible on the primary CANDELS pages at MAST\footnote{https://archive.stsci.edu/prepds/candels}, through the Vizier service\footnote{http://vizier.u-strasbg.fr/viz-bin/VizieR}, and from the CANDELS team project website\footnote{http://candels.ucolick.org}

\begin{ThreePartTable}
\begin{TableNotes}
\footnotesize
\item[$\dagger$] The fluxes and uncertainties in some HST bands and for some of the sources were initially set to -99 even if there was no indication of bad measurement. Columns 19-28 contain the fixed values. Here we include the original version of these columns as such measurements were adopted to estimate photometric redshifts and stellar population parameters. These columns are identified by the suffix \_PHZ (for photo-z). Tests showed that photo-z for most of the sources were not strongly affected by this problem. However, we further OR-flag the FLAGS column with the value 4 to reflect the 160 sources for which $\Delta z/(1+z)>0.1$ and potential less robust stellar population parameters.
\end{TableNotes}

\begin{longtable*}{rll}
\caption{Multi-wavelength photometric catalog entries\label{tab:cat_mw}}\\
\hline
\hline \\[1pt]
& & \\
Col. \# & Name & Description \\
& & \\
\hline \\[1pt]
\endhead
\endfoot

\insertTableNotes 
\endlastfoot

 1 & ID          &   Sequential ID number in the F160W-based \texttt{SExtractor} catalog.         \\[1pt] 
 2 & IAU\_designation                      &  Official IAU designation of the object. \\[1pt]
 3 & RA                       & Right Ascension and Declination (J2000) in the F160W mosaic expressed is decimal degrees, \\[1pt]
 4 & DEC                      & after they have been converted to the CFHTLS astrometric system. The values  in these two \\[1pt]
 & & columns are suggested as the coordinates of the objects in the catalog. \\[1pt]
 5 & RA\_Lotz2008                       &   Original Right Ascension and Declination (J2000) in the F160W mosaic in decimal degrees,                 \\[1pt]
 6 & DEC\_Lotz2008                       &  whose astrometry was calibrated using the \citet{lotz2008} system.                 \\[1pt]
 7 & FLAGS                      &  Flag. A value of 1 corresponds to sources falling in regions of low S/N as it can be \\[1pt]
 & & at the borders of the mosaic; a value of 2 indicates that a source, as identified \\[1pt]
 & & by its footprint in the segmentation map, falls close to a bright star or to its \\ [1pt]
 & & diffraction spikes; a value of 3 indicates that the source suffers from both \\[1pt]
 & & a low S/N and contamination from bright stars. Sources free from any of \\[1pt]
 & & the above effects have a flag of 0.                 \\[1pt]
 8 & CLASS\_STAR                      &    \texttt{SExtractor} parameter \texttt{CLASS\_STAR} from the F160W mosaic.               \\[1pt]
 9 & CFHT\_u\_FLUX                       &  Col. 9-54: Flux densities and associated uncertainties, expressed in $\mu$Jy. A value of -99 has \\[1pt]
10 & CFHT\_u\_FLUXERR                       &    been set to the flux and  associated uncertainty for those objects falling outside the coverage      \\[1pt]
11 & CFHT\_g\_FLUX                      &   of the mosaic in a specific band or when bad pixels within the segmentation   map          \\[1pt]
12 & CFHT\_g\_FLUXERR                       &    contaminate the flux measurement.                         \\[1pt]
13 & CFHT\_r\_FLUX                       &       \hspace{4.5cm}$\dotsc$      \\[1pt]
14 & CFHT\_r\_FLUXERR                        &       \hspace{4.5cm}$\dotsc$                       \\[1pt]
15 & CFHT\_i\_FLUX                       &                \hspace{4.5cm}$\dotsc$              \\[1pt]
16 & CFHT\_i\_FLUXERR                       &         \hspace{4.5cm}$\dotsc$                     \\[1pt]
17 & CFHT\_z\_FLUX                       &                \hspace{4.5cm}$\dotsc$              \\[1pt]
18 & CFHT\_z\_FLUXERR                       &          \hspace{4.5cm}$\dotsc$                    \\[1pt]
19 & ACS\_F606W\_FLUX                       &          \hspace{4.5cm}$\dotsc$                    \\[1pt]
20 & ACS\_F606W\_FLUXERR                       &           \hspace{4.5cm}$\dotsc$                   \\[1pt]
21 & ACS\_F814W\_FLUX                       &             \hspace{4.5cm}$\dotsc$                 \\[1pt]
22 & ACS\_F814W\_FLUXERR                       &         \hspace{4.5cm}$\dotsc$                     \\[1pt]
23 & WFC3\_F125W\_FLUX                       &              \hspace{4.5cm}$\dotsc$                \\[1pt]
24 & WFC3\_F125W\_FLUXERR                       &            \hspace{4.5cm}$\dotsc$                  \\[1pt]
25 & WFC3\_F140W\_FLUX                       &                \hspace{4.5cm}$\dotsc$              \\[1pt]
26 & WFC3\_F140W\_FLUXERR                       &          \hspace{4.5cm}$\dotsc$                    \\[1pt]
27 & WFC3\_F160W\_FLUX                       &                 \hspace{4.5cm}$\dotsc$             \\[1pt]
28 & WFC3\_F160W\_FLUXERR                       &         \hspace{4.5cm}$\dotsc$                     \\[1pt]
29 & WIRCAM\_J\_FLUX                       &               \hspace{4.5cm}$\dotsc$               \\[1pt]
30 & WIRCAM\_J\_FLUXERR                       &         \hspace{4.5cm}$\dotsc$                     \\[1pt]
31 & WIRCAM\_H\_FLUX                       &              \hspace{4.5cm}$\dotsc$                \\[1pt]
32 & WIRCAM\_H\_FLUXERR                       &         \hspace{4.5cm}$\dotsc$                     \\[1pt]
33 & WIRCAM\_K\_FLUX                       &               \hspace{4.5cm}$\dotsc$               \\[1pt]
34 & WIRCAM\_K\_FLUXERR                       &          \hspace{4.5cm}$\dotsc$                    \\[1pt]
35 & NEWFIRM\_J1\_FLUX                       &           \hspace{4.5cm}$\dotsc$                   \\[1pt]
36 & NEWFIRM\_J1\_FLUXERR                       &            \hspace{4.5cm}$\dotsc$                  \\[1pt]
37 & NEWFIRM\_J2\_FLUX                       &               \hspace{4.5cm}$\dotsc$               \\[1pt]
38 & NEWFIRM\_J2\_FLUXERR                       &        \hspace{4.5cm}$\dotsc$                      \\[1pt]
39 & NEWFIRM\_J3\_FLUX                       &            \hspace{4.5cm}$\dotsc$                  \\[1pt]
40 & NEWFIRM\_J3\_FLUXERR                       &        \hspace{4.5cm}$\dotsc$                      \\[1pt]
41 & NEWFIRM\_H1\_FLUX                       &              \hspace{4.5cm}$\dotsc$                \\[1pt]
42 & NEWFIRM\_H1\_FLUXERR                       &        \hspace{4.5cm}$\dotsc$                      \\[1pt]
43 & NEWFIRM\_H2\_FLUX                       &               \hspace{4.5cm}$\dotsc$               \\[1pt]
44 & NEWFIRM\_H2\_FLUXERR                       &        \hspace{4.5cm}$\dotsc$                      \\[1pt]
45 & NEWFIRM\_K\_FLUX                       &                \hspace{4.5cm}$\dotsc$              \\[1pt]
46 & NEWFIRM\_K\_FLUXERR                       &        \hspace{4.5cm}$\dotsc$                      \\[1pt]
47 & IRAC\_CH1\_FLUX                       &              \hspace{4.5cm}$\dotsc$                \\[1pt]
48 & IRAC\_CH1\_FLUXERR                       &         \hspace{4.5cm}$\dotsc$                     \\[1pt]
49 & IRAC\_CH2\_FLUX                       &                \hspace{4.5cm}$\dotsc$              \\[1pt]
50 & IRAC\_CH2\_FLUXERR                       &         \hspace{4.5cm}$\dotsc$                     \\[1pt]
51 & IRAC\_CH3\_FLUX                       &               \hspace{4.5cm}$\dotsc$               \\[1pt]
52 & IRAC\_CH3\_FLUXERR                       &          \hspace{4.5cm}$\dotsc$                    \\[1pt]
53 & IRAC\_CH4\_FLUX                       &             \hspace{4.5cm}$\dotsc$                 \\[1pt]
54 & IRAC\_CH4\_FLUXERR                       &         \hspace{4.5cm}$\dotsc$                     \\[1pt]
55 & ACS\_F606W\_v08\_FLUX                       &   Col. 55-62: Flux densities and associated uncertainties for the AEGIS data set.                 \\[1pt]
56 & ACS\_F606W\_v08\_FLUXERR                       &      \hspace{4.5cm}$\dotsc$                   \\[1pt]
57 & ACS\_F814W\_v08\_FLUX                       &          \hspace{4.5cm}$\dotsc$               \\[1pt]
58 & ACS\_F814W\_v08\_FLUXERR                       &      \hspace{4.5cm}$\dotsc$                   \\[1pt]
59 & WFC3\_F125W\_v08\_FLUX                       &       \hspace{4.5cm}$\dotsc$                  \\[1pt]
60 & WFC3\_F125W\_v08\_FLUXERR                       &       \hspace{4.5cm}$\dotsc$                  \\[1pt]
61 & WFC3\_F160W\_v08\_FLUX                       &      \hspace{4.5cm}$\dotsc$                   \\[1pt]
62 & WFC3\_F160W\_v08\_FLUXERR                       &       \hspace{4.5cm}$\dotsc$                  \\[1pt]
63 & IRAC\_CH3\_v08\_FLUX                       &  Col. 63-66: Flux densities and associated  uncertainties  for the IRAC 5.8 and $8.0\mu$m band                  \\[1pt]
64 & IRAC\_CH3\_v08\_FLUXERR                       &  with the old convolution kernel, which were used for the computation of photometric            \\[1pt]
65 & IRAC\_CH4\_v08\_FLUX                       &   redshifts and stellar masses.                      \\[1pt]
66 & IRAC\_CH4\_v08\_FLUXERR                       &        \hspace{4.5cm}$\dotsc$                 \\[1pt]
67 & ACS\_F606W\_FLUX\_PHZ                       &   Col. 67-76: Flux densities and associated uncertainties for the fluxes adopted to compute \\[1pt]
68 & ACS\_F606W\_FLUXERR\_PHZ                       &     photometric redshifts and stellar population parameters\tnote{$\dagger$}   \\[1pt]
69 & ACS\_F814W\_FLUX\_PHZ                       &          \hspace{4.5cm}$\dotsc$               \\[1pt]
70 & ACS\_F814W\_FLUXERR\_PHZ                       &      \hspace{4.5cm}$\dotsc$                   \\[1pt]
71 & WFC3\_F125W\_FLUX\_PHZ                       &       \hspace{4.5cm}$\dotsc$                  \\[1pt]
72 & WFC3\_F125W\_FLUXERR\_PHZ                       &       \hspace{4.5cm}$\dotsc$                  \\[1pt]
73 & WFC3\_F140W\_FLUX\_PHZ                       &      \hspace{4.5cm}$\dotsc$                   \\[1pt]
74 & WFC3\_F140W\_FLUXERR\_PHZ                       &       \hspace{4.5cm}$\dotsc$                  \\[1pt]
75 & WFC3\_F160W\_FLUX\_PHZ                       &      \hspace{4.5cm}$\dotsc$                   \\[1pt]
76 & WFC3\_F160W\_FLUXERR\_PHZ                       &       \hspace{4.5cm}$\dotsc$                  \\[1pt]
77 & DEEP\_SPEC\_Z                      &   Spectroscopic redshift from the DEEP2 and DEEP3 catalogs. If no match is found, \\[1pt]
& & the value has been set to -99.                \\[1pt]
\hline
\end{longtable*}
\end{ThreePartTable}

\section{Content of the photometric redshift catalog}
\label{appendix:photoz}

\begin{longtable*}{rll}
\caption{Photometric redshift catalog entries\label{tab:cat_photoz}}\\
\hline
\hline \\[1pt]
& & \\
Col. \# & Name & Description \\
& & \\
\hline \\[1pt]
\endhead
 1 & ID & Sequential ID number. \\[1pt]
 2 & Photo\_z\_Median & Median of the photometric redshift measurements.\\[1pt]
 3 & Photo\_z\_Salvato & Col. 3-8: Photmetric redshift measurements from the individual group.\\[1pt]
 4 & Photo\_z\_Mobasher &  \hspace{4.5cm}$\dotsc$     \\[1pt]
 5 & Photo\_z\_Finkelstein &   \hspace{4.5cm}$\dotsc$    \\[1pt]
 6 & Photo\_z\_Barro &   \hspace{4.5cm}$\dotsc$    \\[1pt]
 7 & Photo\_z\_Wiklind &   \hspace{4.5cm}$\dotsc$     \\[1pt]
 8 & Photo\_z\_Wuyts  &   \hspace{4.5cm}$\dotsc$    \\[1pt]
 9 & D95  & Accuracy of the photometric redshifts based on their confidence intervals (Dahlen et al 2012).\\[1pt]
 10 & Spec\_z & Spectroscopic redshift of the control sample (-99 otherwise), from the DEEP3 catalog.\\[1pt]
 11 & Photo\_z\_lower68  & Col. 11-14: 68\% and 95\% confidence intervals for the median of the redshift measurements. For \\[1pt]
 12 & Photo\_z\_upper68  & details on its computation see Section \ref{sect:photoz} and \citet{dahlen2013}.\\[1pt]
 13 & Photo\_z\_lower95  & \hspace{4.5cm}$\dotsc$   \\[1pt]
 14 & Photo\_z\_upper95 & \hspace{4.5cm}$\dotsc$   \\[1pt]
\hline
\end{longtable*}

\section{Content of the stellar mass catalog}
\label{appendix:mstar}
Below is the list of the columns available in the stellar mass catalog.

\begin{longtable*}{rll}
\caption{Stellar mass catalog entries \label{tab:cat_mstar}}\\
\hline
\hline \\[1pt]
& & \\
Col. \# & Name & Description \\
& & \\
\hline \\[1pt]
\endhead
1   &  ID      & Sequential ID number in the F160W-based \texttt{SExtractor} catalog.     \\[1pt]
2   &  RAdeg     & Col. 2-3: Right Ascension and Declination (J2000) in the F160W mosaic.     \\[1pt]
3   &  DECdeg     &       \hspace{4.5cm}$\dotsc$  \\[1pt]
4   &  Hmag       & Magnitude in the F160W band.   \\[1pt]
5   &  PhotFlag    & Flag in the photometric catalog.  \\[1pt]
6   &  Class\_star  & \texttt{SExtractor} \texttt{CLASS\_STAR} parameter.   \\[1pt]
7   &  AGNflag    & = 1 for those objects with a counterpart in the \citet{nandra2015} catalog, 0 otherwise.   \\[1pt]
8   &  zphot      & Photometric redshift measurement.   \\[1pt]
9   &  zspec    &  Spectroscopic redshift from the DEEP3 catalog.   \\[1pt]
10  &  zspec\_q   &  Quality flag for the spectroscopic redshift: 1=Good; 2=Fair; 3=Poor.  \\[1pt]
11  &  zspec\_refer & Source of the spectroscopic redshift catalog. 1=DEEP2/3. \\[1pt]
12  &  zbest     & Best redshift measurement: if spectroscopic redshift is available, then \texttt{zbest=zspec}, otherwise \texttt{zbest=zphot}.    \\[1pt]
13  &  zphot\_l68   & Col. 13-16: 68\% and 95\% confidence intervals on the photometric redshift measurements.  \\[1pt]
14  &  zphot\_u68   &    \hspace{4.5cm}$\dotsc$    \\[1pt]
15  &  zphot\_l95    &    \hspace{4.5cm}$\dotsc$   \\[1pt]
16  &  zphot\_u95   &     \hspace{4.5cm}$\dotsc$   \\[1pt]
17  &  zphotAGN    & Photometric redshifts for the sources with a match in the X-Ray catalog of \citet{nandra2015}, computed  \\[1pt]
& & adopting AGN specific SED templates and priors (see \citealt{nandra2015} for more details). \\[1pt]
18  &  M\_neb\_med  & Col. 18-19: Stellar mass measurement and associated uncertainty from the median of the logarithm of  \\[1pt]
19  &  s\_neb\_med  &  stellar mass measurements obtained taking into account nebular emission contamination.  \\[1pt]
20  &  M\_med   & Col. 20-21: Stellar mass measurement and associated uncertainty from the median of the logarithm of       \\[1pt]
21  &  s\_med       &  stellar mass measurements obtained without considering nebular emission contamination.  \\[1pt]
22  &  M\_14a\_cons & Col. 22-35: Stellar mass measurements from the individual methods. See Table \ref{tab:mstar_par} for details on the  \\[1pt]
23  &  M\_11a\_tau   &  configuration of each method. The number in the name matches that in Table \ref{tab:mstar_par}.  \\[1pt]
24  &  M\_6a\_tau\^{}NEB &     \hspace{4.5cm}$\dotsc$   \\[1pt]
25  &  M\_13a\_tau   &     \hspace{4.5cm}$\dotsc$   \\[1pt]
26  &  M\_12a    &       \hspace{4.5cm}$\dotsc$    \\[1pt]
27  &  M\_6a\_tau &  \hspace{4.5cm}$\dotsc$   \\[1pt]
28  &  M\_2a\_tau  & \hspace{4.5cm}$\dotsc$   \\[1pt]
29  &  M\_15a   &   \hspace{4.5cm}$\dotsc$   \\[1pt]
30  &  M\_10c   &   \hspace{4.5cm}$\dotsc$   \\[1pt]
31  &  M\_14a\_lin &  \hspace{4.5cm}$\dotsc$  \\[1pt]
32  &  M\_14a\_deltau & \hspace{4.5cm}$\dotsc$  \\[1pt]
33  &  M\_14a\_tau &  \hspace{4.5cm}$\dotsc$  \\[1pt]
34  &  M\_14a\_inctau & \hspace{4.5cm}$\dotsc$  \\[1pt]
35  &  M\_14a    &   \hspace{4.5cm}$\dotsc$  \\[1pt]
36  &  M\_neb\_med\_lin & Col 36-37: Stellar mass measurement and associated uncertainty from the median of the linear value  \\[1pt]
37  &  s\_neb\_med\_lin &of the stellar mass measurements obtained taking into account nebular emission contamination. \\[1pt]
38  &  M\_med\_lin   & Col. 38-39: Stellar mass measurement and associated uncertainty from the median of the linear value  \\[1pt]
39  &  s\_med\_lin   &  of the stellar mass measurements obtained without taking into account nebular emission contamination.  \\[1pt]
\hline
\end{longtable*}

\section{Content of the physical parameters catalog}
\label{appendix:spp}
Below is the list of the columns available in the physical parameters catalog. The number in the column name matches that in Table \ref{tab:mstar_par} and  \citet{mobasher2015a}. We refer the reader to Table \ref{tab:mstar_par} and \citet{mobasher2015a} for details on the configuration of each method.

\begin{longtable*}{rll}
\caption{Physical parameters catalog entries\label{tab:cat_spp}}\\
\hline
\hline \\[1pt]
& & \\
Col. \# & Name & Description \\
& & \\
\hline \\[1pt]
\endhead
 1   &  ID               &      Sequential ID number in the F160W-based \texttt{SExtractor} catalog.  \\[1pt]
  2  &  age\_2a\_tau       &    Age from method $2a$ [log(t/yr)]                                                                                \\[1pt]
  3  &  tau\_2a\_tau       &    $\tau$ from method $2a$ [Gyr]                                                                                   \\[1pt]
  4  &  Av\_2a\_tau        &    $A_V$ from method $2a$ [mag]                                                                                    \\[1pt]
  5  &  SFR\_2a\_tau       &    SFR from method $2a$ [$M_\odot/$yr]                                                                             \\[1pt]
  6  &  chi2\_2a\_tau      &    Reduced $\chi^2$ from method $2a$                                                                               \\[1pt]
  7  &  age\_4b           &     Age from method $4b$ [log(t/yr)]                                                                               \\[1pt]
  8  &  EBV\_4b           &     E(B-V) from method $4b$ [mag]                                                                                  \\[1pt]
  9  &  age\_6a\_tau       &    Age from method $6a_\tau$ [log(t/yr)]                                                                           \\[1pt]
 10  &  tau\_6a\_tau       &    $\tau$ from method $6a_\tau$ [Gyr]                                                                              \\[1pt]
 11  &  EBV\_6a\_tau       &    E(B-V) from method $6a_\tau$ [mag]                                                                              \\[1pt]
 12  &  SFR\_6a\_tau       &    SFR from method $6a_\tau$ [$M_\odot/$yr]                                                                        \\[1pt]
 13  &  met\_6a\_tau       &    Gas metallicity from method $6a_\tau$ [$Z_\odot$]                                                               \\[1pt]
 14  &  extlw\_6a\_tau     &    Extinction law from method $6a_\tau$: 1=Calzetti; 2=SMC                                                         \\[1pt]
 15  &  chi2\_6a\_tau      &    Reduced $\chi^2$ from method $6a_\tau$                                                                          \\[1pt]
 16  &  L1400\_6a\_tau     &    Rest-frame luminosity at 1400 Angstrom  from method $6a_\tau$ ($L_\nu(1400$\AA$)$ [erg/s/Hz]                    \\[1pt]
 17  &  L2700\_6a\_tau     &    Rest-frame luminosity at 2700 Angstrom  from method $6a_\tau$ ($L_\nu(2700$\AA$)$ [erg/s/Hz]                    \\[1pt]
 18  &  UMag\_6a\_tau      &    U rest-frame magnitude from method $6a_\tau$  [AB system]                                                       \\[1pt]
 19  &  BMag\_6a\_tau      &    B rest-frame magnitude  from method $6a_\tau$ [AB system]                                                       \\[1pt]
 20  &  VMag\_6a\_tau      &    V rest-frame magnitude from method $6a_\tau$ [AB system]                                                        \\[1pt]
 21  &  RMag\_6a\_tau      &    R rest-frame magnitude from method $6a_\tau$ [AB system]                                                        \\[1pt]
 22  &  IMag\_6a\_tau      &    I rest-frame magnitude from method $6a_\tau$ [AB system]                                                        \\[1pt]
 23  &  JMag\_6a\_tau      &    J rest-frame magnitude from method $6a_\tau$ [AB system]                                                        \\[1pt]
 24  &  KMag\_6a\_tau      &    K rest-frame magnitude from method $6a_\tau$ [AB system]                                                        \\[1pt]
 25  &  age\_10c          &     Age from method $10c$ [log(t/yr)]                                                                              \\[1pt]
 26  &  SFH\_10c          &     SFH from method $10c$ (1=exponentially decreasing ; 2=constant; 3=truncated; 4=no solution)                    \\[1pt]
 27  &  tau\_10c          &     $\tau$ from method $10c$ ($\tau$ =-99 if SFH=2 or 4) [Gyr]                                                     \\[1pt]
 28  &  met\_10c          &     Gas metallicity from method $10c$ [$Z_\odot$]                                                                  \\[1pt]
 29  &  M\_l99\_11a\_tau    &   Lower stellar mass 99\% confidence interval from method $11a_\tau$ [log(M/M$_\odot$)]                            \\[1pt]
 30  &  M\_u99\_11a\_tau    &   Upper stellar mass 99\% confidence interval  from method $11a_\tau$ [log(M/M$_\odot$)]                           \\[1pt]
 31  &  age\_11a\_tau      &    Age  from method $11a_\tau$ [ [log(t/yr)]                                                                       \\[1pt]
 32  &  SFR\_11a\_tau      &    SFR from method $11a_\tau$ [$M_\odot/$yr]                                                                       \\[1pt]
 33  &  M\_l68\_12a        &    Lower stellar mass 68\% confidence interval from method $12a$ [log(M/M$_\odot$)]                                \\[1pt]
 34  &  M\_u68\_12a        &    Upper stellar mass 68\% confidence interval from method $12a$ [log(M/M$_\odot$)]                                \\[1pt]
 35  &  M\_l95\_12a        &    Lower stellar mass 95\% confidence interval from method $12a$ [log(M/M$_\odot$)]                                \\[1pt]
 36  &  M\_u95\_12a        &    Upper stellar mass 95\% confidence interval from method $12a$ [log(M/M$_\odot$)]                                \\[1pt]
 37  &  age\_12a          &     Age  from method $12a$ [log(t/yr)]                                                                             \\[1pt]
 38  &  tau\_12a          &     $\tau$ from method $12a$ [Gyr]                                                                                 \\[1pt]
 39  &  EBV\_12a          &     E(B-V) from method $12a$ [mag]                                                                                 \\[1pt]
 40  &  met\_12a          &     Metallicity from method $12a$ [$Z_\odot$]                                                                      \\[1pt]
 41  &  Lbol\_12a         &     log(L$_\mathrm{bol}$/L$_\odot$), corrected for dust extinction, from method $12a$ [log(L/L$_\odot$)]           \\[1pt]
 42  &  chi2\_12a         &     Reduced $\chi^2$ from method $12a$                                                                             \\[1pt]
 43  &  age\_13a\_tau      &    Age  from method $13a_\tau$ [log(t/yr)]                                                                         \\[1pt]
 44  &  tau\_13a\_tau      &    $\tau$ from method $13a_\tau$ [Gyr]                                                                             \\[1pt]
 45  &  Av\_13a\_tau       &    $A_V$ from method $13a_\tau$ [mag]                                                                              \\[1pt]
 46  &  SFR\_13a\_tau      &    SFR from method $13a_\tau$ [$M_\odot/$yr]                                                                       \\[1pt]
 47  &  chi2\_13a\_tau     &    $\chi^2$ from method $13a_\tau$                                                                                 \\[1pt]
 48  &  age\_14a          &     Age  from method $14a$ [log(t/yr)]                                                                             \\[1pt]
 49  &  SFH\_14a          &     SFH from method $14a$ (1=constant; 2=linearly increasing; 3=delayed; 4=exponentially	decreasing))           \\[1pt]
 50  &  tau\_14a          &     $\tau$ from method $14a$ [Gyr]                                                                                 \\[1pt]
 51  &  EBV\_14a          &     E(B-V) from method $14a$ [mag]                                                                                 \\[1pt]
 52  &  SFR\_14a          &     SFR from method $14a$ [$M_\odot/$yr]                                                                           \\[1pt]
 53  &  q\_14a            &     fit quality (1=Best; 2=Good; 3=Bad/No solution) from method $14a$                                              \\[1pt]
 54  &  age\_6a\_tau\_neb   &   Age from method $6a^{\mathrm{NEB}}_\tau$ [log(t/yr)]                                                             \\[1pt]
 55  &  tau\_6a\_tau\_neb   &   $\tau$ from method $6a^{\mathrm{NEB}}_\tau$ [Gyr]                                                                \\[1pt]
 56  &  EBV\_6a\_tau\_neb   &   E(B-V) from method $6a^{\mathrm{NEB}}_\tau$ [mag]                                                                \\[1pt]
 57  &  SFR\_6a\_tau\_neb   &   SFR from method $6a^{\mathrm{NEB}}_\tau$ [$M_\odot/$yr]                                                          \\[1pt]
 58  &  met\_6a\_tau\_neb   &   Gas metallicity from method $6a^{\mathrm{NEB}}_\tau$ [$Z_\odot$]                                                 \\[1pt]
 59  &  extlw\_6a\_tau\_neb &   Extinction law from method $6a^{\mathrm{NEB}}_\tau$: 1=Calzetti; 2=SMC                                           \\[1pt]
 60  &  chi2\_6a\_tau\_neb  &   Reduced $\chi^2$ from method $6a^{\mathrm{NEB}}_\tau$                                                            \\[1pt]
 61  &  L1400\_6a\_tau\_neb &   Rest-frame luminosity at 1400 Angstrom  from method $6a^{\mathrm{NEB}}_\tau$ ($L_\nu(1400$\AA$)$ [erg/s/Hz]      \\[1pt]
 62  &  L2700\_6a\_tau\_neb &   Rest-frame luminosity at 2700 Angstrom  from method $6a^{\mathrm{NEB}}_\tau$ ($L_\nu(2700$\AA$)$ [erg/s/Hz]      \\[1pt]
 63  &  UMag\_6a\_tau\_neb  &   U rest-frame magnitude from method $6a^{\mathrm{NEB}}_\tau$  [AB system]                                         \\[1pt]
 64  &  BMag\_6a\_tau\_neb  &   B rest-frame magnitude  from method $6a^{\mathrm{NEB}}_\tau$ [AB system]                                         \\[1pt]
 65  &  VMag\_6a\_tau\_neb  &   V rest-frame magnitude from method $6a^{\mathrm{NEB}}_\tau$ [AB system]                                          \\[1pt]
 66  &  RMag\_6a\_tau\_neb  &   R rest-frame magnitude from method $6a^{\mathrm{NEB}}_\tau$ [AB system]                                          \\[1pt]
 67  &  IMag\_6a\_tau\_neb  &   I rest-frame magnitude from method $6a^{\mathrm{NEB}}_\tau$ [AB system]                                          \\[1pt]
 68  &  JMag\_6a\_tau\_neb  &   J rest-frame magnitude from method $6a^{\mathrm{NEB}}_\tau$ [AB system]                                          \\[1pt]
 69  &  KMag\_6a\_tau\_neb  &   K rest-frame magnitude from method $6a^{\mathrm{NEB}}_\tau$ [AB system]                                          \\[1pt]
 70  &  age\_6a\_deltau    &    Age from method $6a_\mathrm{del}$ [log(t/yr)]                                                                   \\[1pt]
 71  &  tau\_6a\_deltau    &    $\tau$ from method $6a_\mathrm{del}$ [Gyr]                                                                      \\[1pt]
 72  &  EBV\_6a\_deltau    &    E(B-V) from method $6a_\mathrm{del}$ [mag]                                                                      \\[1pt]
 73  &  SFR\_6a\_deltau    &    SFR from method $6a_\mathrm{del}$ [$M_\odot/$yr]                                                                \\[1pt]
 74  &  met\_6a\_deltau    &    Gas metallicity from method $6a_\mathrm{del}$ [$Z_\odot$]                                                       \\[1pt]
 75  &  extlw\_6a\_deltau  &    Extinction law from method $6a_\mathrm{del}$: 1=Calzetti; 2=SMC                                                 \\[1pt]
 76  &  chi2\_6a\_deltau   &    Reduced $\chi^2$ from method $6a_\mathrm{del}$                                                                  \\[1pt]
 77  &  L1400\_6a\_deltau  &    Rest-frame luminosity at 1400 Angstrom  from method $6a_\mathrm{del}$ ($L_\nu(1400$\AA$)$ [erg/s/Hz]            \\[1pt]
 78  &  L2700\_6a\_deltau  &    Rest-frame luminosity at 2700 Angstrom  from method $6a_\mathrm{del}$ ($L_\nu(2700$\AA$)$ [erg/s/Hz]            \\[1pt]
 79  &  UMag\_6a\_deltau   &    U rest-frame magnitude from method $6a_\mathrm{del}$  [AB system]                                               \\[1pt]
 80  &  BMag\_6a\_deltau   &    B rest-frame magnitude  from method $6a_\mathrm{del}$ [AB system]                                               \\[1pt]
 81  &  VMag\_6a\_deltau   &    V rest-frame magnitude from method $6a_\mathrm{del}$ [AB system]                                                \\[1pt]
 82  &  RMag\_6a\_deltau   &    R rest-frame magnitude from method $6a_\mathrm{del}$ [AB system]                                                \\[1pt]
 83  &  IMag\_6a\_deltau   &    I rest-frame magnitude from method $6a_\mathrm{del}$ [AB system]                                                \\[1pt]
 84  &  JMag\_6a\_deltau   &    J rest-frame magnitude from method $6a_\mathrm{del}$ [AB system]                                                \\[1pt]
 85  &  KMag\_6a\_deltau   &    K rest-frame magnitude from method $6a_\mathrm{del}$ [AB system]                                                \\[1pt]
 86  &  age\_6a\_invtau    &    Age from method $6a_\mathrm{inv}$ [log(t/yr)]                                                                   \\[1pt]
 87  &  tau\_6a\_invtau    &    $\tau$ from method $6a_\mathrm{inv}$ [Gyr]                                                                      \\[1pt]
 88  &  EBV\_6a\_invtau    &    E(B-V) from method $6a_\mathrm{inv}$ [mag]                                                                      \\[1pt]
 89  &  SFR\_6a\_invtau    &    SFR from method $6a_\mathrm{inv}$ [$M_\odot/$yr]                                                                \\[1pt]
 90  &  met\_6a\_invtau    &    Gas metallicity from method $6a_\mathrm{inv}$ [$Z_\odot$]                                                       \\[1pt]
 91  &  extlw\_6a\_invtau  &    Extinction law from method $6a_\mathrm{inv}$: 1=Calzetti; 2=SMC                                                 \\[1pt]
 92  &  chi2\_6a\_invtau   &    Reduced $\chi^2$ from method $6a_\mathrm{inv}$                                                                  \\[1pt]
  93  &  L1400\_6a\_invtau  &    Rest-frame luminosity at 1400 Angstrom  from method $6a_\mathrm{inv}$ ($L_\nu(1400$\AA$)$ [erg/s/Hz]            \\[1pt]
  94  &  L2700\_6a\_invtau  &    Rest-frame luminosity at 2700 Angstrom  from method $6a_\mathrm{inv}$ ($L_\nu(2700$\AA$)$ [erg/s/Hz]            \\[1pt]
  95  &  UMag\_6a\_invtau   &    U rest-frame magnitude from method $6a_\mathrm{inv}$  [AB system]                                               \\[1pt]
  96  &  BMag\_6a\_invtau   &    B rest-frame magnitude  from method $6a_\mathrm{inv}$ [AB system]                                               \\[1pt]
  97  &  VMag\_6a\_invtau   &    V rest-frame magnitude from method $6a_\mathrm{inv}$ [AB system]                                                \\[1pt]
  98  &  RMag\_6a\_invtau   &    R rest-frame magnitude from method $6a_\mathrm{inv}$ [AB system]                                                \\[1pt]
  99  &  IMag\_6a\_invtau   &    I rest-frame magnitude from method $6a_\mathrm{inv}$ [AB system]                                                \\[1pt]
 100  &  JMag\_6a\_invtau   &    J rest-frame magnitude from method $6a_\mathrm{inv}$ [AB system]                                                \\[1pt]
 101  &  KMag\_6a\_invtau   &    K rest-frame magnitude from method $6a_\mathrm{inv}$ [AB system]                                                \\[1pt]
 102  &  age\_10c\_dust     &    Age from method $10c^\mathrm{dust}$ [log(t/yr)]                                                                 \\[1pt]
 103  &  SFH\_10c\_dust     &    SFH from method $10c^\mathrm{dust}$: 1=exponentially decreasing ; 2=constant; 3=truncated; 4=no solution        \\[1pt]
 104  &  tau\_10c\_dust     &    $\tau$ from method $10c^\mathrm{dust}$ ($\tau$ =-99 if SFH=2 or 4) [Gyr]                                        \\[1pt]
 105  &  met\_10c\_dust     &    Gas metallicity from method $10c^\mathrm{dust}$ [$Z_\odot$]                                                     \\[1pt]
 106  &  age\_14a\_const    &    Age  from method $14a_\mathrm{const}$ [log(t/yr)]                                                               \\[1pt]
 107  &  EBV\_14a\_const    &    E(B-V) from method $14a_\mathrm{const}$ [mag]                                                                   \\[1pt]
 108  &  SFR\_14a\_const    &    SFR from method $14a_\mathrm{const}$ [$M_\odot/$yr]                                                             \\[1pt]
 109  &  q\_14a\_const      &    fit quality (1=Best; 2=Good; 3=Bad/No solution) from method $14a_\mathrm{const}$                                \\[1pt]
 110  &  age\_14a\_lin      &    Age  from method $14a_\mathrm{lin}$ [log(t/yr)]                                                                 \\[1pt]
 111  &  EBV\_14a\_lin      &    E(B-V) from method $14a_\mathrm{lin}$ [mag]                                                                     \\[1pt]
 112  &  SFR\_14a\_lin      &    SFR from method $14a_\mathrm{lin}$ [$M_\odot/$yr]                                                               \\[1pt]
 113  &  q\_14a\_lin        &    fit quality (1=Best; 2=Good; 3=Bad/No solution) from method $14a_\mathrm{lin}$                                  \\[1pt]
 114  &  age\_14a\_deltau   &    Age  from method $14a_\mathrm{del}$ [log(t/yr)]                                                                 \\[1pt]
 115  &  tau\_14a\_deltau   &    $\tau$ from method $14a_\mathrm{del}$ [Gyr]                                                                     \\[1pt]
 116  &  EBV\_14a\_deltau   &    E(B-V) from method $14a_\mathrm{del}$ [mag]                                                                     \\[1pt]
 117  &  SFR\_14a\_deltau   &    SFR from method $14a_\mathrm{del}$ [$M_\odot/$yr]                                                               \\[1pt]
 118  &  q\_14a\_deltau     &    fit quality (1=Best; 2=Good; 3=Bad/No solution) from method $14a_\mathrm{del}$                                  \\[1pt]
 119  &  age\_14a\_tau      &    Age  from method $14a_\tau$ [log(t/yr)]                                                                         \\[1pt]
 120  &  tau\_14a\_tau      &    $\tau$ from method $14a_\tau$ [Gyr]                                                                             \\[1pt]
 121  &  EBV\_14a\_tau      &    E(B-V) from method $14a_\tau$ [mag]                                                                             \\[1pt]
 122  &  SFR\_14a\_tau      &    SFR from method $14a_\tau$ [$M_\odot/$yr]                                                                       \\[1pt]
 123  &  q\_14a\_tau        &    fit quality (1=Best; 2=Good; 3=Bad/No solution) from method $14a_\tau$                                          \\[1pt]
\hline
\end{longtable*}

\end{document}